\documentclass[prb,twocolumn,superscriptaddress,showpacs]{revtex4}

\usepackage{amsmath}
\usepackage{amsfonts}

\usepackage{graphicx}
\usepackage{times}
\usepackage{pstricks}

\newcommand{\be}{\begin{equation} }
\newcommand{\ee}{\end{equation} }
\newcommand{\ba}{\begin{eqnarray} }
\newcommand{\ea}{\end{eqnarray} }

\begin{document}


\title{Quasi-Topological Phases of Matter and Topological Protection}

\author{Parsa Bonderson}
\affiliation{Station Q, Microsoft Research, Santa Barbara, CA 93106-6105}
\author{Chetan Nayak}
\affiliation{Station Q, Microsoft Research, Santa Barbara, CA 93106-6105}
\affiliation{Department of Physics, University of California, Santa Barbara, CA 93106}

\date{\today}

\begin{abstract}
We discuss systems which have some, but not all of the hallmarks of topological phases.
These systems' topological character is not fully captured by a local order parameter,
but they are also not fully described at low energies by topological quantum field theories.
For such systems, we formulate the concepts of quasi-topological phases
(to be contrasted with true topological phases),
and symmetry-protected quasi-topological phases.
We describe examples of systems in each class and discuss the
implications for topological protection of information and operations.
We explain why topological phases and quasi-topological
phases have greater stability than is sometimes appreciated.
In the examples that we discuss, we focus on Ising-type
(a.k.a. Majorana) systems particularly relevant to recent theoretical advances
and experimental efforts.
\end{abstract}

\maketitle


\section{Introduction and Summary}

Topological phases of matter~\cite{Wen90a} have been the subject of intense
interest because they represent new phases of matter which may
support anyons~\cite{Leinaas77,Wilczek82b,Goldin85,Fredenhagen89,Froehlich90} or their generalizations
and could have potential use for fault-tolerant quantum
computation~\cite{Kitaev97,Freedman98,Freedman03b,Nayak08}.
When a system that is in a topological phase is on a manifold $\Sigma$,
it has a set of orthonormal ``pseudo-ground'' states $|a\rangle$, $a=1,2,\ldots,{N_\Sigma}$,
which includes the absolute ground state of the system.
These pseudo-ground states satisfy
\begin{equation}
\delta {E_0} \equiv
\max( |{E_a}-E_{a'}| )=O(e^{-L/\xi})
\end{equation}
(which is why we do not simply refer to them all as ground states),
where $L$ is the system size and $\xi$ is the correlation length of the system,
which is finite in the limit $L\rightarrow \infty$.
These states are separated from the rest of the spectrum by
an energy gap $\Delta$ that remains non-zero in the limit that the
system size $L\rightarrow\infty$.
Thus, these pseudo-ground states form an
${N_\Sigma}$-dimensional degenerate ground state subspace
in the thermodynamic limit. In a topological phase, $N_\Sigma$ depends only
on the topological configuration of the system, e.g. genus, number of boundaries,
and boundary conditions. Furthermore,
for any local operator $\phi$,
\begin{equation}
\label{eqn:top-phase-def}
\langle a| \phi |b\rangle = C \delta_{ab} + O(e^{-L/\xi})
\end{equation}
where $C$ is a constant independent of $a,b$.
Since the Hamiltonian is a sum of local operators, the statement that
the energy splitting between ground states is $O(e^{-L/\xi})$ is redundant,
since it follows from Eq.~(\ref{eqn:top-phase-def}).
Perturbations at frequencies much smaller than $\Delta$ (including, of course,
static perturbations) can be described effectively at long times as
perturbations acting entirely within the pseudo-ground state subspace ${\cal H}_{0} \equiv \text{span}\left\{ |1\rangle, \ldots, | N_\Sigma \rangle \right\}$.
If they are local, such perturbations
cannot, up to exponential accuracy, cause transitions between
states, according to Eq.~(\ref{eqn:top-phase-def}).
At most, the state of the system can acquire an overall phase.
Thus, quantum information encoded in this pseudo-ground state subspace is
``topologically protected'' at zero temperature
in the thermodynamic limit. At non-zero temperature, the error rate
$\Gamma$ is exponentially suppressed by the gap, $\Gamma \sim e^{-\Delta/T}$.

A system that satisfies the properties listed above
is described by a topological quantum field theory (TQFT)~\cite{Witten89}. In particular, the low-energy degrees of freedom at long distances are described by an effective theory in which the degrees of freedom are entirely topological, i.e. they depend only on global properties of the system (with no dynamical degrees of freedom nor geometry/metric dependence).
A TQFT encapsulates the properties of the low-energy states of
a topological phase on an arbitrary manifold in the limit
that $\xi \rightarrow 0$ and $\Delta\rightarrow\infty$
or, equivalently, for $1/\xi$ and $\Delta$ finite (and positive), but
at length scales $x\gg\xi$ and energy scales $\delta {E_0}\ll\omega\ll\Delta$.
These inequalities must be satisfied for the following reasons.
The topological properties dominate short-ranged interactions
only at length scales larger than the correlation length $\xi$. Moreover,
all processes must be performed at frequencies large enough that
the system does not notice the finite-sized splitting between pseudo-ground states,
but small enough that quasiparticles are not excited.
Topological phases and TQFTs can exist in any spacetime dimension, but we will usually
focus on the case of two spatial dimensions (plus one time dimension) in this paper.
A TQFT also describes the topological properties (e.g. braiding statistics) of quasiparticles, which appear in the theory either through Wilson lines (which represent quasiparticles' worldlines) and
higher-dimensional analogues or, equivalently, by considering ground states
on manifolds with punctures/boundaries.

The latter point of view -- that topologically non-trivial excited states can be understood
by studying pseudo-ground states in the presence of punctures
(with appropriate boundary conditions at those punctures
corresponding to the type of quasiparticle) -- is very
similar to Laughlin's original construction of fractionally-charged
quasiparticles in the quantum Hall effect by threading flux through
the hole in an annulus~\cite{Laughlin83,Tao84}.
Thus, the notion of topological degeneracy of the pseudo-ground state subspace
naturally includes the degenerate (nonlocal) anyonic state space of non-Abelian anyons~\cite{Goldin85,Fredenhagen89,Froehlich90}.
In this case, the corrections in Eq.~(\ref{eqn:top-phase-def}) will be $O(e^{-r/\xi})$, where $r$ is the distance separating boundaries/quasiparticles. These exponentially-suppressed corrections are due to non-universal corrections to the topological theory
(i.e. physics beyond the TQFT) that result in tunneling of topological excitations either around non-trivial cycles or between boundaries/quasiparticles. These processes generically result in energy splittings that fully resolve the degeneracies~\cite{Bonderson09b}.
Transformations of the manifold that leave
the topology unchanged are realized by unitary transformations
acting on the (finite-dimensional) Hilbert space of pseudo-ground states.
Transformations that are continuously connected to the identity
are realized trivially, but those which are not connected to the identity
are realized by non-trivial unitary transformations.
For instance, on the plane with $n$ punctures,
these unitary transformations form a representation of the $n$-strand
braid group. Two-dimensional [($2+1$)-D] TQFTs are very tightly constrained by
self-consistency (see, for instance, Appendix E of Ref.~\onlinecite{Kitaev06a}).
Therefore, if a system is in a topological
phase, as defined above, then its properties can be predicted by
a discrete mathematical structure with very few ``free'' parameters. Such
systems have been proven to be stable against small perturbations. In particular, ``Ocneanu rigidity'' establishes that the algebraic structure of a TQFT is rigid~\cite{Ocneanu-unpublished,Etingof05}, which implies that the topological properties are unchanged by small perturbations of the Hamiltonian that do not close the gap. Moreover, it has also been established that the energy gap in a topological phase of any quantum spin Hamiltonian does not close for sufficiently small perturbations~\cite{Klich10,Bravyi10,Bravyi11}.

Many model Hamiltonians are in topological phases
and satisfy the definition given above. However, this
begs the question: do actual physical systems ever satisfy
the definition of a topological phase?
Real solids are never fully gapped. At the very least,
they will have gapless phonons and, usually,
gapless photons. These gapless excitations can modify
some of the predictions of TQFTs while leaving others intact.
Therefore, it is important to exercise caution in applying
the abstract concept of a topological phase and its mathematical
formulation as a TQFT to real systems.

Phonons, photons, and other gapless excitations are usually not included
in model Hamiltonians because they are viewed as inessential
to the basic physics. One might adopt the perspective
that the ``system'' is some subset of the degrees of freedom in a solid
(only the spins, for instance) which satisfies the definition
of a topological phase while the remaining degrees of freedom
constitute the ``environment'' which perturbs the system.
From this perspective, we can state our basic problem as:
what happens when a system that is in a topological phase
is perturbed by coupling it (either at its boundary; at an interior point;
or even everywhere in its interior) to another system that has gapless excitations?
The ``other system'' may be intrinsic to the solid (e.g. phonons);
partly intrinsic and partly extrinsic (e.g. photons); or even completely
external (e.g. a metallic lead or gate). This situation is not addressed by
studies of perturbations which act entirely within the Hilbert
space of the ``system''~\cite{Klich10,Bravyi10,Bravyi11}.
If, instead, we view the entire solid
with all of its degrees of freedom as our system,
we can, alternatively, phrase the problem as:
can a system with gapless excitations retain some
aspects of topological phases that remain stable against
small perturbations, and how should we classify such systems?
This formulation includes a somewhat broader class of systems,
since it includes not only those which can be divided (however unnaturally)
into a gapped topological phase and a gapless ``environment,''
but also systems which necessarily have gapless excitations
as part and parcel of their topological character. As we will see,
such systems can be divided into quasi-topological phases
and symmetry-protected quasi-topological phases, which have different
degrees of stability to perturbations. In this paper, we will discuss these
classes in detail, together with examples, and the extent to which they are
robust and can protect quantum information.
True topological phases are reviewed
in many papers (see, for instance, Ref.~\onlinecite{Nayak08}
and references therein), but we will briefly recapitulate their properties and the extent to which
they are robust before discussing the more subtle cases
of the various types of quasi-topological phases. For other approaches to the stability
of topological properties of some widely-studied many-body systems,
see Refs.~\onlinecite{Nussinov08,Goldstein11,Budich12,Cheng11}.

It is useful to summarize by stating the definitions here.
We have already stated the definition of a topological phase.
It is worth noting that topological phases can also be defined
for systems in the presence of disorder, by appropriately modifying
the definition, which we also include.

{\bf Definition (Topological Phase with Disorder)}:
A system is in a topological phase
if there is an energy $\Delta$ and a length $\xi$,
which, respectively, have a strictly positive limit and a finite,
non-negative limit as $L\rightarrow \infty$, such that the following properties hold.
On a manifold $\Sigma$, there is a set of orthonormal energy eigenstates $|a\rangle$, where $a\in \{1,2,\ldots,{N_\Sigma}\}$, with energies $E_a < \Delta$,
including the absolute ground state $|1\rangle$  of the system
(which, without loss of generality, we take to have energy $E_1=0$).
$N_\Sigma$ depends only on the topological configuration of the system.
For any local operator $\phi$,
\begin{equation}
\label{eqn:top-phase-def-localized-preview}
\langle a| \phi |b\rangle = C \delta_{ab} + O(e^{-L/\xi})
,
\end{equation}
where $C$ is a constant independent of $a,b \in \{1,2,\ldots,{N_\Sigma}\}$. All other states with energies less than $\Delta$ are localized excitations
of states in ${\cal H}_{0} \equiv \text{span}\left\{ |1\rangle, \ldots, | N_\Sigma \rangle \right\}$.

The notion of localized excitations of states will be defined and explained in Section~\ref{sec:dirty}.

We note that it follows (nontrivially) from this definition that the pseudo-ground state subspace of such a system is described by a TQFT (as does that of a non-disordered topological phase).

{\bf Definition (Quasi-Topological Phase)}:
A system is in a quasi-topological phase if there is an energy $\Delta$ and a length $\xi$,
which, respectively, have a strictly positive limit and a finite,
non-negative limit as $L\rightarrow \infty$, such that the following properties hold.

(i) On a manifold $\Sigma$, there is a set of orthonormal energy eigenstates $|a\rangle$, where $a\in \{1,2,\ldots,{N_\Sigma}\}$, including the absolute ground state $|1\rangle$ (with energy $E_1=0$), such that
for any local operator $\phi$,
\begin{equation}
\label{eqn:strong-exponential}
\langle {a}| \, \phi \, | {b}\rangle = {C} \delta_{ab} + O(e^{-L/\xi} )
\end{equation}
for $a, b \in \{1,2,\ldots,{N_\Sigma}\}$, where ${C}$ is independent of $a, b$.
The Hamiltonian is a particular local operator, so it follows that
\begin{equation}
\delta {E_0} \equiv \text{max}(|{E_{a}}|) = O(e^{-L/\xi})
\end{equation}
for $a\in \{1,2,\ldots,{N_\Sigma}\}$, where $E_{a}$ is the energy of the state
$|{a}\rangle$.

(ii) $N_\Sigma$ depends only on the topological configuration of the system.

(iii) We define
\begin{equation}
{\Omega} \equiv \text{min}(|{E_\chi}-E_{a}|)
\end{equation}
for any $a\in \{1,2,\ldots,{N_\Sigma}\}$
and $|\chi\rangle \notin \mathcal{H}_{0} \equiv \text{span}\left\{ |1\rangle, \ldots, | N_\Sigma \rangle \right\}$.
Then, there must exist a finite $z>0$ and an $L$-independent $\kappa>0$ such that
\begin{equation}
\Omega\sim\kappa L^{-z}.
\end{equation}

(iv) There are no states with energy less than $\Delta$ that have topologically nontrivial excitations.

The notion of topologically nontrivial excitations will be defined and explained in Section~\ref{sec:TQFTs}.

When we consider different examples of quasi-topological phases,
we will note that some of them have the property that,
at energy scales $E$ between the pseudo-ground states
and the gapless excitations, $\delta {E_0} \ll E \ll \Omega$,
they exhibit the characteristic behavior of a topological phase.
For any fixed large system size $L$, there is no difference
between such a quasi-topological phase and a topological
phase, with $\Omega$ playing the role of the gap. Furthermore,
at energies less than $\Delta$, the topological and gapless
degrees of freedom are essentially decoupled.
We will adopt the heuristic
term ``strong quasi-topological'' for such phases.
We note that if $z=0$, then the system is in a topological phase (and hence is also ``strong'').

{\bf Definition (Symmetry-Protected Quasi-Topological Phase):}
A system is in a symmetry-protected quasi-topological phase
associated with a symmetry group $G$ if there is an energy $\Delta$
and a length scale $\xi$ which, respectively, have a strictly positive limit and a finite, non-negative limit as $L\rightarrow \infty$, such that the following properties hold.

(i) On a manifold $\Sigma$, there is a set of orthonormal energy eigenstates $|a\rangle$, where $a\in \{1,2,\ldots,{N_\Sigma}\}$, including the absolute ground state $|1\rangle$ (with energy $E_1=0$), such that
for any local operator $\phi$ that is invariant under
the group $G$
\begin{equation}
\label{eqn:sp-exponential}
\langle {a}| \, \phi \, | {b}\rangle = {C} \delta_{ab} + O(e^{-L/\xi} )
\end{equation}
for $a, b \in \{1,2,\ldots,{N_\Sigma}\}$, where ${C}$ is independent of $a, b$.
For operators which are not invariant under $G$, the corrections will be bounded below by $\alpha L^{-\zeta}$ for some nonzero coefficient $\alpha$ and finite non-negative exponent $\zeta$.

The Hamiltonian is a particular local operator that is invariant under
the group $G$, so it follows that
\begin{equation}
\delta {E_0} \equiv \text{max}(|{E_{a}}|) = O(e^{-L/\xi})
\end{equation}
for $a\in \{1,2,\ldots,{N_\Sigma}\}$, where $E_{a}$ is the energy of the state
$|{a}\rangle$.

(ii) $N_\Sigma$ depends only on the topological configuration of the system.

(iii) We define
\begin{equation}
{\Omega} \equiv \text{min}(|{E_\chi}-E_{a}|)
\end{equation}
for any $a\in \{1,2,\ldots,{N_\Sigma}\}$
and $|\chi\rangle \notin \mathcal{H}_{0} \equiv \text{span}\left\{ |1\rangle, \ldots, | N_\Sigma \rangle \right\}$.
Then, there must exist a finite $z>0$ and an $L$-independent $\kappa>0$ such that
\begin{equation}
\Omega\sim\kappa L^{-z}.
\end{equation}

(iv) There are no states with energy less than $\Delta$ that have topologically nontrivial excitations.

The organization of the paper is as follows:

In Section~\ref{sec:thermo-limit}, we define various length scales of system to make precise the notion of keeping these lengths long for the purposes of topological protection and for taking thermodynamic limits.

In Section~\ref{sec:clean}, for the sake of concreteness, we give two
examples of topological phases of a physical system.
We explain why it is nearly impossible to realize such a system
in nature. If one could be realized, however,
it would be incredibly stable against all types of weak local perturbations
and, therefore, would topologically protect quantum information.

In Section~\ref{sec:dirty}, we consider the modification of topological phases for systems that include disorder. We provide examples and a precise definition for a {\it localized excitation} of a state and the resulting definition of a topological phase with disorder.

In Section~\ref{sec:TQFTs}, we explain how the structure of a TQFT arises in the low energy spectrum of topological phases.

In Section~\ref{sec:QTP}, we consider quasi-topological phases. In Section~\ref{sec:strong},
we consider a few examples that will prove to be strong quasi-topological phases.
These phases represent the least impactful modification to a topological phase by the inclusion of gapless degrees of freedom, since they essentially preserve the full TQFT structure. We show that strong quasi-topological phases are nearly as robust as topological phases and protect quantum information nearly as well. The class of strong quasi-topological phases includes all topological phases as a subclass.

In Section~\ref{sec:definition}, we discuss and explain the general definition of a quasi-topological phase (given above). Since virtually all physical systems have gapless excitations of some kind, this is the generic case of a putative topological phase.
A quasi-topological phase does not, in general, exhibit the full TQFT structure of a topological phase (as does a strong quasi-topological phase).
Nevertheless, it protects quantum information against a wider class of perturbations than is often appreciated, including coupling to external gapless electrons. The class of quasi-topological phases includes all strong quasi-topological phases as a subclass.

In Section~\ref{sec:weak}, we consider quantum Hall states in electronic systems and demonstrate that they are quasi-topological phases which are not strong. We find that some quantum Hall states preserve a ``strong subsector,'' i.e. a subsector which retains its TQFT structure (as in a strong quasi-topological phase).

In Section~\ref{sec:electron-parity}, we define and discuss
electron-parity protected and other symmetry-protected
quasi-topological phases, for which topological superconductors are an example.
These phases behave as quasi-topological phases, as long as the symmetry
is preserved. In particular, for superconductors, electron parity
must be preserved, which necessitates isolation from gapless electrons.

In Section~\ref{sec:protection}, we discuss how these various classes of quasi-topological phases can be used to protect quantum information.
We compare the ability to topologically protect quantum information for these various classes and contrast this to the protection given by ``conventional'' qubits.

Finally, in Section \ref{sec:discussion}, we discuss some of the implications.

\section{Macroscopic Length Scales of the System}
\label{sec:thermo-limit}

In this paper, we will be concerned with the scaling of various properties
as the system size is taken to infinity. In discussing this limit,
we will make reference to three classes of length scales characterizing the macroscopic configuration of the system
on the manifold $\Sigma$ (which we assume to be path-connected and compact): ``system sizes,'' ``separations between boundaries,'' and ``winding lengths.''

We define the size ${\cal L}_{\mathcal{M}}$ of a manifold $\mathcal{M}$ to be
\begin{equation}
{\cal L}_{\mathcal{M}} = \max\limits_{x,y \in \mathcal{M}} \left\{ d(x,y) \right\}
\end{equation}
where $d(x,y)$ is the length of the shortest curve in $\mathcal{M}$ connecting points $x$ and $y$.
For the system of interest on manifold $\Sigma$, we will refer to ${\cal L}_{\Sigma}$ as the system size.
We often consider one-dimensional or two-dimensional systems that are embedded in three-dimensional systems, in which case it may be useful to consider other length scales, such as
the size of the three-dimensional manifold
in which the manifold $\Sigma$ is embedded. We will define these as needed.

When the manifold $\Sigma$ has $n$ distinct connected boundary components $\partial_1 \Sigma , \ldots , \partial_n \Sigma$,
we define the separation between boundaries $r_{ij}$ to be
\begin{equation}
r_{ij} = \min\limits_{\substack{x \in \partial_i \Sigma \\ y \in \partial_j \Sigma }} \left\{ d(x,y) \right\}
.
\end{equation}

To define the winding lengths of the system, we consider the fundamental group $\pi_{1}(\Sigma)$,
each element of which is an equivalence class of closed paths in $\Sigma$ that can be continuously deformed into each other.
For each element $p \in \pi_{1}(\Sigma)$ other than the identity $e$,
we define $\ell_p$ to be the length of the shortest path that
belongs to the equivalence class that defines $p$. (We exclude the identity element because it would have length $\ell_e =0$.) There is a large amount of redundancy in using all the winding lengths $\ell_p$ to characterize the system's winding lengths and we will typically only be interested in the shortest few of these, which will contribute most significantly to corrections. Moreover, some of these winding lengths are inconsequential. In particular, the winding length around a single boundary component does not play a role in splitting degeneracies, since the boundary conditions imposed on the system determines a definite boundary state.

In general, there may be many $\ell_p$ and $r_{ij}$ that may be treated, along with ${\cal L}_\Sigma$, as independent parameters. In considering how the system behaves in the infinite-size limit, it is most natural to take all lengths ${\cal L}_{\Sigma},r_{ij},\ell_p \rightarrow \infty$ simultaneously. However, varying these different parameters (${\cal L}_{\Sigma}$, $r_{ij}$, and $\ell_p$) independently will generally produce distinct effects, possibly upon different subspaces of the low-energy states. To simplify notation, we will refer to all these length scales (${\cal L}_{\Sigma}$, $r_{ij}$, and $\ell_p$) of the system collectively as $L$, with the understanding that one should appropriately resolve effects due to a particular length scale based on context. For example, ``$L\rightarrow \infty$'' means all the lengths are taken to infinity, while ``$O(e^{-L/\xi})$ corrections'' means varying any one of ${\cal L}_{\Sigma}$, $r_{ij}$, or $\ell_p$ will, respectively, result in a $O(e^{-{\cal L}_{\Sigma}/\xi})$, $O(e^{-\ell_{p}/\xi})$, or $O(e^{-r_{ij}/\xi})$ correction on a particular subspace of the Hilbert space.
The corresponding lengths for a submanifold $\mathcal{M}\subset \Sigma$ will
similarly be denoted collectively by $L_\mathcal{M}$.

We will assume that the $L\rightarrow \infty$ limit is taken with microscopic quantities
such as the particle density or lattice spacing held fixed, so that the number of particles
or lattice sites also goes to infinity.

\section{Topological Phases}
\label{sec:true}

\subsection{Clean Systems}
\label{sec:clean}



\subsubsection{3D Superconductor}
\label{subsec:3D-super}

Although we will focus primarily on two-dimensional topological phases in this paper,
we begin with a three-dimensional one because, as we shall see,
it is easier to construct a topological phase in three dimensions. Our example is
a clean three-dimensional $s$-wave superconductor. The electronic excitation spectrum
has a gap $\Delta$, and photons are gapped by the Anderson-Higgs mechanism.
However, the system has gapless phonons. Therefore, in order to have a system
which is truly fully-gapped, we will have to assume that the atomic masses
are infinite so that there are no phonons at low energies. (Since superconductivity
is usually due to the electron-phonon interaction, we will have to assume that
the superconductivity is due to some other short-ranged attractive force
which does not require gapless excitations.) Such a system is seemingly
completely boring at low energies (apart from its ability to carry current without
dissipation). However, its topological properties are non-trivial.
Suppose we assume that that the entire universe
is filled by this material and that the universe (which we shall assume to have
three spatial dimensions) is a three-torus, $T^3$. Then, the pseudo-ground states
are $8$-fold degenerate, since fermionic excitations can have either periodic
or anti-periodic boundary conditions around each of the three generators
of $T^3$. A local operator can neither measure fermionic boundary conditions
nor change them. In order to measure a boundary condition, a fermion would have to go all
the way around one of the generators of the torus. This can happen virtually:
a virtual pair can be created, one can go around the torus, and they can annihilate.
Such a process leads to an energy splitting $\sim e^{-L/\xi}$.
In order to change the boundary condition, a flux tube which winds all the
way around one of the generators of the torus (say, the $x$-direction, for concreteness)
would have to slide in the $y$- or $z$-direction until it has gone all the way around that direction. Suppose a pair of such flux tubes is created virtually,
such a sliding process occurs, and then, finally, the flux tubes annihilate.
Then one ground state will tunnel into another.
This causes a splitting $\sim e^{-a L^2}$: each flux tube costs an energy
$\propto L$ and the time required for a flux tube to
slide all the way around the torus is $\propto L$, so the action
for such a process is $\propto L^2$.
Thus, the $8$ pseudo-ground states (which are separated from the rest of the spectrum
by an energy gap) have an energy splitting $\sim e^{-L/\xi}$,
and local operators cannot distinguish them. In other words, it is a topological phase.

\subsubsection{Kitaev's Honeycomb Model Implemented Ideally}

To construct a concrete example of a two-dimensional topological phase,
we begin with Kitaev's honeycomb lattice model~\cite{Kitaev06a}. The Hamiltonian is:
\begin{multline}
\label{eqn:Kitaev-honeycomb}
H = -{J_x}\!\!\!\sum_{x-\text{links}} \!\!{S^x_i}{S^x_{j}}
-{J_y}\!\!\!\sum_{y-\text{links}}\!\! {S^y_i}{S^y_{j}}
-{J_z}\!\!\!\sum_{z-\text{links}} \!\!{S^z_i}{S^z_{j}}\,\\
+ \frac{J'}{2}  \left( \sum_{ \vec{r}_{ij} =  \hat{x},  \vec{r}_{kj}=\hat{y}} \!\!\!
S^x_i S^z_j S^y_k \,+ \,
\mbox{rot. symm. equivalents}\right )
\end{multline}
where the $z$-links are the vertical links on the honeycomb lattice,
and the $x$ and $y$ links are the other two types of links on the honeycomb lattice,
which are at angles $\pm\pi/3$ from the vertical at one sublattice (and $\pm 2 \pi/3$ at the other);
$\hat{x}, \hat{y}, \hat{z}$ are the unit vectors in the directions of the corresponding links.

In order to ensure that no gapless degrees of freedom exist in the system, one must be (unrealistically) careful in designing the system.
To avoid the presence of (gapless) photons, one may embed the honeycomb lattice
inside a three-dimensional superconductor. (Note that, in Kitaev's paper~\cite{Kitaev06a}, the three-spin terms in
Eq.~(\ref{eqn:Kitaev-honeycomb}) were generated by a magnetic field, which we
cannot have if the lattice is embedded in a three-dimensional superconductor. However, we will
assume that these terms are simply present due to some other microscopic reason.
Alternatively, we could replace the Hamiltonian of Eq.~(\ref{eqn:Kitaev-honeycomb}) with
the Hamiltonian described in Ref.~\onlinecite{Yao07}, where each vertex is replaced with a triangle of vertices, which spontaneously breaks time-reversal symmetry and removes the need for a magnetic field to provide a gapped non-Abelian topological phase.)
One must also assume that there are no
impurities in the superconductor and that the atoms
at the vertices of the honeycomb lattice and in the superconductor
have infinite mass so that there are no vibrational excitations (phonons) of the lattice.
(We can assume that the superconductivity is caused by a strong short-ranged
attractive force rather than the electron-phonon interaction.)

This system has two possible topological phases. The ``$A$ phase'' occurs for
$|J_x| + |J_y| < |J_z|$ (and for cyclic permutations of the values of $J_x, J_y, J_z$) and
$J'$ small or zero. The ``$B$ phase'' occurs for $|J_x| + |J_y| \geq |J_z|$,
$|J_x| + |J_z| \geq |J_y|$, $|J_z| + |J_y| \geq |J_x|$ and $J'$ non-zero.
In both phases, the system has a gap $\Delta$ (which is a function of $J, J'$),
below which there are no excitations above the pseudo-ground state space. In particular, there are no phonons
and photons are gapped by the three-dimensional superconductor.
The spin excitations on the honeycomb lattice are gapped.
We emphasize, as motivation for the rest of the paper, the artificial and physically unrealistic conditions required to ensure that the system is fully gapped.

The system described here is truly in a topological phase in either the $A$ or $B$ phase,
and it satisfies Eq.~(\ref{eqn:top-phase-def}). The low-energy state space of the $A$ phase is described by
a TQFT which is known, alternatively, as the ``toric code,''
the quantum double $D({\mathbb{Z}_2})$, or $\mathbb{Z}_2$ gauge theory.
This phase has four quasiparticle types: $I, e, m, \psi$.
These four quasiparticle types can be understood as follows.
In the limit that $|J_z| \gg |J_x| + |J_y|$, the two spins connected by a
link in the $z$-direction will be locked together.
Therefore, we can combine them into a single spin-$1/2$ which we call $\sigma_i$,
and we can merge the corresponding two lattice points into a single lattice point.
The resulting points lie on the midpoints of a square lattice,
and the spins are governed by the following effective Hamiltonian~\cite{Kitaev97}:
\begin{equation}
\label{eqn:toric-code}
H_{\rm TC} =  - J_{\rm eff} \sum_{{\rm vertices}\,\, v} \prod_{i\in {\cal N}(v)} \sigma_i^z -
J_{\rm eff}\sum_{{\rm plaquettes}\,\, p} \prod_{i\in p} \sigma_i^x
\end{equation}
The effective spins in the immediate neighborhood ${\cal N}(v)$ of any vertex $v$ of the square lattice
interact according to the first term in Eq.~(\ref{eqn:toric-code}), where
$J_{\rm eff}={J_x^2}{J_y^2}/(16{|J_z|^3})$. In the ground state, $\prod_{i\in {\cal N}(v)} \sigma_i^z =1$
for all $v$ and $\prod_{i\in p} \sigma_i^x=1$ for all $p$. The $I$ quasiparticle is simply the
absence of any other quasiparticle or, in other words, when the system is in its ground state.
The $e$ quasiparticles are vertices $v$ at which $\prod_{i\in {\cal N}(v)} \sigma_i^z = -1$.
The $m$ quasiparticles are plaquettes $p$ at which $\prod_{i\in p} \sigma_i^x = -1$.
The $\psi$ quasiparticles are plaquettes $p$ at which $\prod_{i\in p} \sigma_i^x = -1$,
one of whose vertices $v$ satisfies $\prod_{i\in {\cal N}(v)} \sigma_i^z = -1$.

The pseudo-ground state degeneracy
on the torus is 4-fold ($N_{T^2}=4$) in the thermodynamic limit. The fusion and braiding properties
(see Appendix~\ref{sec:AlgTQFT}) are determined by the fusion algebra
\begin{equation}
\begin{array}{rrrrrrr}
I \times I = I, & & I \times e = e, & & I \times m = m, & & I \times \psi = \psi, \\
e\times e =I, & & e \times m = \psi, & & e\times \psi =m, & & m \times m = I, \\
m \times \psi = e, & & \psi \times \psi = I , & & & &
\end{array}
\end{equation}
together with the associativity $F$-symbols and braiding $R$-symbols.

The $F$-symbols are defined by the statement that
if the system is in the state in which quasiparticle types
$a$, $b$, and $c$ fuse to $d$ such that $a$ and $b$
have definite fusion channel $e$, then the system has amplitude
$[F^{abc}_d]_{ef}$ to be in the state in which $b$ and $c$ fuse to
$f$ [see Eq.~(\ref{eq:F_move})]. In the toric code (i.e. the $D(\mathbb{Z}_2)$ phase),
all of the $F$-symbols are trivial, i.e. they are $1\times 1$ matrices equal to $1$. (When a fusion process is prohibited by the fusion algebra, the state space is empty, so the corresponding $F$-symbols are trivially equal to $0$ and conventionally left implicit.)

The $R$-symbols are defined such that state of two quasiparticles of types $a$ and $b$ in the definite fusion channel $c$ acquires the phase
$R^{ab}_c$ when the two quasiparticles
are exchanged in a counter-clockwise manner [see Eq.~(\ref{eq:R_move})]. For Ising anyons, the only non-trivial $R$-symbols are:
\begin{equation}
R^{\psi\psi}_I = -1, \qquad \quad R^{m e}_{\psi} = -1 .
\end{equation}
All other $R$-symbols are equal to $1$. The apparent asymmetry
between $R^{e m}_{\psi}=1$ and $R^{m e}_{\psi}=-1$ is not observable
and can be changed by a gauge transformation.
Only the combination $R^{m e}_{\psi} R^{e m}_{\psi}$ is observable.

The $B$ phase is described by the so-called ``Ising TQFT.''
This phase has three quasiparticle types: $I$, $\sigma$, and $\psi$.
These particles can be understood as follows. As usual, the $I$ quasiparticle is
the absence of any other quasiparticle or, simply, when the system is in its ground state.
The following operator commutes with the Hamiltonian:
\be
\label{Eq_Pcons}
W_p = \prod_{i=1}^6 S^{e(i)}(i)
\ee
where the product is over the vertices of a hexagonal plaquette $p$,
and $e(i) = z$ for a vertex which sits between $x$ and $y$ links on the plaquette,
$y$ for a vertex  which sits between $x$ and $z$
links on a plaquette, and $x$ for a vertex which sits between $y$ and $z$ links on a plaquette.
In the ground state, ${W_p} = \frac{1}{2^6}$. A $\sigma$ quasiparticle is a plaquette at which
${W_p} = - \frac{1}{2^6}$. $\psi$ is a neutral fermionic excitation. A pair of
$\psi$ quasiparticles can
be created at neighboring sites $i$ and $j$, connected by an $x$-link,
if we act on the ground state with $S^x_i S^x_j$.

The degeneracy of the pseudo-ground state space on the torus is 3-fold ($N_{T^2} = 3$) in the thermodynamic limit.
The fusion and braiding properties (see Appendix~\ref{sec:AlgTQFT}) are determined by the fusion algebra
\begin{equation}
\label{eq:Ising_fusion_rules}
\begin{array}{rrrrr}
I \times I = I, & & I \times \psi = \psi, & & I \times \sigma = \sigma, \\
\psi\times \psi =I, & & \psi \times \sigma = \sigma, & & \sigma \times \sigma = I+ \psi,
\end{array}
\end{equation}
together with the associativity $F$-symbols and braiding $R$-symbols.
As a result of these fusion rules, the plane with $2n$ $\sigma$-charged punctures
whose combined topological charge is trivial
has $2^{n-1}$-fold degeneracy in the thermodynamic limit.
Consequently, $\sigma$ particles can be understood as supporting Majorana
zero modes -- a {\it pair} of Majorana zero modes is an ordinary fermionic
level, which has two states, occupied or unoccupied. Therefore, we will often
use the terms $\sigma$ particle and Majorana zero modes interchangeably.
However, when the distinction is important, we will use `$\sigma$ particle'
in topological phases or quasi-topological phases
and `Majorana zero modes' in symmetry-protected quasi-topological phases.
Note, however, that we will use the term `Majorana fermion' to refer to
$\psi$ particles -- the defects that carry Majorana fermion zero modes have
different topological charge from gapped Majorana fermion excitations.

For Ising anyons, the nontrivial $F$-symbols are:
\begin{eqnarray}
F^{\psi\sigma\psi}_\sigma &=& F^{\sigma\psi\sigma}_\psi = -1 \cr
\left[F^{\sigma\sigma\sigma}_\sigma \right]_{ef} &=&
\frac{1}{\sqrt{2}}\left[
\begin{array}{rr}
1 & 1 \\
1 & -1
\end{array}
\right]_{ef}
\end{eqnarray}
where $e,f = I$ and $\psi$. The other (trivial) $F$-symbols are $1\times 1$ matrices equal to $1$
if they are allowed by the fusion rules and $0$ if they are not allowed by them.

For Ising anyons, the (nontrivial) $R$-symbols are:
\begin{eqnarray}
R^{\psi\psi}_I &=& -1, \qquad \quad R^{\psi\sigma}_\sigma = R^{\sigma\psi}_\sigma = -i \, , \cr
R^{\sigma\sigma}_I &=& e^{-i \pi /8} \, , \qquad R^{\sigma\sigma}_\psi = e^{i 3\pi /8}
.
\end{eqnarray}
From these fusion rules and $F$-symbols and $R$-symbols, we see that $\sigma$ particles in an Ising
anyon system (such as the $B$ phase) are non-Abelian anyons.

In this section, we have spelled out several examples of topological
phases in (somewhat pedantic) detail in order to use it
as a point of reference for later sections. Other examples of ideal systems that can be similarly constructed include
discrete gauge lattice models (of which the ``toric code'' is an example)~\cite{Kitaev97} and the Levin-Wen lattice models~\cite{Levin05a}.
All of these systems have no bulk gapless excitations (when suitably designed) and satisfy the definition of a topological phase (given in the introduction).
Up to exponentially-small corrections, the $N_\Sigma$ pseudo-ground states are degenerate,
and the unitary transformations associated
with braiding operations are equal to those of a TQFT in the limit that
the quasiparticles are all far apart, the braid operations
are done very slowly, and the temperature $T$ is small.

\subsection{Disordered Systems}
\label{sec:dirty}

It is worth keeping in mind that, in the real world, solids contain impurities
and other defects. As a result, the gap may close. For instance, consider the
toric code Hamiltonian in Eq.~(\ref{eqn:toric-code}) and suppose that the coupling constants
in front of the two terms are random:
\begin{multline}
\label{eqn:toric-code-disorder}
H =  H_{\rm TC}  -  \sum_{{\rm vertices}\,\, v}  {J_1}(v) \prod_{i\in {\cal N}(v)} \sigma_i^z\\
- \sum_{{\rm plaquettes}\,\, p} {J_2}(p)\prod_{i\in p} \sigma_i^x
\end{multline}
For instance, we can consider, as a simple model, the situation in which
${J_1}(v)$ and ${J_2}(p)$ are independently chosen at random from the interval
${J_1}(v), {J_2}(p) \in [-W,W]$. For $W\geq J_{\rm eff}$, the system will have excitations
at arbitrarily low energies. However, the states which are at low energies
will be localized. In order to be more precise, we note that, in the toric code Hamiltonian
in Eq.~(\ref{eqn:toric-code}) or its disordered version in Eq.~(\ref{eqn:toric-code-disorder}),
quasiparticles cannot move, so they are always localized.
However, if we perturb the model even slightly -- for instance, by
adding a weak magnetic field with components in both the $x$- and $z$-directions, then
both $e$ and $m$ quasiparticles will be able to move in the absence of disorder.
In the presence of disorder, all quasiparticles will be localized
(since the system is two-dimensional~\cite{Abrahams79}).
Consequently, even though there will be no gap for $W\geq J_{\rm eff}$, the system is
still in a topological phase. Only a slight modification of the definition (given in the introduction) is needed. We first define a localized state more formally.

\begin{figure}[tbp]
\begin{center}
\includegraphics[width=3.25in,angle=0]{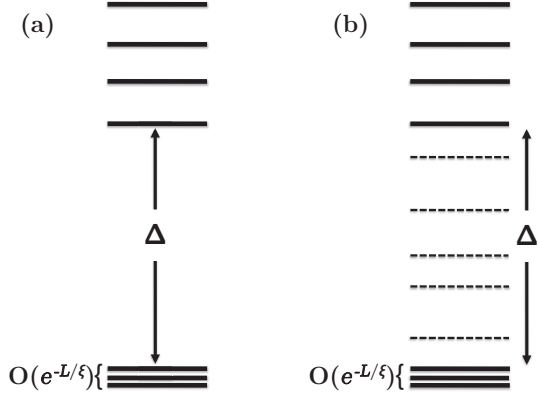}
\caption{(a) The spectrum of a topological phase in a clean system. (b) The spectrum
of a topological phase in a system in which impurities have caused the gap to be filled in
by localized states (depicted with dashed lines). }
\label{fig:top-phase-spectrum}
\end{center}
\end{figure}

A state $|\psi^{\prime}\rangle$ is defined to be a {\it localized excitation} of $|\psi\rangle$
with localization length $\lambda$ if $E_{\psi^{\prime}} > E_\psi$ and the following conditions hold:
(1) There exists a bounded, local operator $\phi$ satisfying
\begin{equation}
\label{eqn:distinguishable}
\left| \langle\psi^{\prime} | \phi |\psi^{\prime}\rangle - \langle\psi | \phi |\psi\rangle \right| > 1
.
\end{equation}
(2) There exists a set of (not-necessarily connected)
regions $\mathcal{A}_{\epsilon} \subset \Sigma$ parameterized by $\epsilon>0$ such that
(a) ${\cal L}_{\mathcal{A}_{\epsilon}} < \lambda \ln(1/\epsilon)$, (b) $\epsilon_1 < \epsilon_2$ implies that $\mathcal{A}_{\epsilon_2} \subset \mathcal{A}_{\epsilon_1}$, and
(c) for any bounded, local operator $\phi$ acting in $\mathcal{A}_{\epsilon}^c = \Sigma \setminus \mathcal{A}_{\epsilon}$ (the complement of $\mathcal{A}_{\epsilon}$)
\begin{equation}
\label{eqn:not-distinguishable}
\left|\text{Tr}\left(\rho_{\mathcal{A}_{\epsilon}^c}^{\prime} \phi\right)
- \text{Tr}\left(\rho_{\mathcal{A}_{\epsilon}^c} \phi \right)\right| < \| \phi \|\,\epsilon
\end{equation}
where $\rho^{\prime}_{\mathcal{A}^c}=\text{Tr}_{\mathcal{A}} \left( |\psi^{\prime}\rangle\langle\psi^{\prime} | \right)$, $\rho_{\mathcal{A}^c}=\text{Tr}_{\mathcal{A}} \left( |\psi \rangle\langle \psi | \right)$, and $\| \phi \|$ is the operator norm of $\phi$.
Intuitively, the two states are distinguishable using local operators,
but are essentially indistinguishable outside some region (the size of which grows logarithmically slowly compared to $1/\epsilon$).

{\bf Definition (Topological Phase):}
A system is in a topological phase if there is an energy $\Delta$ and a length $\xi$,
which, respectively, have a strictly positive limit and a finite, non-negative limit as $L\rightarrow \infty$, such that the following properties hold.
On a manifold $\Sigma$, there is a set of orthonormal energy eigenstates $|a\rangle$, $a\in \{1,2,\ldots,{N_\Sigma}\}$ with energies $E_a < \Delta$,
including the absolute ground state $|1\rangle$  of the system
(which, without loss of generality, we take to have energy $E_1=0$).
$N_\Sigma$ depends only on the topological configuration of the system.
For any local operator $\phi$,
\begin{equation}
\label{eqn:top-phase-def-localized}
\langle a| \phi |b\rangle = C \delta_{ab} + O(e^{-L/\xi})
\end{equation}
for some length scale $\xi$, where $C$ is a constant independent of $a,b \in \{1,2,\ldots,{N_\Sigma}\}$. The Hamiltonian of the system is a particular local operator, so it follows that these states satisfy ${E_a}= O(e^{-L/\xi})$ and form the basis of the pseudo-ground state space $\mathcal{H}_0 = \text{span} \left\{ |1\rangle , \ldots, |{N_\Sigma} \rangle \right\}$ of the system. All other states with energies less than $\Delta$ are localized excitations of states in $\mathcal{H}_0$.

We note that $\xi$ in Eq.~(\ref{eqn:top-phase-def-localized})
may now be interpreted as the localization length for quasiparticles.
The energy $\Delta$ is no longer an energy gap, but is, instead, a mobility gap.
If there are localized states below $\Delta$, all the way down to
zero energy, then we expect that, in general, the first energy
eigenstate above $\mathcal{H}_0$ is at energy $\propto 1/V$, where $d$ is the spatial dimension and $V$ is the volume $\sim L^d$.
The spectrum in a disordered system is compared to the spectrum in a clean system
in Figure~\ref{fig:top-phase-spectrum}.

In the definition above, a key property characterizing the
pseudo-ground states is that local operators cannot distinguish them or
have nontrivial matrix elements between them. However, we have also required
that $N_\Sigma$ depend only on the topological configuration
of the system. Consequently, Haah's three-dimensional model~\cite{Haah11} does not
satisfy our definition; although the ground states are locally indistinguishable,
their number depends on the system size (in a highly
non-trivial manner), not just on the topological configuration.

An alternative definition of a topological phase~\cite{Hastings11} starts, instead, from
the requirement that the states $|a\rangle$ with $a\in \{1,2,\ldots,{N_\Sigma}\}$
not be {\it topologically trivial} in the following sense.
If a state $|\psi\rangle$ is topologically trivial, then for any $\epsilon>0$ there
is a $d<L$ (the system size) and a unitary quantum circuit $U$ of range $d$
such that $|\psi\rangle$ satisfies
\begin{equation}
\left\| \, |\psi\rangle - U |\psi_{\rm prod}\rangle \right\| \leq \epsilon
\end{equation}
where $|\psi_{\rm prod}\rangle$ is a product state (e.g. the
state with all spins up). The range $d$ is the
maximum linear size of the support of unitary operators in the circuit multiplied
by the depth of the circuit. In other words, by acting with a unitary transformation
which only couples spins which are a distance $d$ apart, it is possible
to transform the density matrix of a topologically-trivial system into that
of a system with all spins up. In a topological phase, such
as that described in Eq.~(\ref{eqn:Kitaev-honeycomb}),
the pseudo-ground states are topologically non-trivial, so this is impossible~\cite{Hastings11}.

In closing this section, we mention that many putative topological
phases are more accurately described as $\mathbb{Z}_2$-graded
topological phases and are described by spin field theories~\cite{Dijkgraaf90}, rather than TQFTs.
In particular, this includes electron systems, such as quantum Hall states,
in which the electron is considered akin to the trivial vacuum quasiparticle type.
Since electrons have nontrivial Fermi statistics, which differs from the trivial statistics of the vacuum, one must exercise care and use the mentioned $\mathbb{Z}_2$-grading
and spin field theory structure where treating them as ``vacuum.''
While it is important to be aware of the distinction, we will abuse terminology and continue to use ``TQFT'' when we really mean ``$\mathbb{Z}_2$-graded TQFT'' or ``spin field theory.''

\section{Relation to TQFT}
\label{sec:TQFTs}

We now make the notion of a topologically non-trivial excitation more precise.
Consider, for the sake of concreteness, a system in two spatial dimensions
in a topological phase. An excited state $\rho$ is
topologically-trivial in a bounded region ${\cal R}$ which is homeomorphic
to an open disk if for any $\epsilon>0$, there exists a state
$|\psi \rangle \in \mathcal{H}_0$ and a unitary
transformation $U$ on the degrees of freedom of ${\cal R}$ such that
\begin{equation}
\label{eqn:no-local-unitary}
\left\|  \rho_{{\cal R}} -  (U^\dagger \rho^\psi_{{\cal R}} U)  \right\| \leq \epsilon
\end{equation}
where $\rho_{{\cal R}} = \text{Tr}_{{\cal R}^{c}} ( \rho )$ is the reduced density matrix for the region
${\cal R}$ in the excited state and $\rho^\psi_{{\cal R}} = \text{Tr}_{{\cal R}^{c}} (\left| \psi \right\rangle \left\langle \psi \right|)$ is the reduced
density matrix for the region ${\cal R}$ in the state $|\psi \rangle$.
We emphasize that the unitary transformation $U$ is essentially
arbitrary in the open disk ${\cal R}$.
In less technical terms, this means that an excitation in region ${\cal R}$
is topologically trivial if it can be related to a pseudo-ground state by
unitary operations acting entirely within ${\cal R}$. The requirement that
${\cal R}$ have the topology of an open disk ensures that the unitary
transformation does not simply move a topologically non-trivial excitation
out of ${\cal R}$, which is ensured since $U$ does not act on the boundary $\partial {\cal R}$.
If an excited state is not topologically trivial in a region ${\cal R}$,
then there is a \emph{topological excitation} in ${\cal R}$.

As we discuss in more detail elsewhere~\cite{Bonderson13},
we can further define different types of quasiparticles by the condition
that local unitary transformations cannot transform them into each other,
which can be stated in a manner similar to Eq.~\ref{eqn:no-local-unitary}.
We can also define a quantum number associated with this quasiparticle
type, called topological charge, and ascribe it to the boundary
of the region.
(From this perspective, the definition of a topological excitation given
in the previous paragraph is simply the condition that an excitation
is of a different type from trivial excitation.)
We can then continue and define fusion rules for
these quasiparticle types as follows.
We consider two disjoint open disk-like regions
and determine their respective topological charges.
We then determine the topological charge of
the combined excitation by considering a larger open disk-like region
that contains the two smaller regions.
In a topological phase, we can continue and define braiding, as long as we are
in the adiabatic regime in which braiding is well-defined, i.e. $\delta E_0 \ll \omega \ll \Delta$, where $\omega$ defines the rate at which the adiabatic exchange process is carried out.
If we assume there are a finite number of quasiparticle types, the algebraic fusion and braiding properties are described by a unitary modular tensor category (see Appendix~\ref{sec:AlgTQFT}).

As we mentioned in the introduction, there are two ways to think about
topologically non-trivial localized objects. They can be viewed as excited
states that locally ``look like'' the ground state, except in compact regions
where the reduced density matrix for a region cannot be unitarily transformed into that of a pseudo-ground state, as per the previous paragraphs. A second way to realize a topologically non-trivial object is as a component of the boundary of the system.
In this case, we can discuss the properties of {\em ground states} on manifolds
with boundary (with various possible boundary conditions).
The relation between the two realizations of topologically non-trivial objects
is analogous to that between Abrikosov vortices and
Josephson vortices in superconductors.
In the definitions in this paper, we will adopt the latter perspective
and focus on ground state properties (and gapless excitations above
these ground states) in the presence of boundaries.

Thus far, we have not made any assumption that
the braiding and fusion rules obtained in this manner
should be those of a TQFT. It can be shown that the definition of a topological phase leads to
the conclusion that the properties of the low-energy states of the system are
encapsulated by a TQFT (see Ref.~\onlinecite{Bonderson13} for
a pedagogical discussion). As a consequence, for instance,
the dimension $N_{T^2}$ of the pseudo-ground state subspace on a torus
in a two-dimensional topological phase is equal to the number
of quasiparticle types.

If one considers a system that does not have a gap $\Delta$ above the pseudo-ground state space, then these arguments may break down.
In this case, at least some correlations may not decay exponentially.
Consequently, there may be some quasiparticle types
which are not locally indistinguishable from the ground state.
Thus, these quasiparticle types may not lead to
pseudo-ground states on the torus. We now consider systems with gapless excitations in detail to determine what properties of topological phases are capable of surviving without a proper gap.

\section{Quasi-Topological Phases}
\label{sec:QTP}

\subsection{Kitaev's Honeycomb Model with Phonons and Photons}
\label{sec:strong}

We first reconsider the previous example(s) without the unrealistic assumption that the atomic masses are infinite, thereby introducing phonons.
We can view the (acoustic) phonons as a gapless environment which perturbs the
spin system and determine which aspects of a topological phase survive.
Intuitively, we expect that the spin system is
quantitatively, but not qualitatively affected by the coupling to
phonons. The gap to excitations of the spin system, $\Delta$, will be modified,
but remain non-zero. At energies less
than $\Delta$, the spectrum should essentially decompose into a direct
product of the topological pseudo-ground states with states of gapless phonon excitations.
On the torus, we expect that there will still be $N_{T^{2}}$ pseudo-ground states,
corresponding to distinct topological sectors (quasiparticle types).
Each sector should also have excited states over it corresponding to
gapless phonon excitations.

To see that these expectations are correct, let us make the model
a little more concrete. We will assume, for simplicity, that the three-dimensional
solid has cubic symmetry. Then we can write a low-energy effective action
for the lattice degrees of freedom in the form:
\begin{equation}
{S_{\rm ph}}[\vec{u}]
=\frac{1}{2}\,\int dt {d^3}x \left[\rho {\left({\partial_t}{u_i}\right)^2}
\,-\, 2\mu {u_{ij}}{u_{ij}}\, -\, \lambda {u^2_{kk}}\right]
\label{eqn:elasticaction}
\end{equation}
Here, $\vec{u}(\vec{r})$ is the displacement of
the atom from a lattice position $\vec{r}$;
this atom is located at $\vec{r}+\vec{u}(\vec{r})$. Here, we have used
the abbreviated notation ${u_{ij}}\equiv\left({\partial_i} {u_j} + {\partial_j} {u_i}\right)/2$.
The density is $\rho$, and $\mu$ and $\lambda$ are the Lam\'e coefficients.

The spins are coupled to the phonons. The detailed form of this coupling will
depend on the physics which generates the spin-spin interactions in
Eq.~(\ref{eqn:Kitaev-honeycomb}), but it is reasonable to assume that the interaction
between spins depends on the distance between them so that when the
atoms in the lattice are displaced from their equilibrium positions, the coupling constants
$J_x$, $J_y$, $J_z$ are affected. Therefore, we expect an interaction of the form:
\begin{multline}
\label{eqn:spin-phonon}
H_{\rm spin-phonon} = g_{\rm s-ph}\!\!\!\sum_{x-\text{links}} {S^x_i}  {S^x_{j}} \,\hat{x}\cdot
\left[\vec{u}(\vec{r_i})-\vec{u}(\vec{r_j})\right] \\+
(x \rightarrow y) \,\, + \,\, (x \rightarrow z)
\end{multline}

In the absence of non-Abelian quasiparticles or non-zero genus,
the spin system has a gap, which stabilizes it against
generic weak interactions, including interactions with phonons.
In other words, if the system is in a situation in which there is a non-degenerate
ground state, then the spin model's gap protects it from the effects of the phonons.
Gapless phonons could only have an interesting effect when there are degenerate pseudo-ground states.
In this case, we essentially have the problem of a field theory (for the gapless phonons)
coupled to a quantum-mechanical system with a finite number of degrees of freedom
(the topological pseudo-ground states).

Consider the Ising $B$ phase of Kitaev's honeycomb model. Using Kitaev's representation~\cite{Kitaev06a} of
spin operators in terms of Majorana fermion operators, $S_i^a = i b_i^a c^a$,
with $a=x,y,z$, we can re-write Eq.~(\ref{eqn:spin-phonon}) in the form
\begin{multline}
\label{eqn:spin-phonon-rewritten}
H_{\rm spin-phonon} = - g_{\rm s-ph}\!\!\!\sum_{x-\text{links}} {b^x_i}  {b^x_{j}} c_i c_j \,\hat{x}\cdot
\left[\vec{u}(\vec{r_i})-\vec{u}(\vec{r_j})\right] \\+
(x \rightarrow y) \,\, + \,\, (x \rightarrow z)\\
= - g_{\rm s-ph}\!\!\!\sum_{x-\text{links}} u_{ij} c_i c_j \,\hat{x}\cdot
\left[\vec{u}(\vec{r_i})-\vec{u}(\vec{r_j})\right] \\+
(x \rightarrow y) \,\, + \,\, (x \rightarrow z)
\end{multline}
where $u_{ij}\equiv {b^a_i}  {b^a_{j}}$, where $a$ is the type of link which connects
sites $i$ and $j$. We see from Eq.~(\ref{eqn:spin-phonon-rewritten}) that
the coupling to phonons does not change one of the basic
simplifying features of the Kitaev model: the $u_{ij}$s commute with the Hamiltonian.
Furthermore, the $u_{ij}$s are not gauge-invariant, so the energy depends only on
the gauge-invariant constants of the motion $\prod_{\rm plaquette} u_{ij}$.
Since the three pseudo-ground states on the torus have the same $\prod_{\rm plaquette} u_{ij}$
for all plaquettes, they have the same energy. Hence, the coupling to phonons does
not affect the pseudo-ground state space degeneracy. There can be phonon excitations above the pseudo-ground state space $\mathcal{H}_0$, but the pseudo-ground states are unaffected.

Now, let us consider $\sigma$ quasiparticles in the Ising phase.
Phonons can, in principle, couple to the Majorana zero mode
in the core of a $\sigma$ quasiparticle. Following Kitaev, we represent the
spins by Majorana fermions, $S^a_{{\bf R}_i} = i b^a_{{\bf R}_i} c_{{\bf R}_i}$,
and replace bilinears of $b$s on links by their expectation values (which commute
with the Hamiltonian). Then the Hamiltonian takes the form of a hopping
Hamiltonian for the $c_{{\bf R}_i}$s. In the presence of a $\sigma$
quasiparticle, i.e. a $\mathbb{Z}_2$ vortex, this hopping Hamiltonian has a zero mode
($\mathbb{Z}_2$ vortices must be present in pairs, although one of them may be at infinity).
Let us call such a zero mode $c_0$, which we assume to be localized at
the origin. There is no possible coupling of phonons to such a zero mode.
The obvious guess would be
\begin{equation}
H_{zm-ph} = c_0 \,c_0 \,\,  {\partial_j}{u_j}(\vec{r}=0)
\end{equation}
But, since $c_0^2=1$, this does \emph{not}, in fact, couple phonons to the zero mode.
There {\it can} be a coupling of phonons to a zero mode and a non-zero mode:
\begin{equation}
H_{zm-ph} = i \,c_0 \,c_n \,\,  {\partial_j}{u_j}(\vec{r}=0)
\end{equation}
where $c_n$ is some other, higher-energy, eigenmode of the Hamiltonian
(but, of course, one which is near the origin, since we assume that the coupling is
local). However, since $c_n$ is gapped, we can integrate it out.
This could generate a term in the action of the form:
\begin{equation}
S_{zm-ph} = \int dt\, c_0 \,i {\partial_t}c_0 \,\,  \left[ {\partial_j}{u_j}(\vec{r}=0) \right]^2 \,+ \, ...
\end{equation}
Such a term gives no contribution to the Hamiltonian and is, therefore, unimportant.
One might worry that this term can affect the anti-commutation relation for $c_0$.
If we now integrate out phonons, we will rescale the coefficient of
$c_0 \,i {\partial_t}c_0$. Physically, this just represents the fact that
the coupling to phonons has caused the zero mode $c_0$ to mix with non-zero
modes $c_n$ so that the correct zero energy eigenmode is actually a
linear combination of $c_0$ and $c_n$s.

Therefore, there are two reasons why phonons, though gapless,
have a very small effect. The first is that phonons are Goldstone
bosons (of broken translational symmetry in a crystalline solid)
and, therefore, have irrelevant derivative interactions with quasiparticles.
Second, there is no way, at low energies, to couple phonons to
interesting topological degrees of freedom, such as Majorana zero modes.
As a result, although phonons are gapless, the degeneracy of topological
ground states is split by terms exponentially-small in $L$, rather than by
power-laws in $L$.

The example above is the most innocuous situation which one
can imagine: there are gapless excitations in the system, but the
coupling of these excitations to topological degrees of freedom is irrelevant.
In this case, the gapless degrees of freedom do not qualitatively alter the topological ones,
and so effectively decouple from the topological degrees of freedom
at low energies. The Hilbert space for energies below $\Delta$ decomposes into the
product of the Hilbert space of a TQFT (describing the pseudo-ground state space) and the Hilbert space of the gapless
degrees of freedom.

There is a quantitatively-important effect of the coupling to phonons
which is not addressed by the definition of a quasi-topological phase.
In this definition, quasiparticles are assumed to be fixed. However, if
quasiparticles can move quantum-mechanically, then they can emit or scatter
off phonons. This is a situation which is beyond the definition of a quasi-topological phase
since, once quasiparticles are mobile, the system is no longer in a quasi-topological phase,
i.e. a moving quasiparticle can cause transitions between the different topological sectors.
However, if there is just one or a few moving quasiparticles, then their topological
properties are governed by the quasi-topological phase which gave birth to them.
Consider a $\sigma$ quasiparticle; phonons couple to its translational motion.
As far as this coupling is concerned,
the exotic properties of the $\sigma$ particle are unimportant
and we can write a low-energy effective interaction between quasiparticles
and phonons:
\begin{equation}
\label{eqn:phonon-particle-coupling}
S_{\text{qp-ph}} = g_{\text{qp-ph}} \!\! \int {d^2}x\,dt \, h_{\rm qp}(x,t) \vec{\nabla} \cdot\vec{u}(x,t)
\end{equation}
where $h_{\rm qp}$ is the energy density due to quasiparticles. Therefore,
there will be an effective interaction between quasiparticles of the form:
\begin{equation}
\label{eq:H_eff}
H_{\rm eff} = \int \frac{{d^2}q}{(2\pi)^2}\, \frac{d\omega}{2\pi}\,
h^1_{\rm qp}(q,\omega) \, V(q,\omega)\, h^2_{\rm qp}(q,\omega)
\end{equation}
where $h^{1,2}_{\rm qp}(\vec{q},\omega)$ are the (Fourier transformed) energy densities of the two quasiparticles,
and
\begin{equation}
V(q,\omega) = g_{\text{qp-ph}}^2 \, \frac{q^2}{\omega^2 - v^2 q^2}
\end{equation}
where $v=\sqrt{(2\mu+\lambda)/\rho}$.
If quasiparticle 1 does not move, so that $h^1_{\rm qp}(q,\omega) \propto \delta(\omega)$,
then the effective interaction between the two quasiparticles is $V(q,0) = -g_{\text{qp-ph}}^2/v^2$
or, in real space, $V(r)\propto \delta^{(2)}(\vec{r})$. In other words, there is no interaction
if one particle is motionless while the other one stays far away. Therefore, if one
quasiparticle is taken around another, while remaining away from it, there is no
dynamical phase due to the phonon-mediated interaction.

However, there is a second effect of phonons.
When quasiparticles are moved, phonons will be emitted.
(One can view the phonon-mediated interaction in Eq.~(\ref{eq:H_eff}) as the effect of
off-shell or virtual phonons, while phonon emission is the effect of
real phonons.) Phonon emission does not affect the phase which results when two quasiparticles
are braided. However, if we are trying to observe this phase by interfering
two different quasiparticle trajectories, there is some amplitude that
the change in phonon number (which we presumably do not measure)
is different along the two trajectories. This will reduce the visibility
of the interference: if we try to interfere two trajectories with a combined
length $l$, then interference between the two trajectories will
decay with $l$ as $e^{-l/L_\phi}$ for some coherence length $L_\phi$.
Phonon emission will give a contribution
to $1/L_\phi$ of the form $(1/{L_\phi}_{\rm phonon}) \propto g_{\text{qp-ph}}^2 T^5$
since $g_{\text{qp-ph}}$ is irrelevant by two powers of energy.
By going to sufficiently low temperatures (which may, admittedly, be technically difficult),
we can make the effect of phonons on the visibility
as small as we wish.
This is an important practical consideration,
but is not a fundamental change in the physics of the state.

Now suppose that we remove the superconductor, so there will be
gapless photons present in the system.
They couple to the spin system in the form:
\begin{equation}
H_{\rm spin-photon} = g{\mu_B}\, {\bf S}\cdot {\bf B}
\end{equation}
Again, in the low-energy limit, there is no way to couple
${\bf B}$ to a zero mode $c_0$.
Unlike in the phonon case, there will be a photon contribution to the
non-universal phase associated with quasiparticle exchange simply as a result
of the interaction between the electromagnetic field and the motion
of the quasiparticles; this contribution
depends on the precise trajectories of the quasiparticles
and also on how fast the process occurs.
There will also be a reduction of the visibility of interference
patterns as a result of photon emission (similar to that resulting from phonon emission).
However, the basic physics
of the topological phase will be unaffected, as in the phonon case.
One difference, however, is that there may be photons of arbitrarily low energies
even when the spin system has finite size $L$. This will be the case
generically, but if we place the spin system within a waveguide of size $L_0$
with very high $Q$ factor, there will be a finite-size gap
$\Omega \propto 1/L_0$.
For $L_0 \gg L$, the regime in which the system behaves as if it were in a gapped
topological phase will be even smaller: $\omega,T \ll \Omega \propto 1/L_0$.
Nevertheless, there is a $\Delta$ below which the system is simply the product of the pseudo-ground states with the gapless excitations.

We now imagine connecting our spin system to a metallic lead
at a single point or weakly coupling it to a metallic gate.
In addition to gapless phonons and photons,
there are now gapless electrons that
can, in principle, tunnel into the system.
However, there is an energy gap which prevents this
from occurring, in spite of the gapless phonons and photons.
In order for an electron to tunnel into Kitaev's honeycomb lattice
model, a fermion must be created {\it and} charge $-e$ must be created.
There are zero-energy fermions at $\mathbb{Z}_2$ vortices.
However, there is a hard charge gap.
Kitaev's model is a pure spin model, so the charge gap is infinite
in the model as originally formulated. However, it can be generalized~\cite{Burnell11,You11,Hyart11}
to a model that allows doping. If we decompose the electrons
into spinons $f^\dag_{i \sigma}$ (where $\sigma=\uparrow, \downarrow$ is
the spin index) and holons $b_{i}$
\begin{equation}
c^\dag_{i \sigma }= f^\dag_{i \sigma} b_{i}
\end{equation}
then, when there is one electron per site, the holons $b_{i }$ are gapped and
the spinons form a $p$-wave paired state, so that
we recover the Kitaev honeycomb model.
However, we can change the electron density, thereby introducing a finite
density of mobile holons $b_{i}$s.
As long as the chemical potential lies in the charge gap (i.e. as long as the spin
system is weakly-coupled to the metallic lead or gate, rather than
strongly coupled, which would dope it as a result)
low-energy electrons cannot tunnel into the system since a holon would have
to be created.

In fact, the stability is even better than this. An electron cannot tunnel
into the system at low energy because a charged excitation must be created.
But let us suppose that there are right-handed neutrinos (as neutrino oscillation
experiments imply) and that these right-handed neutrinos do not carry
any weak isospin (nor any conserved quantum numbers other than fermion parity).
Then, they have the same electro-weak quantum numbers
as Majorana quasiparticles in Kitaev's model. Therefore, the
charge gap no longer protects the system. One might worry
that a neutrino could be absorbed by a Majorana zero mode.
However, this process is still not allowed. To see this,
it is helpful to use a representation of the spins
in terms of Majorana fermions which has only two (rather than four)
Majorana fermions per site of the honeycomb lattice~\cite{Feng07}, so as to avoid any
gauge redundancy. We will call the two Majorana fermion operators on
each site $a_1$ and $a_2$, with $S^z_j = i a_{1j} a_{2j}$. Then, we might worry
about a tunneling term such as
\begin{equation}
\label{eqn:tunneling-to-zero-mode}
H_{\text{tun}} = it\,a_{1j} {\nu_R}(\vec{r}=\vec{R}_j)
\end{equation}
Here, $\nu_R$ is the right-handed neutrino operator.
We are interested in the potentially dangerous case in which
the $j$th lattice site $\vec{r}=\vec{R}_j$ is at a Majorana zero mode. However,
the coupling in Eq.~(\ref{eqn:tunneling-to-zero-mode}) is, in fact, not allowed.
If we re-write the Majorana fermion operator in terms of the spin operators then
it takes the form $a_{1j} = (\prod_{i<j} S^z_i) S^x_j$, where the product runs
over all sites that lie along a string between the site $j$ and the edge of the system.
\footnote{If there are other vortices and zero modes in the system, then we
could write an operator that is bilinear in two Majorana fermion operators.
Such an operator would involve a string of spin operators connecting the two vortex locations 
and would cause a splitting between zero modes.}
Consequently, the tunneling term takes the form:
\begin{equation}
\label{eqn:no-tunneling-to-zm}
H_{\text{tun}} = it\,(\prod_{i<j} S^z_i) S^x_j \nu_R(\vec{r}=\vec{R}_j)
\end{equation}
This is clearly not a local coupling, so it cannot occur.
On the other hand, a local coupling with amplitude $t'$ could generate a term such as that
in Eq.~(\ref{eqn:no-tunneling-to-zm}) at order $\sim L$ in $t'$,
i.e. $t \sim (t')^L$. Given that the coupling is a perturbation with $t'$ small, the resulting effect on the zero mode
will be exponentially suppressed by the system size.
This is good news for the stability of this phase,
but bad news if one were hoping to use the phase
as a neutrino detector.

\subsection{Definition}
\label{sec:definition}

\begin{figure}[t!]
\begin{center}
\includegraphics[width=3.25in,angle=0]{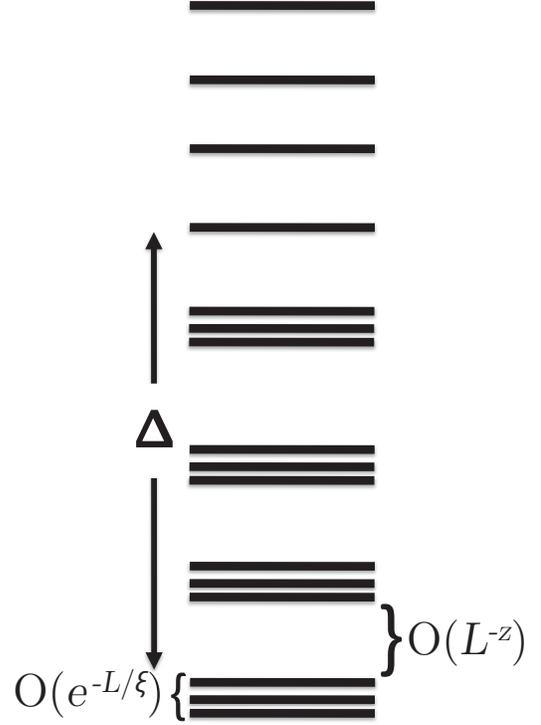}
\caption{The spectrum of a quasi-topological phase.
The pseudo-ground states have exponentially-small splitting,
while their separation from the rest of the spectrum vanishes as
a power-law. Note that the multiplets at energies below $\Delta$
do not necessarily have the same number of states in a generic
quasi-topological phase (as depicted), but they would
in a strong quasi-topological phase.}
\label{fig:strong-q-top-spectrum}
\end{center}
\end{figure}

With the above example in mind, we now define the concept of a
 {\it quasi-topological phase}.

{\bf Definition (Quasi-Topological Phase)}:
A system is in a quasi-topological phase if there is an energy $\Delta$ and a length $\xi$,
which, respectively, have a strictly positive limit and a finite,
non-negative limit as $L\rightarrow \infty$, such that the following properties hold.

(i) On a manifold $\Sigma$, there is a set of orthonormal energy eigenstates $|a\rangle$, where $a\in \{1,2,\ldots,{N_\Sigma}\}$, including the absolute ground state $|1\rangle$ (with energy $E_1=0$), such that
for any local operator $\phi$,
\begin{equation}
\label{eqn:strong-exponential2}
\langle {a}| \, \phi \, | {b}\rangle = {C} \delta_{ab} + O(e^{-L/\xi} )
\end{equation}
for $a, b \in \{1,2,\ldots,{N_\Sigma}\}$, where ${C}$ is independent of $a, b$.
The Hamiltonian is a particular local operator, so it follows that
\begin{equation}
\delta {E_0} \equiv \text{max}(|{E_{a}}|) = O(e^{-L/\xi})
\end{equation}
for $a\in \{1,2,\ldots,{N_\Sigma}\}$, where $E_{a}$ is the energy of the state
$|{a}\rangle$.

(ii) $N_\Sigma$ depends only on the topological configuration of the system.

(iii) We define
\begin{equation}
{\Omega} \equiv \text{min}(|{E_\chi}-E_{a}|)
\end{equation}
for any $a\in \{1,2,\ldots,{N_\Sigma}\}$
and $|\chi\rangle \notin \mathcal{H}_{0} \equiv \text{span}\left\{ |1\rangle, \ldots, | N_\Sigma \rangle \right\}$.
Then, there must exist a finite $z>0$ and an $L$-independent $\kappa>0$ such that
\begin{equation}
\Omega\sim\kappa L^{-z}.
\end{equation}

(iv) There are no states with energy less than $\Delta$ that have topologically nontrivial excitations.

In other words, the topologically-protected pseudo-ground states
approach each other exponentially-fast in the system size, forming a pseudo-ground state space $\mathcal{H}_0$
separated from the rest of the spectrum by a
finite-system-size gap $\Omega\sim\kappa L^{-z}$.

The energy $\Delta$ can be understood to be a ``topological mobility gapp'' in the following sense.
$\Delta$ is the minimum energy required to pair-create topologically
non-trivial excitations (as defined in Section~\ref{sec:dirty})
from a pseudo-ground state. We note that there may be
topologically trivial excitations of energy less than $\Delta$. These may include 
bound states of topologically nontrivial excitations that 
are collectively topologically trivial (i.e. their combined topological charge is trivial).
These excitations below $\Delta$ do not cause a splitting between putative
degenerate states. Only when there are topologically non-trivial
excitations capable of moving to separations on the order of $L$, $r$, or $\ell$
can these excitations cause a splitting between putative
degenerate states. For instance, in a perturbed toric code Hamiltonian
(so that $e$ excitations interact and move),
a pair of $e$ excitations can change the winding number only if they
have enough energy for one of them to encircle a generator of the torus.
Therefore, $\Delta$ is the energy gap to
the creation of a pair of topologically non-trivial
excitations which are capable of moving to separations on the order of $L$, $r$, or $\ell$.

From the preceding discussion, we see that
Kitaev's model in a solid with the concomitant coupling to phonons and photons (as considered in Section~\ref{sec:strong})
is an example of a quasi-topological phase.
It is a phase of matter in which the splitting between ground states $\delta E$
remains exponentially suppressed in spite of gapless excitations with an energy spacing
$\Omega$. However, some features of this example 
are not generic for quasi-topological phases. 
In particular, for this example, there is a regime of energy and temperature,
$\delta E_0 \ll \omega, T \ll \Omega$, in which the system behaves
as a gapped topological phase with the pseudo-ground state degeneracies,
braiding and fusion rules expected from the Ising TQFT.
Moreover, there is a much larger regime, $\delta E_0 \ll \omega, T \ll \Delta$,
in which the Hilbert space of the system can be understood in terms of topological sectors,
each of which contains a topological ground state and gapless excitations
above that ground state. In other words, the low (compared to $\Delta$)
energy effective theory of the system is essentially a TQFT
tensored with a set of relatively benign gapless theory.
However, this is not necessarily the case in all quasi-topological phases,
as we will demonstrate in the next subsection.

We will use the term {\it strong quasi-topological phases}
to refer to phases which satisfy the property that they
show the characteristic behavior of a topological phase in the regime
$\delta E_0 \ll \omega, T \ll \Omega$. The topological
properties encapsulated in $N_\Sigma$, $F$- and $R$-matrices,
etc. are all precisely the same as in some TQFT (which
was the Ising TQFT in the example considered in this section).

Other examples of strong quasi-topological phases
include the toric code~\cite{Kitaev97} and Levin-Wen \cite{Levin05a} models,
when either one is realized as a spin system in a solid; and also
fractional quantum Hall states realized in
ultra-cold bosonic atoms in rotating traps (see, e.g. Refs.~\onlinecite{Wilkin98,Cooper01})
and in optical lattices with effective gauge fields (see, e.g. Refs.~\onlinecite{Jimenez-Garcia12}). In fact, some well-known phases actually satisfy our definition of strong quasi-topological phases. Ordinary insulators are trivial
strong quasi-topological phases: they have a
gap $\Delta$ and a unique ground state for any manifold topology
(precisely as in the trivial TQFT), but they also support gapless phonons and photons.
Meanwhile, 3D superconductors with finite atomic masses have the topological properties
of the TQFT described in Section~\ref{subsec:3D-super}, but with gapless phonon excitations.

Strong quasi-topological phases are the most ideal cases one could hope for when
introducing gapless excitations into a topological phase, since all of the topological
properties survive, and the system below $\Delta$ is simply the product of a
TQFT with the gapless degrees of freedom. More generally, gapless degrees
of freedom will be able to couple non-trivially to the topological degrees of freedom.
As discussed at the end of Section~\ref{sec:TQFTs}, gapless degrees of freedom
may generally introduce long-ranged correlations that cause the TQFT structure
to break down. In the next subsection, we will see examples of quasi-topological
phases that are not strong.

\subsection{Fractional Quantum Hall Effect}
\label{sec:weak}

For examples of a quasi-topological phases that are not ``strong,''
we turn now to fractional quantum Hall (FQH)
states~\cite{Tsui82,Laughlin83} in electronic systems, which are
the most-commonly given examples of topological phases~\cite{Girvin87,Read89a,Zhang89,Wen92a}. However,
there are gapless phonons and photons in a GaAs
quantum well or heterostructure, so the (entire) system does not have
a gap. The phonons are relatively harmless in these phases, and if these were the only gapless degrees of freedom included, the quantum Hall systems would be in strong quasi-topological phases. On the other hand, photons have a non-trivial effect that breaks the TQFT structure.
We will begin by demonstrating this in the case of a Laughlin state~\cite{Laughlin83}. The extension of this analysis to other FQH states, such as the Haldane-Halperin hierarchy/composite fermion states~\cite{Haldane83,Halperin84,Jain89}, is straightforward, but we explicitly consider Ising-type FQH states~\cite{Moore91,Lee07,Levin07,Bonderson07d,Bonderson12a} and Read-Rezayi states~\cite{Read99}, as these are particularly physically relevant examples that also highlight the varying levels of TQFT structure that can survive in quasi-topological phases.
Other examples of quasi-topological phases include
quantum Hall ferromagnets~\cite{Sondhi93}.

\subsubsection{Laughlin States}

The effective theory for a $\nu=1/m$ Laughlin state (where $m$ is an odd integer for fermionic particles) is the U$(1)_m$ Chern-Simons theory with action:
\begin{equation}
S = \int {d^2}x\,dt\, \frac{m}{4\pi}\epsilon_{\mu\nu\lambda} {a_\mu} {\partial_\nu} {a_\lambda}
.
\end{equation}
This TQFT has $m^{g}$-fold degenerate ground states on a manifold $\Sigma$ of genus $g$.
We consider the effect that coupling this TQFT to gapless degrees of freedom has on its ground states, in particular when the system is on a torus ($g=1$).

First, let us consider coupling to phonons. The lowest-order possible coupling of this effective theory to phonons is:
\begin{equation}
\label{FQH-phonons}
S_{\rm FQH-ph} = \lambda \int {d^2}x\,dt\,
\epsilon_{ij} {\partial_i} {a_j} \, \partial_k u_k
\end{equation}
where $i,j=1,2$ are spatial indices. This interaction is irrelevant by
one power of energy.

For the (pseudo-)ground states, we can focus on zero momentum modes. Then the action (including phonons) takes the form:
\begin{equation}
S = \int dt\, \frac{m}{4\pi}\left( {\tilde{a}_2} {\partial_t} {\tilde{a}_1} -
{\tilde{a}_1} {\partial_t} {a_2} \right)
+ \frac{1}{2}\rho \int dt \, (\partial_t \tilde{u}_i)^2
\end{equation}
where $\tilde{a}_i$ and $\tilde{u}_i$ are the $p=0$ Fourier components
of $a_i$ and $u_i$.
The interaction between the gauge field and phonons in Eq.~(\ref{FQH-phonons})
vanishes for the zero-momentum modes, so the Chern-Simons field decouples from
phonons. (Phonons, like all Goldstone bosons, have derivative couplings,
so they decouple in the zero-momentum limit.) Therefore, the ground state degeneracy
of the U$(1)_m$ Chern-Simons theory is unaffected by the coupling to phonons and
the low-energy theory is simply a produce of the TQFT and phonon degrees of freedom.

Any quantum Hall state will have an Abelian U$(1)$
Chern-Simons field as part of its effective field theory, in order to account for the
charged degrees of freedom (and the Hall conductance), so our conclusion
about the decoupling of phonons holds from the U$(1)$ sector
holds for any quantum Hall state. If the quantum Hall state has other Abelian sectors,
then the lowest order possible coupling to phonons is of the same form.
Therefore, the zero-momentum modes decouple. Non-Abelian
sectors cannot couple to phonons in this way, since the field strength
of a non-Abelian gauge field is not gauge invariant; therefore, the lowest order
coupling is between $\partial_k u_k$ and $(F_{ij}^a)^2$ which also vanishes
in the zero momentum limit.

So far, everything is just as we would expect in a strong quasi-topological phase.
Next, however, we consider the coupling of a $\nu=1/m$ Laughlin state to
the electromagnetic field, focusing on a system on a torus.
The effective field theory takes the form
\begin{multline}
S = \int {d^2}x\,dt\, \left(\frac{m}{4\pi}\epsilon_{\mu\nu\lambda} {a_\mu} {\partial_\nu} {a_\lambda}
+ \frac{1}{2\pi} A^{\text{2D}}_{\mu} \right)\epsilon_{\mu\nu\lambda} {\partial_\nu} {a_\lambda}\\
+ \frac{1}{2 e^2}\int {d^3}x\,dt\, F_{MN} F^{MN}
\end{multline}
We have used Greek letters for $(2+1)$D indices, $\mu=t,x,y$
and uppercase Roman letters for $(3+1)$D indices, $M=t,x,y,z$.
We will use lowercase Roman letters for 2D spatial coordinates,
$i=x,y$. Here, $A_M$ is the electromagnetic gauge field, $A^{\text{2D}}_\mu$
is its restriction to the torus, and $-e$ is the charge of an electron.
Again, to study ground states we focus on the zero-momentum modes.
For simplicity, we will assume that the 3D world is
a torus $T^3$ on which we have coordinates $0\leq {x_1}, {x_2} \leq L$ and
$0\leq {x_3} \leq L_z$ with ${x_1} \equiv {x_1}+L$, ${x_2} \equiv {x_2} +L$, and
${x_3}\equiv {x_3}+L_z$, and furthermore that the 2D quantum Hall system is on the torus $T^2$
defined by ${x_3}={L_z}/2$, $0\leq {x_1}, {x_2} \leq L$. (This is merely
a convenience in order to avoid dealing with curved coordinates on a torus
embedded in $\mathbb{R}^3$.)
We must now exercise some care in order to normalize the fields properly in finite volumes.
We write ${a_\mu}(x,t) = \frac{1}{L}\sum_p e^{ip\cdot x} {\tilde{a}_\mu}(p,t)$ and
${A_\mu}(x,t) = \frac{1}{L\,{L_z}^{1/2}}\sum_p e^{ip\cdot x} {\tilde{A}_\mu}(p,t)$
[since $a_\mu$ is a (2+1)D field and $A_\mu$ is a (3+1)D field].
With these normalizations, the effective action for zero momentum
modes takes the form:
\begin{multline}
S = \int dt\, \frac{m}{4\pi}\left( {\tilde{a}_2} {\partial_t} {\tilde{a}_1} -
{\tilde{a}_1} {\partial_t} {\tilde{a}_2} \right)
+ \frac{1}{2 e^2}\int dt\, (\partial_t \tilde{A}_M)^2\\
+ \frac{1}{L_z^{1/2}} \int dt \, {\tilde{A}_i}\epsilon_{ij} {\partial_t}{\tilde{a}_j}
\end{multline}
Here, $\tilde{a}_i$ and $\tilde{A}_i$ are the $p=0$ Fourier components of these fields
and, in this equation, $M=x,y,z$.
We have normalized the Fourier transforms so that the first two terms
do not depend on $L$ and $L_z$. At zero momentum, $a_t$ and $A_t$ drop out
of the theory. (However, in the full theory, we could take the gauge $a_t=A_t=0$
and also an additional gauge condition, such as $\partial_i a_i = \partial_i A_i =0$.)
As a result of the decoupling of $A_t$, we see that the long-ranged Coulomb interaction
does not affect the ground state degeneracy on the torus in the infrared limit. Indeed, any
static interaction will leave the ground state degeneracy topologically protected (i.e. the degeneracy splitting will be exponentially suppressed), so long as the quantum Hall state remains stable. However, dynamical photons can have a more
interesting effect, as we will demonstrate.

Now, we integrate out $\tilde{A}_i$ to obtain an effective action for $\tilde{a}_i$ alone:
 \begin{equation}
S = \int dt\, \left[\frac{m}{4\pi}\left( {\tilde{a}_2} {\partial_t} {\tilde{a}_1} -
{\tilde{a}_1} {\partial_t} {\tilde{a}_2} \right)
-  \frac{1}{2}\frac{e^2}{L_z} {\tilde{a}_i^2}\right]
\end{equation}
This is the lowest Landau level action on the torus
(with $\tilde{a}_1$ and $\tilde{a}_2$ playing the
roles of the $x$ and $y$ coordinates of a particle in the lowest Landau level)
in the presence of a harmonic oscillator potential.
Therefore, the energy levels have spacing $\delta E_0 \propto \frac{e^2}{L_z}
= \alpha \frac{c}{L_z}$. Here, we have restored the speed of light $c$
and the fine structure constant $\alpha=e^2/c$. The splitting caused by the electromagnetic field
is small due to the smallness of the fine structure constant $\alpha \approx 1/137$.
But it vanishes, nevertheless, as a power of $L_z$.

Incompressibility is a hallmark of quantum Hall states: the electron
density is locked to the magnetic field and, therefore, cannot fluctuate.
However, when the magnetic field itself has quantum fluctuations, the density
also fluctuates. This results in the $O(L_{z}^{-1})$ splitting found above.

From this analysis, we see that the $m$ would-be pseudo-ground states (corresponding to the ground states of the U$(1)_m$ Chern-Simons TQFT)
actually exhibit a power-law energy splitting themselves when (3+1)D electromagnetism is included, rather than exponential splitting.
Therefore, the pseudo-ground state subspace $\mathcal{H}_0$ has dimension $N_{T^2} =1$ (i.e. it only includes the single true ground state),
since it only includes states that are degenerate with the ground state to within exponential accuracy $O(e^{-L/\xi})$. It is clear that this will also give $N_{\Sigma} =1$ for an arbitrary 2D manifold $\Sigma$.
We note that the trivial TQFT (which has no nontrivial topological excitations) also has $N_{\Sigma} =1$ for all $\Sigma$, so one might wonder whether the result is simply a strong quasi-topological phase corresponding to a trivial TQFT. To see this is not the case, we consider the braiding statistics of quasiparticles.

Quasiparticle excitations of quantum Hall states can be electrically-charged
and can, therefore, couple to photons more strongly
than spin excitations in Kitaev's model. There will be
a $V(r)=q_1 q_2 /r$ interaction between two quasiparticles with
electric charges $q_1$ and $q_2$, respectively. Therefore, there will be a non-universal dynamical
contribution to the phase which does not decay with the separation
between the quasiparticles. (This can be seen microscopically
from the wavefunctions, as in Refs.~\onlinecite{Sondhi92,Simon08}.) The dynamical
phase $\phi$ satisfies
\begin{equation}
\label{eqn:dyn-phase-FQHE}
\phi = {\int_0^\tau} dt \, E(t) \sim \frac{\tau}{r} \sim  \frac{1}{v}
> \frac{1}{v_{\rm max}}
.
\end{equation}
Therefore, we cannot make the dynamical phase arbitrarily small.
In addition, there will be an Aharonov-Bohm phase~\cite{Aharonov59}, $\gamma_{AB}$,
which depends on the area enclosed in quasiparticle trajectories
(as well as the quasiparticles' charge and the value of the background magnetic field).
Both of these terms are due to non-universal physics which is beyond the TQFT description.
Consequently, exchanging quasiparticles leads to the unitary transformation
\begin{equation}
U = e^{i\gamma_{AB}} \,e^{-i\phi} R
\end{equation}
where $R$ is the (topological) quasiparticle braiding statistics described by the U$(1)_m$ TQFT.
For a counterclockwise exchange of a pair of charge $e/m$ fundamental quasiholes in a $\nu=1/m$ Laughlin state, the braiding statistics factor is $R= e^{i \pi /m}$, corresponding to that of U$(1)_m$ flux $1$ objects~\cite{Arovas84}.
Although neither one of the non-universal phases is small {\it a priori}, in an interference experiment,
we can vary the area in order to isolate the Aharonov-Bohm contribution
and, in principle, we can make the dynamical phase the same for two different
interfering trajectories. Therefore, the topological braiding phases predicted by the original U$(1)_m$ Chern-Simons TQFT
do have a well-defined meaning, even though it may be difficult to measure it
in practice.

Thus, the Laughlin FQH states physically realized in electronic systems are quasi-topological states that are not ``strong.''
They retain the braiding statistics of quasiparticles of a U$(1)_m$ Chern-Simons TQFT, but they do not possess the TQFT structure for surfaces of genus $g>0$.
The basic underlying reason for this is that the different quasiparticle types in these states are locally distinguishable,
even far away, as a result of the electromagnetic fields which they generate.

\subsubsection{Ising-type States}
\label{sec:Ising-typeFQH}

In a non-Abelian quantum Hall state, the situation can be more interesting.
Consider, for instance, the Moore-Read (MR) Pfaffian state at filling fraction $\nu= \frac{1}{m}$,
where $m$ is an even integer for fermionic particles (and one can always add integers to the filling, since they may be treated as ``inert'', so that the $\nu=1/2$ MR state gives a candidate for the $\nu=5/2$ FQH plateau).
The corresponding TQFT for this state can be written as a spectrum restriction of the product of a U$(1)_m$ TQFT and an Ising TQFT, where the spectrum restriction requires the $I$ and $\psi$ Ising topological charges to be paired with integer U$(1)_m$ flux values, and the $\sigma$ Ising charge to be paired with half-integer U$(1)_m$ flux values (for more details, see e.g. Ref.~\onlinecite{Bonderson11b}). The ground state degeneracy on the torus for this TQFT is $N_{T^{2}}=3m$, with the familiar factor of $m$ coming from the Laughlin-like U$(1)_m$ sector corresponding to the charged degrees of freedom and
the factor of $3$ from the non-Abelian Ising sector's degrees of freedom.

We can now introduce coupling to photons as we did for the Laughlin states.
The electromagnetic field couples only to the charged degrees of freedom,
so one might naively conclude that the degeneracy on the torus is reduced
to $3$. However, quasiparticles carrying non-Abelian topological charge $\sigma$
must carry electrical charge charge $(n+\frac{1}{2})e/m$ for integer $n$ (corresponding to half-integer fluxes), which differentiates
them electrically from quasiparticles carrying Ising topological charge $I$ or $\psi$, which must carry electric charge $ne/m$ for integer $n$ (corresponding to integer fluxes).
Therefore, the corresponding ground states on the torus will have their energy degeneracy
lifted by the same mechanism that split the ground state degeneracy of the Laughlin states.
One might now expect that ground states corresponding to
the electrically-neutral topological charges $I$ and $\psi$ around one
non-trivial cycle of the torus remain degenerate, but this is not the case. The
modular transformations of the MR TQFT indicate that these states may be rewritten as a linear combination of states which have
definite values of topological charge around any other generator of the torus. In this way, the states with the electrically-neutral
topological charges $I$ or $\psi$ with respect to a given generator of the torus will have a superposition of states that include ones with topological charge $\sigma$ with respect to a different generator of the torus.
This can be see directly from the MR TQFT's modular $S$-matrix, which
cannot be written as a direct product of an $S$-matrix for the neutral quasiparticle subsector and an $S$-matrix for the charge subsector. For example, the $\nu=1/2$ MR state has
\begin{equation}
S=\frac{1}{\sqrt{8}}\left[
\begin{array}{cccccc}
1 & 1 & \sqrt{2} & 1 & 1 & \sqrt{2} \\
1 & 1 & -\sqrt{2} & 1 & 1 & -\sqrt{2} \\
\sqrt{2} & -\sqrt{2} & 0 & i\sqrt{2} & -i\sqrt{2} & 0 \\
1 & 1 & i\sqrt{2} & -1 & -1 & -i\sqrt{2} \\
1 & 1 & -i\sqrt{2} & -1 & -1 & i\sqrt{2} \\
\sqrt{2} & -\sqrt{2} & 0 & -i\sqrt{2} & i\sqrt{2} & 0 \\
\end{array}%
\right]
\end{equation}
where the columns and rows are labeled, in order, by the quasiparticle types carrying Ising$\times$U$(1)_{2}$ topological charge labels $I_0$, $\psi_0$, $\sigma_{\frac{1}{2}}$, $I_1$, $\psi_1$, and $\sigma_{\frac{3}{2}}$, where the subscripts indicate the U$(1)_{2}$ Chern-Simons flux value.
Because of this, the degeneracy between states with topological charge $I_0$ and $\psi_0$ (or $I_1$ and $\psi_1$) with respect to a given generator of the torus is broken as well, leaving $N_{T^2}=1$. Similar arguments also result in $N_{\Sigma}=1$ for any manifold without boundaries.
Hence, the coupling to the electromagnetic field only directly affects
the charged degrees of freedom, but in some topological phases, such as the MR states, the
charged and neutral degrees of freedom are inseparably linked through the selection rules
of allowed topological charges and the modular transformations.
Consequently, topological degeneracies associated
with seemingly neutral degrees of freedom are lifted as well.

Turning now towards the braiding statistics of this state, we consider systems with boundaries or quasiparticles. We find that the pseudo-ground state degeneracy due to boundaries and/or quasiparticles that carry non-Abelian Ising charge $\sigma$ is unaffected by the inclusion of gapless photons. Just as we showed for Laughlin states, the effect of including electromagnetism appears in an overall dynamical phase. This phase does not alter the quasiparticle braiding statistics, though perhaps makes it very difficult to experimentally extract. Hence, exchanging quasiparticles leads to the unitary transformation
\begin{equation}
U = e^{i\gamma_{AB}} \,e^{-i\phi} R
\end{equation}
where $R = R_{\text{Ising}} R_{\text{U}(1)_m}$ is the (topological) quasiparticle braiding statistics described by the MR state's TQFT, which takes the form of a product of the braiding statistics due to the Ising and the U$(1)_m$ sectors
(see Ref.~\onlinecite{Bonderson11b} for
more details and a proof of the braiding statistics of these states).

While the non-universal phases will make it difficult to observe the absolute phase
factors of any braiding operation, it does not in any way disturb the relative phases acquired by different fusion channels of the non-Abelian anyons. In particular, the ratio of $R^{\sigma\sigma}_I$ and $R^{\sigma\sigma}_\psi$, for braiding two quasiparticles carrying an Ising $\sigma$ topological charge each, is robust. The underlying reason is that the $\sigma$ topological charges (Majorana
zero modes), which are responsible for the non-Abelian statistics of
quasiparticles in these systems, do not couple to photons or phonons,
as we saw in the previous section in the context of Kitaev's honeycomb
lattice model. The non-Abelian statistics is ``mediated'' by the neutral fermion $\psi$, which only
couples to the Ising topological degrees of freedom in these quasi-topological phases.

The behavior of other Ising-type quantum Hall states is very similar under coupling to gapless degrees of freedom, but, in some cases, have interesting differences. The anti-Pfaffian state~\cite{Lee07,Levin07} is constructed by taking the particle-hole conjugate of the $\nu=1/2$ MR state. As such, its TQFT is simply the complex conjugate of the MR state's TQFT. The analysis for the MR state applies exactly the same, except the braiding statistics factors $R$ are complex conjugated.

The Bonderson-Slingerland (BS) hierarchy states~\cite{Bonderson07d,Bonderson12a} built on the MR and anti-Pfaffian states have similar TQFTs, except the U$(1)_m$ sector is replaced by multiple U$(1)$s coupled together in a fashion described by a coupling $K$-matrix, which we therefore write as U$(1)_K$ [and $\sigma$ Ising charges are paired with half-integer flux values in the $1$st U$(1)$]. These states will have certain filling fractions $\nu=p/q$ (with $p$ and $q$ mutually prime), which can be determined from the $K$-matrix (for example, an experimentally prominent sequence of BS states is given by filling fractions $\nu= \frac{n}{3n-1}$). The ground state degeneracy on the torus for a BS state's TQFT is $N_{T^2} = 3 \det K$, the factor of $3$ coming from the Ising sector and the factor of $\det K$ coming from the U$(1)_K$ sector.

When the numerator $p$ of the filling fraction is odd for these BS states, the analysis for coupling to phonons and photons is similar to that of the MR state. The TQFT and hence the $S$-matrix cannot be written as a product of a neutral sector TQFT (or $S$-matrix) with a charged sector TQFT (or $S$-matrix), so the TQFT's ground state degeneracy due to genus $g>0$ is fully split by the presence of (gapless) photons in the system coupling to the quasiparticles' electric charges. Hence, the resulting quasi-topological phase has $N_{\Sigma}=1$ for all manifolds $\Sigma$ without boundaries or quasiparticles. The pseudo-ground state degeneracy due to boundaries or quasiparticles carrying Ising $\sigma$ topological charge is preserved. Though there are non-universal phases similarly occurring for quasiparticle exchange processes, the braiding statistics of the quasiparticles is again preserved, in this case being described by $R = R_{\text{Ising}} R_{\text{U}(1)_K}$.

In the case when the numerator $p$ of the filling fraction is even for these BS states, something more interesting occurs. Namely, the neutral quasiparticle sector will contain a quasiparticle that carries Ising topological charge $\sigma$ (in addition to the two neutral quasiparticles that carry Ising topological charges $I$ and $\psi$, respectively). Moreover, an even numerator BS state's TQFT takes the form of a product of a TQFT describing the neutral sector and a TQFT describing the charged sector. In particular, the charge sector will be a $\mathbb{Z}_q$ TQFT, and the neutral sector TQFT will be one of the eight Galois conjugates of the Ising TQFT, meaning it has the same Ising fusion rules listed in Eq.~(\ref{eq:Ising_fusion_rules}), but the $F$-symbols and $R$-symbols may be different (see, e.g. Table 1 of Ref.~\onlinecite{Kitaev06a}). These eight different Ising TQFT Galois conjugates can be distinguished entirely by the topological twist factor $\theta_\sigma = e^{i \frac{2n+1}{8} \pi}$ of the $\sigma$ quasiparticle, with the integer $n$ (mod 8) labeling the different TQFTs. For example, the Ising TQFT has $\theta_\sigma = e^{i \frac{\pi}{8}}$ and the $\nu=2/5$ BS state built over the MR state has $\theta_\sigma = e^{-i \frac{3}{8} \pi}$ in its neutral sector.

Since these even numerator BS states have this product structure, the coupling of the TQFT to (gapless) photons to electric charge will split the degeneracy associated with the charge sectors, but not that of the neutral sectors. The neutral sector (Ising Galois conjugate) TQFT structure is fully preserved in this quasi-topological phase. The pseudo-ground state degeneracy on any manifold is that of the Ising TQFT. Despite this, the quasi-topological phase is still not quite strong (though it is close). This is because the braiding statistics of quasiparticles is not quite described by the neutral sector TQFT, as there are additional Abelian braiding statistics phase factors that arise from the charged sector, i.e. $R = R_{\text{Ising}} R_{\text{U}(1)_K}$. We describe this situation by saying the the neutral topological
degrees of freedom form a ``strong subsector'' of the quasi-topological phase.
There are, of course, also non-universal (dynamical and Aharonov-Bohm) phase factors that arise for quasiparticle exchange operations, just as for the other cases.

The stability of a quasi-topological phase is greater than one might
naively expect -- not just the ``strong'' ones, which have a strong family
resemblance to (true) topological phases, but even more generic quasi-topological phases.
Suppose, for instance, that an Ising-type FQH state
is coupled to gapless electrons. This could be due to a metallic lead.
Or, for instance, suppose that there is a parallel
conduction channel in a GaAs quantum well (e.g. in the dopant layer),
and that there is tunneling from the parallel conduction channel
to the two-dimensional electron gas (2DEG).
Even if the tunneling were to occur at a $\sigma$ quasiparticle,
where there is a Majorana zero mode, there would be a charge gap protecting
the bulk system against such tunneling at low energies. For an electron to tunnel
into an Ising-type state, it must create a neutral fermion (which costs zero
energy at a $\sigma$ quasiparticle) {\em and} a charge $-e$, which
costs a finite energy (roughly twice the transport gap). However, a
gapless Majorana fermion edge excitation is very different from an electron in this respect.
A gapless Majorana fermion from the edge {\it can} tunnel to a zero mode in the bulk
at low energies. However, this tunneling will be exponentially suppressed with distance of the bulk quasiparticle from the edge.
Therefore, it is important to keep $\sigma$ quasiparticles
far from the edge to prevent this from happening~\cite{Bishara09b,Rosenow09,Clarke11}.

As we noted in the case of the Kitaev honeycomb lattice model,
neutrinos are not a problem, either. Although no charged excitations
would be created if a neutrino were to absorbed by the Majorana zero mode
of a $\sigma$ quasiparticle, such a tunneling operator would be suppressed
because it is a non-local operator when written in terms of electrons.
To see this, note that the Majorana zero mode operator is a Bogoliubov-de Gennes operator
for a composite fermion $p+ip$ superconductor. Since the composite
fermion operator is non-local in terms of the electron operator,
$\psi_{\text{cf}}(x)=\psi_{\text{el}}(x)\,e^{i\int_{x_0}^{x}a}$ (the gauge field
$a$ is determined by the Chern-Simons constraint), where $x_0$ is on the boundary of the system, a term that is linear in such an operator is non-local in terms of electron operators.

Therefore, the quasiparticle degeneracy and braiding statistics
of an Ising-type FQH state is robust against phonons, photons, and (weak) coupling
to a metal. However, only the projective part of the braid group representation
is likely to be readily measurable because the dynamical phase is never arbitrarily small,
since the gapless excitations in the system do not decouple
in the low-energy limit. Moreover, the pseudo-ground state degeneracy
on higher-genus surfaces exhibits an $O(L_z^{-1})$
splitting due to the interaction with photons (though some states retain protection of the degeneracy and other TQFT properties of their neutral sector).

In such a quasi-topological phase, no fine-tuning is required
to keep the system in this phase.
Consequently, it is not fatal to have mobile
electrons in the dopant layer. Even though they can, in principle, tunnel into
the quantum Hall droplet, they are prevented from doing so at low energies.
Even when there are degenerate states due to Majorana zero modes,
it costs a non-zero energy for an electron to tunnel into the system because
charged excitations must be created, and they have non-zero energy cost.
Therefore, as we discuss in Section \ref{sec:protection},
topological qubits in the Ising-type FQH states
(or any other non-Abelian FQH state) can serve as
quantum memories which are perfect in the limit that the temperature
approaches zero and all distances (the size of the system, the separation
between quasiparticles) go to infinity. However, the overall phase associated with
braiding is not protected. This is not important because we are generally
interested in relative phases.

\subsubsection{Read-Rezayi States}

The $\mathbb{Z}_k$-Parafermionic Read-Rezayi (RR) FQH states (which have filling fraction $\nu=\frac{k}{Mk+2}$, where $M$ is odd for fermionic systems) exhibit behavior similar to the Ising-type FQH state when coupling to gapless degrees of freedom. These states similarly split into two classes. For all $k$ even, the RR state's corresponding TQFT does not have the form of a product of TQFTs (the neutral and charged quasiparticles are inseparably linked through modular transformations). For these states, the TQFT's ground state degeneracy due to genus is completely broken by the introduction of gapless photons ($N_{\Sigma}=1$ for manifolds $\Sigma$ without boundaries), while the degeneracy associated with boundaries and non-Abelian quasiparticles, as well as the quasiparticle braiding statistics are described by the RR state's TQFT.

For all $k$ odd, the RR state's corresponding TQFT is the product of a neutral sector TQFT and a charged sector TQFT~\cite{Bonderson07b}. In particular, the neutral sector TQFT is given by SU$(2)_k / \mathbb{Z}_2$ (which is the restriction of the SU$(2)_k$ Chern-Simons TQFT to integer spin quasiparticles). For example, in the $k=3$ RR state, the neutral sector is described by the Fibonacci TQFT [SU$(2)_3 / \mathbb{Z}_2$]. Again, the charged sector TQFT experiences similar effects due to the presence of photons, but the neutral sector is unaffected. The neutral sector is, thus, a strong subsector and is describe by a fully preserved SU$(2)_k / \mathbb{Z}_2$ TQFT structure. However, these quasi-topological phases are not strong, because the charge sector still contributes to the braiding statistics of quasiparticles, which is described by the full RR state's TQFT.

\section{Electron Parity-Protected and other Symmetry-Protected
Quasi-Topological Phases}
\label{sec:electron-parity}

\subsection{2D Superfluids and Superconductors}

In the previous examples, we have seen how the
topological character of a phase is modified by
coupling to Goldstone bosons, gapless gauge field excitations (such
as photons), and even gapless fermions. However, the Goldstone bosons
which we considered -- phonons -- were, in a sense, ``external'' to the topological system.
We now consider what happens when the Goldstone bosons are more
intimately linked to the topological degrees of freedom.

2D paired fermion superfluids and superconductors~\cite{Volovik99,Read00,Ivanov01} share many properties with
topological phases and are often treated as though they actually
topological phases. An $s$-wave paired fermion superfluid has vortex excitations
and unpaired fermions. Since vortices have logarithmic interactions with
each other, it is difficult to define a regime in which the braiding phase dominates
the dynamical phase when vortices are moved. However, in a two-dimensional superconductor,
the vortex-vortex interaction falls off as $1/R^2$. Therefore, for large enough
separation, we can define braiding. Vortices are bosonic with respect to each other,
but when a vortex encircles an unpaired fermion or vice versa,
the state of the system acquires a minus sign. In this respect, vortices and unpaired fermions
are similar to the $e$ and $m$ quasiparticle types in the toric code phase.
Furthermore, one can consider a collective formed by a vortex and an unpaired
fermion, and it is analogous to an $\psi$ quasiparticle. The fusion rules also follow
those of the quasiparticle types of the toric code. A chiral $p$-wave paired fermion
superfluid has Majorana zero modes in its vortex cores. As a result, it has
the fusion rules expected for $\sigma$ particles in the Ising topological phase.
Braiding can only be defined up to a phase, as result of the logarithmic
interaction between vortices noted above; the relative braiding phases (e.g.
the ratio $R^{\sigma\sigma}_{I}/R^{\sigma\sigma}_{\psi}$) are those expected
for Ising anyons. In a chiral $p$-wave superconductor, the braiding phases
are well-defined and equal to those of Ising anyons, with an unpaired fermion
identified with the $\psi$ quasiparticle. Despite these similarities between
superfluid/superconducting states and topological phases, there are important
differences, as we will see.

Consider, for example, the 2D paired superfluid
or superconductor on a torus.
For simplicity, we first consider the example of $s$-wave pairing
and then discuss the more interesting example of $p$-wave pairing.
We begin with the case of neutral superfluids, which have gapless Goldstone modes.
The order parameter $\theta$ will have winding numbers ${w_m}$ and ${w_l})$ around the meridian and longitude of the torus, respectively.
We assume that the torus has modular parameter $i$
or, in other words, that it is an $L\times L$ square with periodic boundary conditions,
so that we have $\theta= 2\pi {w_m}x/L + 2\pi {w_l}y/L$.
As a result of the Goldstone mode, a state with
winding numbers $({w_m},{w_l})$ on this torus will have energy
\begin{equation}
\label{eqn:order-param-energy}
E = \frac{\rho_s^{2D}}{2} \int {d^2}x \, (\nabla \theta)^2 =  {2\pi^2 \rho_s^{2D}}({w^2_m}+{w^2_l})
\end{equation}
We assume that the torus is very thin,
and $\rho_s^{2D}$ is the superfluid density per unit area.
Therefore, the $(0,0)$ state is the ground state
and the gap to non-zero winding states is a constant, independent of $L$.
(A four-fold degenerate ground state,
corresponding to winding numbers $(0,0), (1,0), (0,1), (1,1)$, would be
expected if a two-dimensional $s$-wave paired superfluid were in a toric code
topological phase, as one might have anticipated,
for reasons discussed above.)

In the case of a chiral $p_x \pm i p_y$ paired superfluid, a similar analysis holds, although
the situation is more complicated because gauge symmetry and rotational symmetry
are intertwined by the ordering. As a result, the $(1,1)$
state has odd fermion number and there are only three pseudo-ground states with
even fermion number. However, the basic observation in Eq.~(\ref{eqn:order-param-energy})
remains unchanged: the splitting between would-be ground states
on the torus is a constant, independent of $L$.
A chiral $p_x \pm i p_y$ paired superfluid has an additional feature which
is not present in the $s$-wave case. Consider a chiral $p_x \pm i p_y$ paired superfluid with $2n$ vortices
(i.e. $2n$ punctures with a vortex at each puncture). There will be a
$2^{n-1}$-dimensional space of pseudo-ground states, due to the existence of
Majorana (nearly) zero modes localized at the vortices. This degeneracy has
exponentially-decaying splittings caused by hybridization between the (nearly) zero
modes; fluctuations of the order parameter do not change this. Therefore, in the chiral
$p_x \pm i p_y$ case, the stiffness of the order parameter
causes some energy splittings to be independent of the system
size while others still decay exponentially.

Now suppose that we couple a chiral $p_x \pm i p_y$ superconductor to gapless fermions.
A fermion can tunnel into a Majorana zero mode because its fermion number can be absorbed by
the condensate. Suppose, for instance, that we have a metallic lead which
is brought into contact with a vortex. The coupling takes the form
\begin{equation}
\label{eqn:zm-lead-coupling}
H_{\rm tun} = \lambda \,c^\dagger(\vec{r}=0)\,\, \gamma \, e^{-i\theta/2} +
\lambda^* \gamma \,\,c(\vec{r}=0)\,\, e^{i\theta/2}
\end{equation}
Here, $\gamma$ is the Majorana zero mode operator and $c(\vec{r}=0)$ is the electron
annihilation operator in the lead at $\vec{r}=0$, which is assumed to be the location
of the vortex. In the superfluid state, $\theta$ can be taken to be a constant.
Whereas $\langle\gamma(\omega) \gamma(-\omega)\rangle = 1/(\omega+i\delta)$
(with $\delta \rightarrow 0$) in the absence of the lead,
in the presence of the lead we, instead, have
$\langle\gamma(\omega) \gamma(-\omega)\rangle = 1/(\omega+i\Gamma)$,
with $\Gamma=\lambda^2 N_F$, where $N_F$ is the density of states at the
Fermi energy in the lead. Consequently, the zero mode is absorbed by the lead,
and there is no longer any ground state degeneracy, as evidenced by
the absence of a pole at $\omega=0$.
It is, of course, obvious that this would happen,
but it is instructive to see how the case of a superfluid is different from
the Kitaev honeycomb lattice model or an FQH state.
In the latter two cases, the charged mode is gapped, so the analogue of the
operator $e^{-i\theta/2}$ (which we can treat as a constant in a superfluid)
creates a gapped excitation.

We now consider the case of a two-dimensional {\it superconductor}
on a torus. In the $s$-wave case, this system would be in a toric code
topological phase if the electromagnetic gauge field were
purely two-dimensional. However, the three-dimensionality
of the electromagnetic field leads to important differences, as we now see.

A non-zero winding number state will have magnetic flux
threading through the corresponding generator of the torus.
If the superconducting surface is thicker than the London penetration
depth, then the flux will be $hc/2e$. If the torus has longitude $\sim L$, then
the magnetic field strength will be $|B| \propto 1/L^2$. If the torus also has meridian
$\sim L$, then the total magnetic field energy will be $\propto B^2 L^3 \propto 1/L$.
In addition, there will be some gradient or superflow energy coming from the
toroidal shell which is within a London penetration depth $\lambda$ of the
inner and/or outer surface (depending on which hole has flux penetrating through it).
For $L\gg \lambda$, the superflow or gradient energy will be
$E_{\rm sf} \propto {\rho_s^{3D}} \lambda ({w^2_m}+{w^2_l})$,
where $\rho_s^{3D}$ is the superfluid density per unit volume.
Therefore, the field energy gives a contribution to the splitting
between ground states which is $\propto 1/L$ while the
order parameter gradient energy gives a contribution which is independent
of $L$ for the (realistic) case of $\rho_s^{3D}$ which is independent of the thickness
$L'$ of the torus.
(However, when expressed in terms of $\rho_s^{2D}$, the gradient energy
takes the form $E_{\rm sf} \propto ({\rho_s^{2D}}/L') \lambda\cdot({w^2_m}+{w^2_l})$,
from which we see that the gradient energy is reduced from the neutral superfluid case
by a factor of $\lambda/L'$, where $L'$ is the thickness of the torus.)
On the other hand, if the superconducting surface is thinner
than the London penetration depth, $L < \lambda$ then the flux can be much smaller.
In the limit of an infinitely-thin superconducting surface, there will be no magnetic field,
and there will be an energy cost for non-zero winding due to the gradient
of the order parameter given in Eq.~(\ref{eqn:order-param-energy}),
precisely as in the case of a neutral superfluid.
On the other hand, the splitting of multi-vortex states
caused by coupling to gapless fermions as in Eq.~(\ref{eqn:zm-lead-coupling})
is unchanged by the electromagnetic field because
Eq.~(\ref{eqn:zm-lead-coupling}) is already gauge-invariant, so there is
a constant splitting, regardless of whether we are dealing with a neutral
or charged superfluid.

From the preceding discussion, we see that
in 2D superconductors and superfluids, there is no pseudo-ground state
degeneracy on the torus ($N_{T^2}=1$) and, if there is a coupling to gapless electrons,
there is no degeneracy of multi-vortex states.
These two lifted degeneracies are related. The latter is due to the fact that an electron can tunnel into
a Majorana zero mode without having to overcome an energy gap (if we can neglect fluctuations of the order parameter, which
is the case in the ordered superfluid or superconducting states). The connection
to the case of the torus follows from the fact that an electron
can be converted into a Majorana fermion in the superconductor by a coupling
such as that in Eq.~(\ref{eqn:zm-lead-coupling}).
In two of the three would-be ground states on the torus,
a Majorana fermion operator $\gamma$ must change sign in going around one of the generators of the torus.
But, since an electron creation/annihilation operator ${c^\dagger}$ or $c$ must be single-valued and
$H_{\rm tun}$ must be single-valued, the phase of the order parameter $\theta$
must also wind around the corresponding generator.
This winding leads to the energy splitting on the torus in Eq. (\ref{eqn:order-param-energy}).

An underlying reason that a 2D superconductor is not in a topological
phase is because the 3D electromagnetic field does not
give an Anderson-Higgs gap to the 2D Goldstone mode.
The 3D electromagnetic field gives a gap to the would-be Goldstone mode
of a 3D superconductor, and a {\it 2D gauge field} (which would
have to be an emergent gauge field)
would give a gap to the would-be Goldstone mode
of a 2D superconductor (the latter is the scenario considered in Ref.~\onlinecite{Hansson04}).
However, the 3D electromagnetic field leaves a gapless
would-be Goldstone mode with dispersion relation $\omega_p \propto \sqrt{p}$.
Therefore, a 2D superconductor could, at best, be in a quasi-topological phase,
and even this is only possible if electron parity is maintained.
There is a second effect of the three-dimensionality of the gauge field:
the gauge field configuration which cancels the order parameter gradient,
${\bf A} = {\bf \nabla} \theta$ must necessarily have a non-zero magnetic field
somewhere in 3D space, i.e. there is no pure gauge extension
of ${\bf A} = {\bf \nabla} \theta$ to the entire 3D world. As a result, there is
magnetic field energy associated with non-zero windings and even some gradient energy
due to the non-zero penetration depth of the magnetic field. If, on the other hand,
the universe were a 3D torus $T^3$ and the superconducting torus $T^2$
wrapped around two of the directions of the universe, then the gauge field ${\bf A} = {\bf \nabla} \theta$ could be trivially extended to the whole universe and there would be no energy splitting
between different ground states on the torus. Therefore, one can understand
the order parameter gradient energy splitting between different would-be ground states
on the torus as resulting from a mismatch between the topology of the 2D system and the topology
of the 3D space. (Note that a 2D gauge field would perfectly
screen gradients in the order parameter, rather than leave an unscreened
layer at the surface of thickness equal to the penetration depth.)
Furthermore, the embedding of the torus
in a real 3D world leaves open the possibility that different
2D pseudo-ground states might be {\it locally} distinguishable by measurements
in three dimensions, which is the case when there is magnetic flux
threading the torus.

\subsection{Definition}

A chiral 2D $p$-wave superconductor is an example of an
{\it electron parity-protected quasi-topological phase}: a phase which
is quasi-topological so long as electron parity is conserved.
If we generalize this, by replacing electron parity by any symmetry,
then we have the notion of a {\it symmetry-protected quasi-topological phase}.

{\bf Definition (Symmetry-Protected Quasi-Topological Phase):}
A system is in a symmetry-protected quasi-topological phase
associated with a symmetry group $G$ if there is an energy $\Delta$
and a length scale $\xi$ which, respectively, have a strictly positive limit and a finite, non-negative limit as $L\rightarrow \infty$, such that the following properties hold.

(i) On a manifold $\Sigma$, there is a set of orthonormal energy eigenstates $|a\rangle$, where $a\in \{1,2,\ldots,{N_\Sigma}\}$, including the absolute ground state $|1\rangle$ (with energy $E_1=0$), such that
for any local operator $\phi$ that is invariant under
the group $G$
\begin{equation}
\label{eqn:sp-exponential2}
\langle {a}| \, \phi \, | {b}\rangle = {C} \delta_{ab} + O(e^{-L/\xi} )
\end{equation}
for $a, b \in \{1,2,\ldots,{N_\Sigma}\}$, where ${C}$ is independent of $a, b$.
For operators which are not invariant under $G$, the corrections will be bounded below by $\alpha L^{-\zeta}$ for some nonzero coefficient $\alpha$ and finite non-negative exponent $\zeta$.

The Hamiltonian is a particular local operator that is invariant under
the group $G$, so it follows that
\begin{equation}
\delta {E_0} \equiv \text{max}(|{E_{a}}|) = O(e^{-L/\xi})
\end{equation}
for $a\in \{1,2,\ldots,{N_\Sigma}\}$, where $E_{a}$ is the energy of the state
$|{a}\rangle$.

(ii) $N_\Sigma$ depends only on the topological configuration of the system.

(iii) We define
\begin{equation}
{\Omega} \equiv \text{min}(|{E_\chi}-E_{a}|)
\end{equation}
for any $a\in \{1,2,\ldots,{N_\Sigma}\}$
and $|\chi\rangle \notin \mathcal{H}_{0} \equiv \text{span}\left\{ |1\rangle, \ldots, | N_\Sigma \rangle \right\}$.
Then, there must exist a finite $z>0$ and an $L$-independent $\kappa>0$ such that
\begin{equation}
\Omega\sim\kappa L^{-z}.
\end{equation}

(iv) There are no states with energy less than $\Delta$ that have topologically nontrivial excitations.

Note that, as in the case of quasi-topological phases,
the pseudo-ground state space of a symmetry-protected
quasi-topological phase does not need to be described by a TQFT
(though it is for strong quasi-topological phases).
In the case of a 2D superconductor, $N_\Sigma = 1$ for manifolds without boundaries, but the quasiparticle braiding statistics (and pseudo-ground state degeneracy for 2D chiral $p$-wave superconductors) is nontrivial.
These features definitely cannot be described by a TQFT, since $N_\Sigma = 1$ for all manifolds without boundaries only for a trivial TQFT.

Symmetry-protected quasi-topological phases are similar in
spirit to {\it symmetry-protected topological phases}
\cite{Chen11a,Chen11b,Kitaev-unpub,Lu12}.
Topological insulators are an example of the latter: they
cannot be continuously deformed into trivial insulators if
time-reversal symmetry is respected. However,
they have no non-trivial quasiparticles and if one is allowed to
violate time-reversal symmetry, then topological insulators can
be continuously deformed into trivial insulators. Thus, when
viewed as topological phases, subject to {\it arbitrary} perturbations,
they are trivial. Similarly, the following systems are trivial
topological phases in the absence of symmetry protection:
$p_x \pm i p_y$ superconductors \cite{Read00,DasSarma06a};
three-dimensional topological insulator--$s$-wave superconductor
heterojunctions with $hc/2e$ vortices present \cite{Fu08};
superconductor--spin-orbit-coupled semiconductor--
ferromagnetic insulator heterojunctions \cite{Sau10}; and spin-orbit-coupled
quantum wires with superconductivity induced through the
proximity effect \cite{Kitaev01,Lutchyn10,Oreg10,Alicea11}.
However, if {\it electron parity} is preserved, then they all have interesting
topological character, which fits the definition above.

For instance, in the case of three-dimensional topological
insulator--$s$-wave superconductor heterojunctions, it is crucial
that the topological insulator have an insulating three-dimensional bulk.
As a matter of principle, the insulating phase is stable and
exists over a range of parameters, so fine-tuning is not
required. However, as a practical matter, in insulators with small
band gaps (such as all of the known topological insulators), even
a small density of impurities will lead to metallic behavior. Therefore,
some effort is needed to eliminate this bulk electrical conduction.
In the case of spin-orbit-coupled quantum wires with proximity-induced
superconductivity, metallic gates may be used to bring the chemical
potential into the necessary regime. This does not require fine-tuning
since there is a range of chemical potentials in which the wire
will support Majorana zero modes. However, the leakage of
electrons from the metallic gates to the quantum wire must be suppressed,
or else electron parity will not be conserved in the wire.

In order to guarantee that electron parity is preserved, some effort may
be required, as in the case of conventional qubits, as described above
in the case of ions and superconducting qubits. However, once this
this is accomplished -- which amounts to the preservation
of a (discrete) symmetry -- then Majorana
fermion zero modes can serve as topologically protected quantum memory
and other sources of error will be exponentially suppressed in the temperature
or system size.

However, as we have seen from the discussion of the previous
section, electron parity-protected quasi-topological phases differs from topological
phases very drastically when it comes to the splitting between would-be
ground states on (for instance) the torus: in the former case, this splitting remains finite,
even for large system size, while in the latter case, it vanishes exponentially fast.
The only way to make the splitting small is to fine-tune it -- essentially by
suppressing the superconducting long-ranged order.

\section{Protecting Quantum Information}
\label{sec:protection}

\subsection{Conventional Systems}

To understand the protection of quantum information
afforded by topological and quasi-topological phases of matter,
it is helpful to contrast it with ``conventional'' qubits based on
local degrees of freedom. As examples, we consider
trapped ions and superconducting charge qubits.

In ion trap qubits, the two-level system which
stores quantum information is the ground state and
a long-lived excited state of an ion. The excited state has
a long lifetime because the decay rate through photon emission
is proportional to the fine structure constant $\alpha\approx 1/137$, which is small,
and to the square of the transition matrix element, which is small if the
transition is electric dipole-forbidden. The energy splitting between the
two states is very accurately measured and stable, so long as the ion
is isolated. Thus trapped ions form excellent qubits, for which gates
have been designed~\cite{Cirac95}. However, although the lifetime of
the excited state is very long, it is fixed once and for all, once the
ion and the excited state have been chosen.
Said differently, the ionic qubit is long-lived because it interacts
weakly with the environment. Though this interaction is weak,
it cannot be made arbitrarily weak.
Therefore, for computations longer than the lifetime of the excited
state, it will be necessary to use software-based error correction
(see, e.g. Ref.~\onlinecite{Nielsen00}).

A superconducting charge qubit~\cite{Bouchiat98}
(or ``Cooper pair box'') is a pair of superconducting islands whose
charging energy is tuned so that the state with $N$ Cooper pairs on one island
is degenerate with the state with $N+1$ Cooper pairs on that island. This qubit
is stable as long as the temperature is much lower than the gap,
so that the system cannot be in a state with $N+\frac{1}{2}$ Cooper pairs
due to thermally-excited quasiparticles. However, the energy difference
between the states with $N$ and $N+1$ Cooper pairs can vary if the electrostatic
potential varies. Gate voltage noise or stray charges can easily cause such
potential variations, so Cooper pair boxes can only operate as qubits
if these sources of electrostatic potential fluctuations are kept small.
In a ``transmon'' qubit~\cite{Koch07}, the Josephson coupling between the islands
is sufficiently large that the energy difference between the two states
is insensitive to variations in the electrostatic potential. In fact,
the sensitivity is exponentially-small in the ratio between the Josephson
and charging energies, $E_J/E_c$. However, this ratio cannot be
made arbitrarily large because, if $E_J/E_c$ is too large,
then quantum gate operations are likely to cause leakage errors
to higher excited states. Therefore, once again, one must choose
a ratio $E_J/E_c$ to give a small, but fixed, error rate.
For longer computations, software error correction is necessary.

\subsection{Topological Protection}

In a topological phase, quantum information is protected for
a different reason: it is encoded in degrees of freedom that
do not couple to local perturbations (up to corrections that are
exponentially suppressed in $L$). Although some tuning may be necessary
to bring a system into a topological phase, the phase will be stable
throughout a region of parameter space and no system parameters
need to be at precise values to optimize the protection from errors.
A topological qubit's lifetime $\tau$ increases without bound
as the temperature is decreased and the system's length scales increased ($\tau \sim \Delta/T , L/\xi$).
In contrast, for conventional qubits (such as those described in the previous paragraphs), it is not possible, even in principle, to make the error rate arbitrarily small as a function of some parameter. Of course, in practice, there may be very little difference between a long-lived topological qubit and a long-lived conventional qubit, and topological qubits may ultimately require software error correction, but there is, at least, a difference in principle.

In addition to protecting the quantum information encoded in qubits,
it is also important to manipulate quantum information without introducing errors.
This provides another point of contrast between the conventional and topological approaches. Similar to the situation for qubit lifetimes, the fidelities of gates and measurements for conventional qubits are fixed at some optimized value by the system parameters, whereas, for topological qubits, these fidelities can be made arbitrarily close to $1$, with exponentially suppressed errors as $L$ is increased and $\omega$ and $T$ are decreased.
One might na\"{i}vely worry that qubits which are truly topologically protected from noise will also be incapable of being read out or acted upon with computational gates. However, topological qubits can be read out either by performing an operation that makes the (formerly non-locally encoded) quantum information local, e.g. by moving quasiparticles close to each other, followed by a local measurement, or by performing a non-local measurement of the topological degrees of freedom, e.g. anyonic interferometry~\cite{Bonderson07c}. Such measurements enjoy topological protection in the sense that they are measuring topological quantum numbers which are not destroyed by the measurement nor local operations, so they can be repeatedly measured to improve the measurement fidelity.

In a topological phase, unitary transformations may be generated by performing topological operations on the system,
such as adiabatic quasiparticle exchanges~\cite{Kitaev97,Freedman98}, topological charge measurements~\cite{Bonderson08a,Bonderson08b}, or changing the system's topology~\cite{Bravyi00,Freedman06,Bonderson10}. The action of such unitary transformations on the pseudo-ground Hilbert space will similarly be topologically protected, i.e. have
exponentially-suppressed errors, as long as the system's lengths are long compared to the correlation length and the temperature, energies, and inverse time scales are small compared to $\Delta$. However, the set of topologically-protected gates that can be generated may not be computationally universal, meaning it may not enable arbitrary quantum computations to be performed.
For two-dimensional systems, the ideal scenario is when braiding operations are computationally universal, which is the case for many
TQFTs, such as SU$(2)_k$ for $k=3$ and $k\ge 5$~\cite{Freedman02b}. For these cases, one only needs planar geometries with quasiparticles/boundaries and the ability to perform (braiding) exchanges or topological charge measurements of pairs of quasiparticles/boundaries to generate computationally universal gates.

Unfortunately, in a topological phase of Ising anyons [which is computationally equivalent to SU$(2)_2$], braiding operations alone cannot generate universal gate sets. This is because, working in the standard (one qubit in four $\sigma$ quasiparticles) encoding, braiding only generates the single qubit Clifford gates. To make this gate set universal, it must be supplemented by an entangling gate, such as the controlled-NOT or controlled-$Z$ gate, and a phase gate outside the set, such as the $\pi/8$-phase gate (or any $\theta$-phase gate with $\theta \neq n \frac{\pi}{4}$). A topologically-protected entangling gate can be implemented given the ability to measure the collective topological charge/total fermion parity of up to (at least) four Ising $\sigma$ quasiparticles~\cite{Bravyi00b}. Both of these missing gates can be implemented with topological protection for the Ising TQFT given the ability to perform certain topology changing operations~\cite{Bonderson10}, following the protocol of Bravyi and Kitaev~\cite{Bravyi00} (see Ref.~\onlinecite{Freedman06} for a recapitulation of their construction). For these topology-changing protocols, the topological protection of degeneracies in the pseudo-ground state subspace due to higher genus plays a crucial role.
onsequently, topology-changing protocols do not produce topologically-protected gates in those
quasi-topological Majorana phases that do not preserve these degeneracies.
Without topology changing operations, one must resort to topologically unprotected operations to generate the missing phase gate, such as using quasiparticle separation to explicitly split the energy degeneracies, interference techniques~\cite{Bonderson10a}, or coupling to conventional qubit systems~\cite{Jiang11,Bonderson11}. Fortunately, the other
topologically-protected Ising gates allow for error-correction of a noisy $\pi/8$-phase gate (which is sufficient supplementation to achieve universality) with a remarkably high threshold of approximately $14\%$~\cite{Bravyi05}.

To summarize, a topological phase will always have some gates (necessarily including the identity gate) that are topologically-protected, which means they can be made arbitrarily accurate by lowering the temperature and increasing the systems' length scales. The contrast with conventional qubits, which enjoy no topological protection, is starkest when the protected gates are computationally universal. When some unprotected gates are needed, the situation vis-a-vis these gates is similar, in principle, to conventional qubits, although error-correction thresholds may be significantly less strict. We now consider how the situation differs for quasi-topological phases (reiterating some of the previous discussions).

\subsubsection{Strong quasi-topological phases}

Strong quasi-topological phases behave essentially the same as topological phases. The low-energy ($E<\Delta$) degrees of freedom are a direct product of a TQFT and the gapless degrees of freedom. All topological degeneracies and operations given by the TQFT are preserved and topologically protected. In particular, topology changing protocols can be employed (at least in principle), if necessary. The visibility of interference processes may be reduced by the presence of the gapless degrees of freedom, but this is not a fundamental distinction.

Strong quasi-topological phases with universal braiding could potentially be realized by certain lattice models~\cite{Kitaev97,Levin05a}, however, these are likely too difficult to practically implement. Kitaev's honeycomb model in the $B$ phase exhibits the Ising TQFT, which, as previously mentioned, requires topology changing operations to achieve topologically protected universal quantum computation.

\subsubsection{Quasi-topological phases}

Quasi-topological phases that are not ``strong'' have a pseudo-ground state subspace that is not described by a TQFT. This pseudo-ground state subspace may still exhibit exponentially-suppressed splitting of degeneracies and, hence, topological-protection of qubits and topological gate operations acting within this subspace. The algebraic structure of these protected subspaces and operations within them will depend on the details of the system. Typically, the pseudo-ground states will be those of the non-local anyonic state space of non-Abelian anyons (boundaries/quasiparticles) in the system, rather than states associated with non-trivial topology (e.g. genus). However, it is possible for a system to have a strong subsector, i.e. a subspace of the pseudo-ground state subspace that is actually described by a TQFT and behaves like a strong quasi-topological phase. Within this smaller subspace, topology changing operations would still be topologically-protected. We emphasize that topologically protected quantum information and computational gates are better protected than is sometimes appreciated for quasi-topological phases. For example, these phases provide robust protection from gapless electrons.

As described in Section~\ref{sec:Ising-typeFQH}, Ising-type FQH states~\cite{Moore91,Lee07,Levin07,Bonderson07d,Bonderson11b}
are likely to occur in the second Landau level, notably at $\nu=5/2$ and $12/5$. These quasi-topological phases have topological protection associated with their non-Abelian anyons,
which carry Ising $\sigma$ topological charge.
Thus, qubits encoded in the non-Abelian quasiparticles and gates produced from braiding or measurement operations will be topologically-protected for all of these states.
As previously discussed, these operations alone are not computationally universal for Ising (or Ising-type) quasiparticles.
Topology changing operations will only be topologically protected within a strong subsector of a quasi-topological
phase (assuming is has one).
For the (fermionic) states with odd numerator (not counting fully filled Landau levels), such as the $\nu=5/2$ MR and anti-Pfaffian states, there is no strong subsector and none of the degeneracies due to genus remain protected.
In the (fermionic) states with even numerator (not counting fully filled Landau levels), such as the $\nu=12/5$ BS states, the neutral sector forms a strong subsector in which the TQFT structure retains topological protection.
We also note that, if the long-ranged Coulomb interactions could somehow be eliminated from FQH states,
so that topology-changing operations could be implemented with topological protection for all the Ising-type FQH states,
only the states with odd denominators have modular transformations capable of enhancing the gate set to make it computationally universal.

The (particle-hole conjugate\cite{Bishara08c} of the) $\mathbb{Z}_3$-parafermionic Read-Rezayi state~\cite{Read99} is also a strong candidate for the $\nu=12/5$ FQH state. This quasi-topological phase has topological protection associated with its non-Abelian anyons (which carry Fibonacci $\varepsilon$ topological charge). Moreover, the neutral sector is a strong subsector described
by the Fibonacci TQFT. Qubits encoded in the non-Abelian quasiparticles and gates produced from braiding or measurement operations will be topologically-protected, and quasiparticle braiding is computationally universal. Topology changing operations are protected within the strong pseudo-ground subspace and described by the Fibonacci TQFT. Actually, these properties are common to all $\mathbb{Z}_k$-parafermionic Read-Rezayi states with $k$ odd (and $k>1$), except that the strong subspace is described by the SU$(2)_k /\mathbb{Z}_2$ TQFT [the restriction of SU$(2)_k$ to integer spins].

\subsubsection{Symmetry protected quasi-topological phases}

Symmetry-protected quasi-topological phases behave as quasi-topological phases, with the prerequisite that the symmetry of the system is not violated by an external effect. The extent to which qubits and gates are topologically protected is bounded by the extent to which this symmetry is protected. Fortunately, it is not necessarily difficult to provide excellent protection of this symmetry. For example, fermion parity protected quasi-topological phases can be protected from fermion parity violating effects simply by introducing insulating layers between the symmetry-protected quasi-topological system and dangerous electron sources (e.g. metallic leads) capable of causing the topological protection to fail. This can be very strong protection since electron tunneling through an insulating barrier is exponentially suppressed as a function of inverse temperature and layer thickness.

The $p_x \pm i p_y$ paired superconducting systems~\cite{Read00,Kitaev01,DasSarma06a,Fu08,Sau10,Lutchyn10,Oreg10,Alicea11,Fidkowski11} are fermion parity protected quasi-topological phases. As long as the system is protected from fermion parity violations, these phases have topological protection associated with their Majorana zero modes (which are akin to Ising $\sigma$ topological charge), but do not maintain any topological protection for degeneracies associated with higher genus. Thus, qubits encoded in the Majorana zero modes and gates produced from braiding or measurement operations will be topologically-protected. As previously discussed, quasiparticle braiding operations alone are not computationally universal. Topology changing operations will not be protected.

It is worth mentioning that the potential drawback of needing to protect fermion parity in these superconductivity-based Majorana systems could also prove to be advantageous. Their sensitivity
to fermion parity is the basis for proposals to coherently couple topological qubits in these systems with conventional qubits~\cite{Jiang11,Bonderson11,Hassler11}. For example, the fermionic parity of a Majorana qubit
can be measured by the Aharonov-Casher effect with ordinary flux
$hc/2e$ superconducting vortices which may, themselves,
be outside the Majorana system.
Consequently, an $hc/2e$ vortex tunneling into a flux qubit
or around a quantum dot will be sensitive to the fermionic parity
of a Majorana qubit in an electron parity-protected
quasi-topological phase. This enables one to coherently couple and transfer
quantum information between the topologically protected qubits and
conventional qubits.

\section{Discussion}
\label{sec:discussion}

It is virtually unimaginable for a system to be in a topological
phase, according to the usual definition given at the beginning of
this paper. If we were to take an $L\times L \times L$ region of space
and shield it perfectly from the rest of the universe, it would still be the case
that the Hamiltonian for the interior of this region would have gapless
excitations. If it is not filled by a superconducting solid, then there will
be gapless photons (or, to be more precise, photons with a gap $c/L$).
If it is filled by a solid, superconducting or otherwise, then there will be gapless
phonons. In order to give an example of a topological phase in Section
\ref{sec:true}, we had to consider the unphysical limit of a superconducting
solid composed of infinitely-massive atoms.

It follows that TQFTs cannot give a complete description of real systems.
This is, perhaps, somewhat surprising because
the mathematical theory of TQFTs does not give any clues
about its limitations, unlike most quantum field theories, which
would suffer from ultraviolet divergences in the absence of a high-energy
cutoff and, therefore, are only applicable below this cutoff. All
predictions of TQFTs are perfectly finite and physically reasonable.

Therefore, the best that one can hope for in the
real world is a situation in which there are gapless excitations, but they decouple
at low energies from the topological degrees of freedom -- or, in the terminology
introduced in this paper, the system is in a {\it ``strong'' quasi-topological phase}.
In this case, a TQFT will apply to some of the degrees of
freedom of a system -- but only in the low-energy, low-temperature,
long-distance limit and, even in this regime, the
degrees of freedom which they do not encompass decouple only
as a power law in the temperature or system size.
(It would be interesting to see if the methods used to prove the stability
of topological phases \cite{Bravyi10,Bravyi11} can be extended to prove the stability
of quasi-topological phases as well.)

In many other cases, the situation may be even worse.
And yet, this does not mean that one should discard TQFTs
altogether. In fact some of their predictions hold (possibly with
modifications) in a wide variety of systems.
Therefore, it is important to distinguish between the different possible types of
discrepancies between a physical system and the predictions of a TQFT.

In this paper, we have classified such systems as quasi-topological phases
and symmetry-protected quasi-topological phases.
They offer successively less protection for quantum information.
Ironically, it appears that they are successively easier to realize in nature.
It is, therefore, important to understand if there is
a fundamental reason for this or whether it is merely an accident
due, ultimately, to the fact that we understand the physics of
topological superconductors, which are in an electron-parity protected quasi-topological
phase, better than we understand quasi-topological phases.
Another interesting question for further research is whether
there are more quasi-topological phases than topological phases
or, in other words, whether the loosening of the requirements
of a true topological phase allows for more possibilities.
We also note that a parallel classification may apply to critical points
at which a topological degeneracy coexists with gapless excitations,
such as the one introduced and analyzed in Ref. \onlinecite{Mulligan10}.

\acknowledgements
We would like to thank M. Hastings, M. Freedman, R. Lutchyn,
and Z. Wang for useful discussions.
C.N. is supported by the DARPA QuEST program
and the AFOSR under grant FA9550-10-1-0524.

\appendix
\section{Algebraic Properties of a $(2+1)$D TQFT/Modular Tensor Category}
\label{sec:AlgTQFT}

In this appendix, we briefly review the algebraic properties of fusion and braiding of quasiparticles, as described by unitary braided tensor categories (BTC). When a BTC has unitary $S$-matrix, it is a modular tensor category and corresponds to a TQFT. For additional details, see Refs.~\onlinecite{Bakalov01,Turaev94,Kitaev06a,Bonderson07b,Bonderson07c}.

There is a finite set of topological charges $\mathcal{C}$, which obey a commutative, associative fusion algebra
\begin{equation}
a \times b = \sum\limits_{c \in \mathcal{C}} N_{ab}^{c} \, c
\end{equation}
where $N_{ab}^{c}$ are positive integers indicating the number of distinct ways charges $a$ and $b$ can be combined to produce charge $c$. There is a unique ``vacuum'' charge, denoted $0$ or $I$, which has trivial fusion (and braiding) with all other charges, i.e. $N_{a 0}^{c} = \delta_{ac}$, and which defines the unique conjugate $\bar{a}$ of each topological charge $a$ via $N_{a b}^{0} = \delta{\bar{a} b}$.

Each fusion product has an associated vector space $V_{ab}^{c}$ with $\dim{V_{ab}^{c}} = N_{ab}^{c}$, and its dual (splitting) space $V^{ab}_{c}$. The states in these fusion and splitting spaces are assigned to trivalent vertices with the appropriately corresponding topological charges:
\begin{equation}
\pspicture[shift=-0.6](-0.1,-0.2)(1.5,-1.2)
  \small
  \psset{linewidth=0.9pt,linecolor=black,arrowscale=1.5,arrowinset=0.15}
  \psline(0.7,0)(0.7,-0.55)
  \psline(0.7,-0.55) (0.25,-1)
  \psline(0.7,-0.55) (1.15,-1)	
  \rput[tl]{0}(0.4,0){$c$}
  \rput[br]{0}(1.4,-0.95){$b$}
  \rput[bl]{0}(0,-0.95){$a$}
 \scriptsize
  \rput[bl]{0}(0.85,-0.5){$\mu$}
  \endpspicture
=\left\langle a,b;c,\mu \right| \in
V_{ab}^{c} ,
\label{eq:bra}
\end{equation}
\begin{equation}
\pspicture[shift=-0.6](-0.1,-0.2)(1.5,1.2)
  \small
  \psset{linewidth=0.9pt,linecolor=black,arrowscale=1.5,arrowinset=0.15}
  \psline(0.7,0)(0.7,0.55)
  \psline(0.7,0.55) (0.25,1)
  \psline(0.7,0.55) (1.15,1)	
  \rput[bl]{0}(0.4,0){$c$}
  \rput[br]{0}(1.4,0.8){$b$}
  \rput[bl]{0}(0,0.8){$a$}
 \scriptsize
  \rput[bl]{0}(0.85,0.35){$\mu$}
  \endpspicture
=\left| a,b;c,\mu \right\rangle \in
V_{c}^{ab},
\label{eq:ket}
\end{equation}
where $\mu=1,\ldots ,N_{ab}^{c}$. Most TQFTs of interest have no fusion multiplicities, i.e. $N_{ab}^{c}=0,1$, in which case the vertex labels $\mu$ are trivial and can be left implicit (as is done elsewhere in this paper).

General states and operators are described using fusion/splitting trees constructed by connecting lines with the same topological charge. The trivalent vertices written here are orthonormal, in that the inner product is defined as
\begin{equation}
\label{eq:inner_product}
\left\langle a,b;c,\mu | a,b;c',\mu' \right\rangle =
  \pspicture[shift=-1](-0.2,-0.35)(1.2,1.75)
  \small
  \psarc[linewidth=0.9pt,linecolor=black,border=0pt] (0.8,0.7){0.4}{120}{240}
  \psarc[linewidth=0.9pt,linecolor=black,border=0pt] (0.4,0.7){0.4}{-60}{60}
  \psset{linewidth=0.9pt,linecolor=black,arrowscale=1.5,arrowinset=0.15}
  \psline(0.6,1.05)(0.6,1.55)
  \psline(0.6,-0.15)(0.6,0.35)
  \rput[bl]{0}(0.07,0.55){$a$}
  \rput[bl]{0}(0.94,0.55){$b$}
  \rput[bl]{0}(0.26,1.25){$c$}
  \rput[bl]{0}(0.24,-0.05){$c'$}
 \scriptsize
  \rput[bl]{0}(0.7,1.05){$\mu$}
  \rput[bl]{0}(0.7,0.15){$\mu'$}
  \endpspicture
=\delta _{c c ^{\prime }}\delta _{\mu \mu ^{\prime }}
  \pspicture[shift=-1](0.15,-0.35)(0.8,1.75)
  \small
  \psset{linewidth=0.9pt,linecolor=black,arrowscale=1.5,arrowinset=0.15}
  \psline(0.6,-0.15)(0.6,1.55)
  \rput[bl]{0}(0.75,1.25){$c$}
  \endpspicture
.
\end{equation}%
(It is often convenient to weight the trivalent vertices with factors that provide an isotopy invariant formulation of diagrams, though we do not do so here.) The identity operator on a pair of anyons with charges $a$ and $b$ is written diagrammatically as
\begin{equation}
\label{eq:Id}
\openone_{ab} =
\pspicture[shift=-0.65](0,-0.5)(1.1,1.1)
  \small
  \psset{linewidth=0.9pt,linecolor=black,arrowscale=1.5,arrowinset=0.15}
  \psline(0.3,-0.45)(0.3,1)
  \psline(0.8,-0.45)(0.8,1)
  \rput[br]{0}(1.05,0.8){$b$}
  \rput[bl]{0}(0,0.8){$a$}
  \endpspicture
 = \sum\limits_{c,\mu } \,\,
 \pspicture[shift=-0.65](-0.1,-0.45)(1.4,1)
  \small
  \psset{linewidth=0.9pt,linecolor=black,arrowscale=1.5,arrowinset=0.15}
  \psline(0.7,0)(0.7,0.55)
  \psline(0.7,0.55) (0.25,1)
  \psline(0.7,0.55) (1.15,1)
  \rput[bl]{0}(0.38,0.2){$c$}
  \rput[br]{0}(1.4,0.8){$b$}
  \rput[bl]{0}(0,0.8){$a$}
  \psline(0.7,0) (0.25,-0.45)
  \psline(0.7,0) (1.15,-0.45)
  \rput[br]{0}(1.4,-0.4){$b$}
  \rput[bl]{0}(0,-0.4){$a$}
\scriptsize
  \rput[bl]{0}(0.85,0.4){$\mu$}
  \rput[bl]{0}(0.85,-0.03){$\mu$}
  \endpspicture
\;
.
\end{equation}

Associativity of fusion is represented in the state space by the $F$-symbols, which (similar to $6j$-symbols) provide a unitary isomorphism relating states written in different bases distinguished by the order of fusion. Diagrammatically, this is represented as
\begin{equation}
\label{eq:F_move}
  \pspicture[shift=-1.0](0,-0.45)(1.8,1.8)
  \small
  \psset{linewidth=0.9pt,linecolor=black,arrowscale=1.5,arrowinset=0.15}
  \psline(0.2,1.5)(1,0.5)
  \psline(1,0.5)(1,0)
  \psline(1.8,1.5) (1,0.5)
  \psline(0.6,1) (1,1.5)
   \rput[bl]{0}(0.05,1.6){$a$}
   \rput[bl]{0}(0.95,1.6){$b$}
   \rput[bl]{0}(1.75,1.6){${c}$}
   \rput[bl]{0}(0.5,0.5){$e$}
   \rput[bl]{0}(0.9,-0.3){$d$}
 \scriptsize
   \rput[bl]{0}(0.3,0.8){$\alpha$}
   \rput[bl]{0}(0.7,0.25){$\beta$}
  \endpspicture
= \sum_{f,\mu,\nu} \left[F_d^{abc}\right]_{(e,\alpha,\beta)(f,\mu,\nu)}
 \pspicture[shift=-1.0](0,-0.45)(1.8,1.8)
  \small
  \psset{linewidth=0.9pt,linecolor=black,arrowscale=1.5,arrowinset=0.15}
  \psline(0.2,1.5)(1,0.5)
  \psline(1,0.5)(1,0)
  \psline(1.8,1.5) (1,0.5)
  \psline(1.4,1) (1,1.5)
   \rput[bl]{0}(0.05,1.6){$a$}
   \rput[bl]{0}(0.95,1.6){$b$}
   \rput[bl]{0}(1.75,1.6){${c}$}
   \rput[bl]{0}(1.25,0.45){$f$}
   \rput[bl]{0}(0.9,-0.3){$d$}
 \scriptsize
   \rput[bl]{0}(1.5,0.8){$\mu$}
   \rput[bl]{0}(0.7,0.25){$\nu$}
  \endpspicture
.
\end{equation}

The counterclockwise braiding exchange operator of topological charges $a$ and $b$ is represented diagrammatically as
\begin{equation}
\label{eq:braid}
R_{ab}=
\pspicture[shift=-0.6](-0.1,-0.2)(1.3,1.05)
\small
  \psset{linewidth=0.9pt,linecolor=black,arrowscale=1.5,arrowinset=0.15}
  \psline(0.96,0.05)(0.2,1)
  \psline(0.24,0.05)(1,1)
  \psline[border=2pt](0.24,0.05)(0.92,0.9)
  \rput[bl]{0}(-0.02,0){$a$}
  \rput[br]{0}(1.2,0){$b$}
  \endpspicture
=\sum\limits_{c,\mu ,\nu }\,\left[
R_{c}^{ab}\right] _{\mu \nu }
 \pspicture[shift=-0.65](-0.1,-0.45)(1.4,1)
  \small
  \psset{linewidth=0.9pt,linecolor=black,arrowscale=1.5,arrowinset=0.15}
  \psline(0.7,0)(0.7,0.55)
  \psline(0.7,0.55) (0.25,1)
  \psline(0.7,0.55) (1.15,1)
  \rput[bl]{0}(0.38,0.2){$c$}
  \rput[br]{0}(1.4,0.8){$a$}
  \rput[bl]{0}(0,0.8){$b$}
  \psline(0.7,0) (0.25,-0.45)
  \psline(0.7,0) (1.15,-0.45)
  \rput[br]{0}(1.4,-0.4){$b$}
  \rput[bl]{0}(0,-0.4){$a$}
\scriptsize
  \rput[bl]{0}(0.85,0.4){$\nu$}
  \rput[bl]{0}(0.85,-0.03){$\mu$}
  \endpspicture
,
\end{equation}
where the $R$-symbols represent the unitary operator for exchanging two anyons in a specific fusion channel, i.e.
\begin{equation}
\label{eq:R_move}
\pspicture[shift=-0.65](-0.1,-0.2)(1.5,1.2)
  \small
  \psset{linewidth=0.9pt,linecolor=black,arrowscale=1.5,arrowinset=0.15}
  \psline(0.7,0)(0.7,0.5)
  \psarc(0.8,0.6732051){0.2}{120}{240}
  \psarc(0.6,0.6732051){0.2}{-60}{35}
  \psline (0.6134,0.896410)(0.267,1.09641)
  \psline(0.7,0.846410) (1.1330,1.096410)	
  \rput[bl]{0}(0.4,0){$c$}
  \rput[br]{0}(1.35,0.85){$a$}
  \rput[bl]{0}(0.05,0.85){$b$}
 \scriptsize
  \rput[bl]{0}(0.82,0.35){$\mu$}
  \endpspicture
= \sum\limits_{\nu }\left[ R_{c}^{ab}\right] _{\mu \nu}
\pspicture[shift=-0.65](-0.1,-0.2)(1.5,1.2)
  \small
  \psset{linewidth=0.9pt,linecolor=black,arrowscale=1.5,arrowinset=0.15}
  \psline(0.7,0)(0.7,0.55)
  \psline(0.7,0.55) (0.25,1)
  \psline(0.7,0.55) (1.15,1)	
  \rput[bl]{0}(0.4,0){$c$}
  \rput[br]{0}(1.4,0.8){$a$}
  \rput[bl]{0}(0,0.8){$b$}
 \scriptsize
  \rput[bl]{0}(0.82,0.37){$\nu$}
  \endpspicture
.
\end{equation}

A BTC is defined entirely by its $N_{ab}^{c}$, $F$-symbols, and $R$-symbols. The $N_{ab}^{c}$ must provide an associative and commutative algebra. The $F$-symbols and $R$-symbols are constrained by the ``coherence conditions'' (also known as the ``polynomial equations''), which ensure that any two series of $F$ and/or $R$ transformations are equivalent if they start in the same state space and end in the same state space~\cite{MacLane63}. Physically, these consistency conditions are interpreted as enforcing locality in fusion and braiding processes.

Distinct sets of $F$-symbols and $R$-symbols describe equivalent theories if they can be related by a gauge transformation given by unitary transformations acting on the fusion/splitting state spaces $V_{c}^{ab}$ and $V^{c}_{ab}$ (which are just redefinitions of the basis states). Ocneanu rigidity demonstrates that the only infinitesimal deformations of a UBTC that continue to satisfy the consistency conditions are infinitesimal gauge transformations of this sort~\cite{Ocneanu-unpublished,Etingof05}. Consequently, for a given fusion algebra, there are a finite number of distinct (up to gauge transformations) UBTCs that satisfy the coherence conditions. The physical consequence of Ocneanu rigidity is that the UBTC describing a system is unchanged by perturbations to the system.

{}From the quantum dimensions
\begin{equation}
d_{a} = d_{\bar{a}} = \left| \left[F_a^{a \bar{a} a}\right]_{00} \right|^{-1}
\end{equation}
and topological twist factors
\begin{equation}
\theta_{a} = \theta_{\bar{a}} =\sum_{c\in \mathcal{C}} \frac{d_c}{d_a} R^{aa}_{c} = d_a \left[F_a^{a \bar{a} a}\right]_{00} \left(R^{\bar{a} a}_{0}\right)^{\ast}
\end{equation}
(which are roots of unity) one can compute the topological $S$-matrix
\begin{equation}
S_{ab} = \frac{1}{\mathcal{D}} \sum_{c\in \mathcal{C}} N_{ab}^{c} d_c \frac{  \theta_c }{ \theta_a \theta_b}
\end{equation}
where $\mathcal{D} = \sqrt{\sum_{a\in \mathcal{C}} d_a^2}$ is the total quantum dimension. A theory is modular (and associated with a TQFT) iff $S$ is unitary. In this case, the $S$-matrix together with the $T$-matrix, $T_{ab} = \theta_a \delta_{ab}$, and the charge conjugation matrix $C_{ab} =\delta_{\bar{a} b}$ obey the modular relations up to a phase
\begin{equation}
(ST)^3 = \Theta C, \quad S^2 = C, \quad C^2 = \openone
\end{equation}
where
\begin{equation}
\Theta = \frac{1}{\mathcal{D}} \sum_{a\in \mathcal{C}} d_a^2 \theta_a
\end{equation}
is a root of unity.

\section{Some Phases of Matter with Algebraically-Protected Topological Degeneracy}

We now discuss two systems which have topological degeneracies,
with all splittings decaying only as power laws in the system size.
It is interesting to note that one example is one-dimensional and the other is
three-dimensional, unlike all of the other examples given in this paper.

Recently, it was shown that a three-dimensional system in
which topological insulating behavior can coexist and compete with
superconductivity could support hedgehog-like defects with Majorana
fermion zero modes \cite{Teo10,Freedman11a}. This system is most definitely not
in a topological phase. The ground state on the $3$-torus in non-degenerate.
There is a particular, highly-symmetric Hamiltonian for which the system
has a U(N)/O(N) symmetry group and Goldstone bosons. Away from this
point, there is a particular preferred ground state. In either case,
hedgehogs interact through a linear confining force and, therefore, are not the low-energy
quasiparticle excitations of the system. However, if one is willing and able to pay
the energy cost associated with keeping the hedgehogs apart, they
have interesting topological properties. When there are $2n$ hedgehogs
at fixed positions, the system has a $2^{n-1}$-dimensional Hilbert space
of degenerate states up to exponentially-small splittings.
This is a three-dimensional system, so there is
no braiding {\it per se}, but the transformations associated with
exchanging hedgehogs and then healing the order parameter
comprise the {\it ribbon permutation group}. This group is represented
only projectively on the Hilbert space of the theory. The unitary transformations
which result are precisely the same (up to phases) as the unitary representations
of the braid group for Ising anyons \cite{Teo10,Freedman11a}.

To get a system in which these hedgehogs are quasiparticles of
the system, we could increase quantum fluctuations of the order
parameter until its stiffness disappears at a quantum phase transition
into a new phase \cite{Freedman11b}. In this phase, hedgehogs are deconfined.
However, there is an emergent gauge field which has gapless
fluctuations. This gauge field mediates a $1/r$ potential between
hedgehogs, similar to the photon-mediated interaction between
charged quasiparticles in the FQH systems. In addition, there are gapless
fermionic excitations called Hopfions.

This phase can be described most transparently in terms of a
$(4+1)d$ effective field theory. In this theory, there is an SU(2) fermion doublet
$c_i$ (where $i$ is a lattice point) in $4+1$-dimensions
which fills a band with second Chern number equal to 1.
The system is assumed to be a slab of infinite extent in three spatial dimensions
$x,y,z$ and finite extent in the fourth dimension $0<u<1$. At the two three-dimensional boundaries of the slab,
there are gapless fermionic excitations. The fermions interact with
an SU(2) gauge field $a_{i\hat{\alpha}}$ which lives on the links of the lattice,
e.g. between lattice point $i$ and $i+\hat{\alpha}$. Finally, there is a Higgs field
$\vec{\Phi}$ which transforms in the triplet representation of SU(2).
The lattice model for the fermions has the form:
\begin{multline}
S'_{4+1}=\sum_i\int d\tau\left[c_i^\dagger\left(\partial_\tau-ia_0 \right)c_i+
(M+4B)c_i^\dagger\Gamma^0c_i\right.\\
\left.-\sum_{\hat{\alpha}=1}^4 \left(c_i^\dagger\frac{i\Gamma^\alpha+B\Gamma^0}{2}e^{ia_{i\hat{\alpha}}}c_{i+\hat{\alpha}}+h.c.\right)\right.\\
+\left.\left(m \Phi_k c_i^\dagger\left(\mathcal{T}\otimes \tau_y\tau_k\right){c_i^\dagger}^T+h.c.\right)\right]+S_{\rm M}[a_\mu]+S_{\rm \sigma}[\Phi]
\label{LatticeAction}
\end{multline}
Here, $c_i$ is the fermion annihilation operator. The $\tau_k$ are Pauli matrices.
The coupling constant $m$ determines the strength of the interaction between the
fermions and the order parameter. The parameters $M$ and $B$ determine the band structure
of the fermions. So long as they are in the range $-2<M/B<0$, the Chern number is equal to $1$.
The Hermitian Dirac $\Gamma$ matrices $\Gamma^{0,1,..,4}$ satisfy
${\Gamma^a}^\dagger=\Gamma^a$ and $\left\{\Gamma^a,\Gamma^b\right\}=\delta^{ab}$.
$S_{\rm M}[a_\mu]$ is the Maxwell term for $a_\mu$ and $S_{\sigma}[\Phi]$ is the lattice version
of the sigma-model action $\frac1{2g}D_\mu \vec{\Phi}\cdot D^\mu \vec{\Phi}$.
$\mathcal{T}$ is the time-reversal matrix analogous to $\sigma_y$ for the two-component fermion.
We assume that the Higgs field $\vec{\Phi}$ condenses for $u<1/2$ while
$\langle \vec{\Phi} \rangle = 0$ for $u>1/2$. Then,
we have a gapped system at the $u=0$ boundary with magnetic monopoles
which interact through a U(1) gauge field (which is the unbroken subgroup when
$\vec{\Phi}$ condenses). These magnetic monopoles support Majorana fermion
zero modes. At the $u=1$ boundary, on the other hand,
there is an SU(2) doublet of gapless Weyl fermions.

These gapless fermions coexist with localized Majorana zero
modes because, as a result of the pseudo-relativistic spectrum,
their coupling to Majorana zero modes is irrelevant. The scaling dimension
of this coupling is $-1/2$, which reflects the
limited phase space for a Majorana zero mode to decay into
the bulk. As a result, the splitting between the two states of a pair
of hedgehogs a distance $r$ apart decays as $1/r^2$ at $T=0$.
In a finite-sized system, we can have topological degeneracy,
up to $O(1/{r^2})$ splittings, while gapless gauge field excitations and
fermions have energies $O(1/L)$. Therefore, exchanges which occur
faster than $O(1/{r^2})$ but slower than $O(1/L)$ are in the adiabatic regime.

Such a system already has gapless Hopfions in the bulk, so connecting it to metallic
leads would seem, at first glance, to be unimportant. However, a metallic
lead will have a finite density-of-states at zero energy, and therefore
the phase space restrictions which save the Majorana zero modes from decaying into
Hopfions will not save it from coupling to a metal. However, there is, once again,
a charge gap protecting the system against electron tunneling. This state of matter
is an insulator with a charge gap. While the Fermi statistics of an electron can
be accommodated at zero energy by a Majorana zero mode, its charge
will cost a finite energy, so such processes cannot occur at low energies.

Therefore, this is an example of a phase in which the splitting
between ground state decays as a power of the quasiparticle separation
($O(1/{r^2})$ in this case) and the splitting between the ground states and the excited states
decays as a smaller power of the system size ($O(1/{L})$ in this case).

One can imagine such a phase in a three-dimensional wire
network \cite{Alicea11,Halperin12} in which the wires in the network can fluctuate
quantum-mechanically between the topological and non-topological
phases. In such a system, as described in Section
\ref{sec:weak}, fine-tuning is not required because an electron
cannot tunnel into the system at low-energy. In this sense, qubits in such a phase
protect quantum information against more types of perturbations
than superconductivity-based Ising anyons do. However, there are
gapless fermions in the bulk, so error rates only go to zero as a power
of the temperature or inverse system size. (Furthermore, there aren't
any known protocols for performing a $\pi/8$ phase gate in such a system.)

We now discuss a spin-orbit coupled nanowire in contact with an
algebraically-ordered superconducting wire \cite{Fidkowski11}. Such a system has
Majorana zero modes but it it gapless and the Majorana zero modes
have finite-size splittings which decays as a power-law in the system size.
The simplest setup exhibiting the key physics is one with
two semiconducting nanowires in contact with the same
power-law-ordered superconducting wire. The semiconducting wires
have strong spin-orbit coupling, and a magnetic field is applied along the
wires so that there is only a single branch of excitations in each direction.
The superconducting wire has length $L$ and each semiconducting wire has length $\ell$.
We introduce a coordinate $-L/2\leq x \leq L/2$ along the superconducting wire.
One semiconducting wire lies along the $-L/2\leq x \leq -L/2+\ell$ part of the superconducting wire
and the other one lies along the $L/2 - \ell \leq x \leq L/2$ part of the superconducting wire.
The superconducting and semiconducting wires are coupled via electron tunneling
and short-ranged density-density interactions. The former are suppressed by the gap
in the superconducting wire, so that only pair tunneling is important at low energies
and temperatures. The effective Hamiltonian is:
\begin{multline}
\label{eq:2wires}
H_{\rm 2 \, wires} =\int_{-L/2}^{-L/2+\ell} dx \biggl(
  \frac{v_1}{2\pi}\left[{K_1} (\partial_x \theta_1)^2 +
K_1^{-1} (\partial_x \phi_1)^2 \right]\\
- \frac{\Delta_{P1}}{(2\pi a)}\,\sin(\sqrt{2}\theta_\rho
- 2{\theta_1})
\biggr)\\
+ \int_{L/2-\ell}^{L/2} dx \biggl(\frac{v_2}{2\pi}\left[{K_2} (\partial_x \theta_1)^2 +
K_2^{-1} (\partial_x \phi_1)^2 \right]\\
- \frac{\Delta_{P2}}{(2\pi a)}\,\sin(\sqrt{2}\theta_\rho
- 2{\theta_2})\biggr)\\
+\int_{-L/2}^{L/2} dx \,
 \frac{v_\rho}{2\pi}\left[{K_\rho} (\partial_x \theta_\rho)^2 +
K_\rho^{-1} (\partial_x \phi_\rho)^2 \right] \\
\end{multline}
Here, $K_1$, $K_2$, and $K_\rho$ are Luttinger parameters for the
two semiconducting wires and the superconducting wire. The semiconducting
wires are effectively spinless Luttinger liquids, due to spin-orbit coupling
and the magnetic field, and are described by the bosonic fields
${\theta_1}, {\phi_1}$ and ${\theta_2}, {\phi_2}$.
The superconducting wire has a spin gap and is described the bosonic fields
${\theta_\rho}, {\phi_\rho}$  for its charge mode.

This model has Majorana zero modes at $x = \pm L/2$ and $x=\pm L/2-\ell$,
which correspond to the different minima of the sinusoidal potentials
in Eq. (\ref{eq:2wires}), as discussed in Ref. \onlinecite{Fidkowski11}.
As is further discussed there, impurities in the region $-L/2 + \ell \leq x \leq L/2 - \ell$
of the superconducting wire cause backscattering/phase slip processes
which split the ground states. A single impurity of strength $v$ causes a splitting
\begin{equation}
\label{eqn:impurity-splitting}
\Delta E \propto \bigl\langle v\cos\bigl(\sqrt{2}\phi_\rho\bigr) \bigr\rangle
\propto \frac{|v|}{L^{{K_\rho}/2}}
\end{equation}
while a random distribution of impurities of variance $W$ causes a splitting
\begin{equation}
\label{eqn:random-splitting}
\Delta E \propto  \frac{W}{L^{{K_\rho}-2}}
\end{equation}

Therefore, this is a system in which the nearly degenerate ground states have
a splitting which decays as a power-law $1/{L^{{K_\rho}/2}}$ or ${L^{{K_\rho}-2}}$
in system size. Meanwhile, the overall total charge mode of the system
is a gapless relativistic free field in the thermodynamic limit, so the bulk
gap in a finite-sized system is $\propto 1/L$. This is, therefore, another system
in which gapless excitations coexist with a power-law split topological degeneracy.

Unlike in the previous 3D example, however, this system is not impervious
to coupling to gapless excitations. If a metallic lead were coupled
to the end of such a wire, then there would be a term in the effective Lagrangian
(without loss of generality, we have coupled the lead to the $x=-L/2$ point),
\begin{equation}
H_{\rm lead} = \Gamma {\Psi^\dagger}(0) e^{i{\theta_1}(0)\sqrt{2}} + \text{h.c.}
\end{equation}
where $\Psi(x)$ is the electron annihilation operator in the lead.
Since ${\theta_1} = ({\theta_+}+{\theta_-})/2$ and $\theta_-$ is fixed to
be at one of the minima of the sinusoidal potential in Eq. (\ref{eq:2wires}),
the scaling dimension of $\Gamma$ is determined by the scaling dimension
of the total charge mode ${\theta_+}$. If the Luttinger parameters in the semiconducting
and superconducting wires satisfy ${K_1}={K_2}=2{K_\rho}$ (which can accommodate
the case in which the semiconducting nanowire has repulsive interactions while the
superconducting nanowire has attractive interactions), then $e^{i{\theta_+}/\sqrt{2}}$
has scaling dimension $1/(4{K_\rho})$. Consequently, ${\Gamma^2}$ has scaling
dimension $1 - 1/(2{K_\rho})$. Since $K_\rho < 1$ if the superconducting wire has
attractive interactions, tunneling into the Majorana zero mode is a
highly relevant perturbation. Therefore, this system
is only stable if fermion parity is enforced.

\bibliography{quasi-topological}

\begin{thebibliography}{104}
\expandafter\ifx\csname natexlab\endcsname\relax\def\natexlab#1{#1}\fi
\expandafter\ifx\csname bibnamefont\endcsname\relax
  \def\bibnamefont#1{#1}\fi
\expandafter\ifx\csname bibfnamefont\endcsname\relax
  \def\bibfnamefont#1{#1}\fi
\expandafter\ifx\csname citenamefont\endcsname\relax
  \def\citenamefont#1{#1}\fi
\expandafter\ifx\csname url\endcsname\relax
  \def\url#1{\texttt{#1}}\fi
\expandafter\ifx\csname urlprefix\endcsname\relax\def\urlprefix{URL }\fi
\providecommand{\bibinfo}[2]{#2}
\providecommand{\eprint}[2][]{\url{#2}}

\bibitem[{\citenamefont{Wen}(1990)}]{Wen90a}
\bibinfo{author}{\bibfnamefont{X.~G.} \bibnamefont{Wen}},
  \bibinfo{journal}{Int. J. Mod. Phys. B} \textbf{\bibinfo{volume}{4}},
  \bibinfo{pages}{239} (\bibinfo{year}{1990}).

\bibitem[{\citenamefont{Leinaas and Myrheim}(1977)}]{Leinaas77}
\bibinfo{author}{\bibfnamefont{J.~M.} \bibnamefont{Leinaas}} \bibnamefont{and}
  \bibinfo{author}{\bibfnamefont{J.}~\bibnamefont{Myrheim}},
  \bibinfo{journal}{Nuovo Cimento B} \textbf{\bibinfo{volume}{37}},
  \bibinfo{pages}{1} (\bibinfo{year}{1977}).

\bibitem[{\citenamefont{Wilczek}(1982)}]{Wilczek82b}
\bibinfo{author}{\bibfnamefont{F.}~\bibnamefont{Wilczek}},
  \bibinfo{journal}{Phys. Rev. Lett.} \textbf{\bibinfo{volume}{49}},
  \bibinfo{pages}{957} (\bibinfo{year}{1982}).

\bibitem[{\citenamefont{Goldin et~al.}(1985)\citenamefont{Goldin, Menikoff, and
  Sharp}}]{Goldin85}
\bibinfo{author}{\bibfnamefont{G.~A.} \bibnamefont{Goldin}},
  \bibinfo{author}{\bibfnamefont{R.}~\bibnamefont{Menikoff}}, \bibnamefont{and}
  \bibinfo{author}{\bibfnamefont{D.~H.} \bibnamefont{Sharp}},
  \bibinfo{journal}{Phys. Rev. Lett.} \textbf{\bibinfo{volume}{54}},
  \bibinfo{pages}{603} (\bibinfo{year}{1985}).

\bibitem[{\citenamefont{Fredenhagen et~al.}(1989)\citenamefont{Fredenhagen,
  Rehren, and Schroer}}]{Fredenhagen89}
\bibinfo{author}{\bibfnamefont{K.}~\bibnamefont{Fredenhagen}},
  \bibinfo{author}{\bibfnamefont{K.~H.} \bibnamefont{Rehren}},
  \bibnamefont{and} \bibinfo{author}{\bibfnamefont{B.}~\bibnamefont{Schroer}},
  \bibinfo{journal}{Commun. Math. Phys.} \textbf{\bibinfo{volume}{125}},
  \bibinfo{pages}{201} (\bibinfo{year}{1989}).

\bibitem[{\citenamefont{Fr\"{o}hlich and Gabbiani}(1990)}]{Froehlich90}
\bibinfo{author}{\bibfnamefont{J.}~\bibnamefont{Fr\"{o}hlich}}
  \bibnamefont{and} \bibinfo{author}{\bibfnamefont{F.}~\bibnamefont{Gabbiani}},
  \bibinfo{journal}{Rev. Math. Phys.} \textbf{\bibinfo{volume}{2}},
  \bibinfo{pages}{251} (\bibinfo{year}{1990}).

\bibitem[{\citenamefont{Kitaev}(2003)}]{Kitaev97}
\bibinfo{author}{\bibfnamefont{A.~Y.} \bibnamefont{Kitaev}},
  \bibinfo{journal}{Annals of Physics} \textbf{\bibinfo{volume}{303}},
  \bibinfo{pages}{2} (\bibinfo{year}{2003}), \eprint{quant-ph/9707021}.

\bibitem[{\citenamefont{Freedman}(1998)}]{Freedman98}
\bibinfo{author}{\bibfnamefont{M.~H.} \bibnamefont{Freedman}},
  \bibinfo{journal}{Proc. Natl. Acad. Sci. USA} \textbf{\bibinfo{volume}{95}},
  \bibinfo{pages}{98} (\bibinfo{year}{1998}).

\bibitem[{\citenamefont{Freedman et~al.}(2003)\citenamefont{Freedman, Kitaev,
  Larsen, and Wang}}]{Freedman03b}
\bibinfo{author}{\bibfnamefont{M.~H.} \bibnamefont{Freedman}},
  \bibinfo{author}{\bibfnamefont{A.}~\bibnamefont{Kitaev}},
  \bibinfo{author}{\bibfnamefont{M.~J.} \bibnamefont{Larsen}},
  \bibnamefont{and} \bibinfo{author}{\bibfnamefont{Z.}~\bibnamefont{Wang}},
  \bibinfo{journal}{Bull. Amer. Math. Soc. (N.S.)}
  \textbf{\bibinfo{volume}{40}}, \bibinfo{pages}{31} (\bibinfo{year}{2003}),
  \eprint{quant-ph/0101025}.

\bibitem[{\citenamefont{Nayak et~al.}(2008)\citenamefont{Nayak, Simon, Stern,
  Freedman, and Sarma}}]{Nayak08}
\bibinfo{author}{\bibfnamefont{C.}~\bibnamefont{Nayak}},
  \bibinfo{author}{\bibfnamefont{S.~H.} \bibnamefont{Simon}},
  \bibinfo{author}{\bibfnamefont{A.}~\bibnamefont{Stern}},
  \bibinfo{author}{\bibfnamefont{M.}~\bibnamefont{Freedman}}, \bibnamefont{and}
  \bibinfo{author}{\bibfnamefont{S.~D.} \bibnamefont{Sarma}},
  \bibinfo{journal}{Rev. Mod. Phys.} \textbf{\bibinfo{volume}{80}},
  \bibinfo{pages}{1083} (\bibinfo{year}{2008}), \eprint{arXiv:0707.1889}.

\bibitem[{\citenamefont{Witten}(1989)}]{Witten89}
\bibinfo{author}{\bibfnamefont{E.}~\bibnamefont{Witten}},
  \bibinfo{journal}{Comm. Math. Phys.} \textbf{\bibinfo{volume}{121}},
  \bibinfo{pages}{351} (\bibinfo{year}{1989}).

\bibitem[{\citenamefont{Laughlin}(1983)}]{Laughlin83}
\bibinfo{author}{\bibfnamefont{R.~B.} \bibnamefont{Laughlin}},
  \bibinfo{journal}{Phys. Rev. Lett.} \textbf{\bibinfo{volume}{50}},
  \bibinfo{pages}{1395} (\bibinfo{year}{1983}).

\bibitem[{\citenamefont{Tao and Wu}(1984)}]{Tao84}
\bibinfo{author}{\bibfnamefont{R.}~\bibnamefont{Tao}} \bibnamefont{and}
  \bibinfo{author}{\bibfnamefont{Y.-S.} \bibnamefont{Wu}},
  \bibinfo{journal}{Phys. Rev. B} \textbf{\bibinfo{volume}{30}},
  \bibinfo{pages}{1097} (\bibinfo{year}{1984}).

\bibitem[{\citenamefont{Bonderson}(2009)}]{Bonderson09b}
\bibinfo{author}{\bibfnamefont{P.}~\bibnamefont{Bonderson}},
  \bibinfo{journal}{Phys. Rev. Lett.} \textbf{\bibinfo{volume}{103}},
  \bibinfo{pages}{110403} (\bibinfo{year}{2009}), \eprint{arXiv:0905.2726}.

\bibitem[{\citenamefont{Kitaev}(2006)}]{Kitaev06a}
\bibinfo{author}{\bibfnamefont{A.~Y.} \bibnamefont{Kitaev}},
  \bibinfo{journal}{Annals of Physics} \textbf{\bibinfo{volume}{321}},
  \bibinfo{pages}{2} (\bibinfo{year}{2006}), \bibinfo{note}{cond-mat/0506438}.

\bibitem[{\citenamefont{Ocneanu}()}]{Ocneanu-unpublished}
\bibinfo{author}{\bibfnamefont{A.}~\bibnamefont{Ocneanu}},
  \bibinfo{note}{unpublished}.

\bibitem[{\citenamefont{Etingof et~al.}(2005)\citenamefont{Etingof, Mikshych,
  and Ostrik}}]{Etingof05}
\bibinfo{author}{\bibfnamefont{P.}~\bibnamefont{Etingof}},
  \bibinfo{author}{\bibfnamefont{D.}~\bibnamefont{Mikshych}}, \bibnamefont{and}
  \bibinfo{author}{\bibfnamefont{V.}~\bibnamefont{Ostrik}},
  \bibinfo{journal}{Ann. Math.} \textbf{\bibinfo{volume}{162}},
  \bibinfo{pages}{581} (\bibinfo{year}{2005}).

\bibitem[{\citenamefont{Klich}(2010)}]{Klich10}
\bibinfo{author}{\bibfnamefont{I.}~\bibnamefont{Klich}},
  \bibinfo{journal}{Annals of Physics} \textbf{\bibinfo{volume}{325}},
  \bibinfo{pages}{2120} (\bibinfo{year}{2010}),
  \bibinfo{note}{arXiv:0912.0945}.

\bibitem[{\citenamefont{{Bravyi} et~al.}(2010)\citenamefont{{Bravyi},
  {Hastings}, and {Michalakis}}}]{Bravyi10}
\bibinfo{author}{\bibfnamefont{S.}~\bibnamefont{{Bravyi}}},
  \bibinfo{author}{\bibfnamefont{M.~B.} \bibnamefont{{Hastings}}},
  \bibnamefont{and}
  \bibinfo{author}{\bibfnamefont{S.}~\bibnamefont{{Michalakis}}},
  \bibinfo{journal}{J. Math. Phys.} \textbf{\bibinfo{volume}{51}},
  \bibinfo{pages}{093512} (\bibinfo{year}{2010}), \eprint{arXiv:1001.0344}.

\bibitem[{\citenamefont{{Bravyi} and {Hastings}}(2011)}]{Bravyi11}
\bibinfo{author}{\bibfnamefont{S.}~\bibnamefont{{Bravyi}}} \bibnamefont{and}
  \bibinfo{author}{\bibfnamefont{M.~B.} \bibnamefont{{Hastings}}},
  \bibinfo{journal}{Comm. Math. Phys.} \textbf{\bibinfo{volume}{307}},
  \bibinfo{pages}{609} (\bibinfo{year}{2011}), \eprint{arXiv:1001.4363}.

\bibitem[{\citenamefont{{Nussinov} and {Ortiz}}(2008)}]{Nussinov08}
\bibinfo{author}{\bibfnamefont{Z.}~\bibnamefont{{Nussinov}}} \bibnamefont{and}
  \bibinfo{author}{\bibfnamefont{G.}~\bibnamefont{{Ortiz}}},
  \bibinfo{journal}{\prb} \textbf{\bibinfo{volume}{77}}, \bibinfo{eid}{064302}
  (\bibinfo{year}{2008}), \eprint{arXiv:0709.2717}.

\bibitem[{\citenamefont{{Goldstein} and {Chamon}}(2011)}]{Goldstein11}
\bibinfo{author}{\bibfnamefont{G.}~\bibnamefont{{Goldstein}}} \bibnamefont{and}
  \bibinfo{author}{\bibfnamefont{C.}~\bibnamefont{{Chamon}}},
  \bibinfo{journal}{\prb} \textbf{\bibinfo{volume}{84}}, \bibinfo{eid}{205109}
  (\bibinfo{year}{2011}), \eprint{arXiv:1107.0288}.

\bibitem[{\citenamefont{{Budich} et~al.}(2012)\citenamefont{{Budich}, {Walter},
  and {Trauzettel}}}]{Budich12}
\bibinfo{author}{\bibfnamefont{J.~C.} \bibnamefont{{Budich}}},
  \bibinfo{author}{\bibfnamefont{S.}~\bibnamefont{{Walter}}}, \bibnamefont{and}
  \bibinfo{author}{\bibfnamefont{B.}~\bibnamefont{{Trauzettel}}},
  \bibinfo{journal}{\prb} \textbf{\bibinfo{volume}{85}}, \bibinfo{eid}{121405}
  (\bibinfo{year}{2012}), \eprint{arXiv:1111.1734}.

\bibitem[{\citenamefont{{Cheng} et~al.}(2012)\citenamefont{{Cheng}, {Lutchyn},
  and {Das Sarma}}}]{Cheng11}
\bibinfo{author}{\bibfnamefont{M.}~\bibnamefont{{Cheng}}},
  \bibinfo{author}{\bibfnamefont{R.~M.} \bibnamefont{{Lutchyn}}},
  \bibnamefont{and} \bibinfo{author}{\bibfnamefont{S.}~\bibnamefont{{Das
  Sarma}}}, \bibinfo{journal}{\prb} \textbf{\bibinfo{volume}{85}},
  \bibinfo{pages}{165124} (\bibinfo{year}{2012}), \eprint{arXiv:1112.3662}.

\bibitem[{\citenamefont{{Yao} and {Kivelson}}(2007)}]{Yao07}
\bibinfo{author}{\bibfnamefont{H.}~\bibnamefont{{Yao}}} \bibnamefont{and}
  \bibinfo{author}{\bibfnamefont{S.~A.} \bibnamefont{{Kivelson}}},
  \bibinfo{journal}{Physical Review Letters} \textbf{\bibinfo{volume}{99}},
  \bibinfo{pages}{247203} (\bibinfo{year}{2007}), \eprint{arXiv:0708.0040}.

\bibitem[{\citenamefont{Levin and Wen}(2005)}]{Levin05a}
\bibinfo{author}{\bibfnamefont{M.~A.} \bibnamefont{Levin}} \bibnamefont{and}
  \bibinfo{author}{\bibfnamefont{X.-G.} \bibnamefont{Wen}},
  \bibinfo{journal}{Phys. Rev. B} \textbf{\bibinfo{volume}{71}},
  \bibinfo{pages}{045110} (\bibinfo{year}{2005}), \eprint{cond-mat/0404617}.

\bibitem[{\citenamefont{Abrahams et~al.}(1979)\citenamefont{Abrahams, Anderson,
  Licciardello, and Ramakrishnan}}]{Abrahams79}
\bibinfo{author}{\bibfnamefont{E.}~\bibnamefont{Abrahams}},
  \bibinfo{author}{\bibfnamefont{P.~W.} \bibnamefont{Anderson}},
  \bibinfo{author}{\bibfnamefont{D.~C.} \bibnamefont{Licciardello}},
  \bibnamefont{and} \bibinfo{author}{\bibfnamefont{T.~V.}
  \bibnamefont{Ramakrishnan}}, \bibinfo{journal}{Phys. Rev. Lett.}
  \textbf{\bibinfo{volume}{42}}, \bibinfo{pages}{673} (\bibinfo{year}{1979}).

\bibitem[{\citenamefont{{Haah}}(2011)}]{Haah11}
\bibinfo{author}{\bibfnamefont{J.}~\bibnamefont{{Haah}}},
  \bibinfo{journal}{\pra} \textbf{\bibinfo{volume}{83}},
  \bibinfo{pages}{042330} (\bibinfo{year}{2011}), \eprint{arXiv:1101.1962}.

\bibitem[{\citenamefont{{Hastings}}(2011)}]{Hastings11}
\bibinfo{author}{\bibfnamefont{M.~B.} \bibnamefont{{Hastings}}},
  \bibinfo{journal}{Phys. Rev. Lett.} \textbf{\bibinfo{volume}{107}},
  \bibinfo{pages}{210501} (\bibinfo{year}{2011}), \eprint{arXiv:1106.6026}.

\bibitem[{\citenamefont{Dijkgraaf and Witten}(1990)}]{Dijkgraaf90}
\bibinfo{author}{\bibfnamefont{R.}~\bibnamefont{Dijkgraaf}} \bibnamefont{and}
  \bibinfo{author}{\bibfnamefont{E.}~\bibnamefont{Witten}},
  \bibinfo{journal}{Commun. Math. Phys.} \textbf{\bibinfo{volume}{129}},
  \bibinfo{pages}{393} (\bibinfo{year}{1990}).

\bibitem[{\citenamefont{Bonderson and Nayak}()}]{Bonderson13}
\bibinfo{author}{\bibfnamefont{P.}~\bibnamefont{Bonderson}} \bibnamefont{and}
  \bibinfo{author}{\bibfnamefont{C.}~\bibnamefont{Nayak}}, \bibinfo{note}{in
  preparation}.

\bibitem[{\citenamefont{{Burnell} and {Nayak}}(2011)}]{Burnell11}
\bibinfo{author}{\bibfnamefont{F.~J.} \bibnamefont{{Burnell}}}
  \bibnamefont{and} \bibinfo{author}{\bibfnamefont{C.}~\bibnamefont{{Nayak}}},
  \bibinfo{journal}{\prb} \textbf{\bibinfo{volume}{84}},
  \bibinfo{pages}{125125} (\bibinfo{year}{2011}), \eprint{arXiv:1104.5485}.

\bibitem[{\citenamefont{{You} et~al.}(2012)\citenamefont{{You}, {Kimchi}, and
  {Vishwanath}}}]{You11}
\bibinfo{author}{\bibfnamefont{Y.-Z.} \bibnamefont{{You}}},
  \bibinfo{author}{\bibfnamefont{I.}~\bibnamefont{{Kimchi}}}, \bibnamefont{and}
  \bibinfo{author}{\bibfnamefont{A.}~\bibnamefont{{Vishwanath}}},
  \bibinfo{journal}{Phys. Rev. B} \textbf{\bibinfo{volume}{86}},
  \bibinfo{pages}{085145} (\bibinfo{year}{2012}), \eprint{arXiv:1109.4155}.

\bibitem[{\citenamefont{{Hyart} et~al.}(2012)\citenamefont{{Hyart}, {Wright},
  {Khaliullin}, and {Rosenow}}}]{Hyart11}
\bibinfo{author}{\bibfnamefont{T.}~\bibnamefont{{Hyart}}},
  \bibinfo{author}{\bibfnamefont{A.~R.} \bibnamefont{{Wright}}},
  \bibinfo{author}{\bibfnamefont{G.}~\bibnamefont{{Khaliullin}}},
  \bibnamefont{and}
  \bibinfo{author}{\bibfnamefont{B.}~\bibnamefont{{Rosenow}}},
  \bibinfo{journal}{Phys. Rev. B} \textbf{\bibinfo{volume}{85}},
  \bibinfo{pages}{140510} (\bibinfo{year}{2012}), \eprint{arXiv:1109.6681}.

\bibitem[{\citenamefont{{Feng} et~al.}(2007)\citenamefont{{Feng}, {Zhang}, and
  {Xiang}}}]{Feng07}
\bibinfo{author}{\bibfnamefont{X.-Y.} \bibnamefont{{Feng}}},
  \bibinfo{author}{\bibfnamefont{G.-M.} \bibnamefont{{Zhang}}},
  \bibnamefont{and} \bibinfo{author}{\bibfnamefont{T.}~\bibnamefont{{Xiang}}},
  \bibinfo{journal}{Phys. Rev. Lett.} \textbf{\bibinfo{volume}{98}},
  \bibinfo{pages}{087204} (\bibinfo{year}{2007}), \eprint{cond-mat/0610626}.

\bibitem[{\citenamefont{Wilkin et~al.}(1998)\citenamefont{Wilkin, Gunn, and
  Smith}}]{Wilkin98}
\bibinfo{author}{\bibfnamefont{N.~K.} \bibnamefont{Wilkin}},
  \bibinfo{author}{\bibfnamefont{J.~M.~F.} \bibnamefont{Gunn}},
  \bibnamefont{and} \bibinfo{author}{\bibfnamefont{R.~A.} \bibnamefont{Smith}},
  \bibinfo{journal}{Phys. Rev. Lett.} \textbf{\bibinfo{volume}{80}},
  \bibinfo{pages}{2265} (\bibinfo{year}{1998}), \eprint{cond-mat/9705050}.

\bibitem[{\citenamefont{Cooper et~al.}(2001)\citenamefont{Cooper, Wilkin, and
  Gunn}}]{Cooper01}
\bibinfo{author}{\bibfnamefont{N.~R.} \bibnamefont{Cooper}},
  \bibinfo{author}{\bibfnamefont{N.~K.} \bibnamefont{Wilkin}},
  \bibnamefont{and} \bibinfo{author}{\bibfnamefont{J.~M.~F.}
  \bibnamefont{Gunn}}, \bibinfo{journal}{Phys. Rev. Lett.}
  \textbf{\bibinfo{volume}{87}}, \bibinfo{pages}{120405}
  (\bibinfo{year}{2001}), \eprint{cond-mat/0107005}.

\bibitem[{\citenamefont{Jim\'enez-Garc\'ia
  et~al.}(2012)\citenamefont{Jim\'enez-Garc\'ia, LeBlanc, Williams, Beeler,
  Perry, and Spielman}}]{Jimenez-Garcia12}
\bibinfo{author}{\bibfnamefont{K.}~\bibnamefont{Jim\'enez-Garc\'ia}},
  \bibinfo{author}{\bibfnamefont{L.~J.} \bibnamefont{LeBlanc}},
  \bibinfo{author}{\bibfnamefont{R.~A.} \bibnamefont{Williams}},
  \bibinfo{author}{\bibfnamefont{M.~C.} \bibnamefont{Beeler}},
  \bibinfo{author}{\bibfnamefont{A.~R.} \bibnamefont{Perry}}, \bibnamefont{and}
  \bibinfo{author}{\bibfnamefont{I.~B.} \bibnamefont{Spielman}},
  \bibinfo{journal}{Phys. Rev. Lett.} \textbf{\bibinfo{volume}{108}},
  \bibinfo{pages}{225303} (\bibinfo{year}{2012}), \eprint{arXiv:1201.6630}.

\bibitem[{\citenamefont{Tsui et~al.}(1982)\citenamefont{Tsui, Stormer, and
  Gossard}}]{Tsui82}
\bibinfo{author}{\bibfnamefont{D.~C.} \bibnamefont{Tsui}},
  \bibinfo{author}{\bibfnamefont{H.~L.} \bibnamefont{Stormer}},
  \bibnamefont{and} \bibinfo{author}{\bibfnamefont{A.~C.}
  \bibnamefont{Gossard}}, \bibinfo{journal}{Phys. Rev. Lett.}
  \textbf{\bibinfo{volume}{48}}, \bibinfo{pages}{1559} (\bibinfo{year}{1982}).

\bibitem[{\citenamefont{Girvin and MacDonald}(1987)}]{Girvin87}
\bibinfo{author}{\bibfnamefont{S.~M.} \bibnamefont{Girvin}} \bibnamefont{and}
  \bibinfo{author}{\bibfnamefont{A.~H.} \bibnamefont{MacDonald}},
  \bibinfo{journal}{Phys. Rev. Lett.} \textbf{\bibinfo{volume}{58}},
  \bibinfo{pages}{1252} (\bibinfo{year}{1987}).

\bibitem[{\citenamefont{Read}(1989)}]{Read89a}
\bibinfo{author}{\bibfnamefont{N.}~\bibnamefont{Read}}, \bibinfo{journal}{Phys.
  Rev. Lett.} \textbf{\bibinfo{volume}{62}}, \bibinfo{pages}{86}
  (\bibinfo{year}{1989}).

\bibitem[{\citenamefont{Zhang et~al.}(1989)\citenamefont{Zhang, Hansson, and
  Kivelson}}]{Zhang89}
\bibinfo{author}{\bibfnamefont{S.~C.} \bibnamefont{Zhang}},
  \bibinfo{author}{\bibfnamefont{T.~H.} \bibnamefont{Hansson}},
  \bibnamefont{and} \bibinfo{author}{\bibfnamefont{S.}~\bibnamefont{Kivelson}},
  \bibinfo{journal}{Phys. Rev. Lett.} \textbf{\bibinfo{volume}{62}},
  \bibinfo{pages}{82} (\bibinfo{year}{1989}).

\bibitem[{\citenamefont{Wen and Zee}(1992)}]{Wen92a}
\bibinfo{author}{\bibfnamefont{X.~G.} \bibnamefont{Wen}} \bibnamefont{and}
  \bibinfo{author}{\bibfnamefont{A.}~\bibnamefont{Zee}},
  \bibinfo{journal}{Phys. Rev. B} \textbf{\bibinfo{volume}{46}},
  \bibinfo{pages}{2290} (\bibinfo{year}{1992}).

\bibitem[{\citenamefont{Haldane}(1983)}]{Haldane83}
\bibinfo{author}{\bibfnamefont{F.~D.~M.} \bibnamefont{Haldane}},
  \bibinfo{journal}{Phys. Rev. Lett.} \textbf{\bibinfo{volume}{51}},
  \bibinfo{pages}{605} (\bibinfo{year}{1983}).

\bibitem[{\citenamefont{Halperin}(1984)}]{Halperin84}
\bibinfo{author}{\bibfnamefont{B.~I.} \bibnamefont{Halperin}},
  \bibinfo{journal}{Phys. Rev. Lett.} \textbf{\bibinfo{volume}{52}},
  \bibinfo{pages}{1583} (\bibinfo{year}{1984}).

\bibitem[{\citenamefont{Jain}(1989)}]{Jain89}
\bibinfo{author}{\bibfnamefont{J.~K.} \bibnamefont{Jain}},
  \bibinfo{journal}{Phys. Rev. Lett.} \textbf{\bibinfo{volume}{63}},
  \bibinfo{pages}{199} (\bibinfo{year}{1989}).

\bibitem[{\citenamefont{Moore and Read}(1991)}]{Moore91}
\bibinfo{author}{\bibfnamefont{G.}~\bibnamefont{Moore}} \bibnamefont{and}
  \bibinfo{author}{\bibfnamefont{N.}~\bibnamefont{Read}},
  \bibinfo{journal}{Nucl. Phys. B} \textbf{\bibinfo{volume}{360}},
  \bibinfo{pages}{362} (\bibinfo{year}{1991}).

\bibitem[{\citenamefont{Lee et~al.}(2007)\citenamefont{Lee, Ryu, Nayak, and
  Fisher}}]{Lee07}
\bibinfo{author}{\bibfnamefont{S.-S.} \bibnamefont{Lee}},
  \bibinfo{author}{\bibfnamefont{S.}~\bibnamefont{Ryu}},
  \bibinfo{author}{\bibfnamefont{C.}~\bibnamefont{Nayak}}, \bibnamefont{and}
  \bibinfo{author}{\bibfnamefont{M.~P.~A.} \bibnamefont{Fisher}},
  \bibinfo{journal}{Phys. Rev. Lett.} \textbf{\bibinfo{volume}{99}},
  \bibinfo{pages}{236807} (\bibinfo{year}{2007}), \eprint{arXiv:0707.0478}.

\bibitem[{\citenamefont{Levin et~al.}(2007)\citenamefont{Levin, Halperin, and
  Rosenow}}]{Levin07}
\bibinfo{author}{\bibfnamefont{M.}~\bibnamefont{Levin}},
  \bibinfo{author}{\bibfnamefont{B.~I.} \bibnamefont{Halperin}},
  \bibnamefont{and} \bibinfo{author}{\bibfnamefont{B.}~\bibnamefont{Rosenow}},
  \bibinfo{journal}{Phys. Rev. Lett.} \textbf{\bibinfo{volume}{99}},
  \bibinfo{pages}{236806} (\bibinfo{year}{2007}), \eprint{arXiv:0707.0483}.

\bibitem[{\citenamefont{Bonderson and Slingerland}(2008)}]{Bonderson07d}
\bibinfo{author}{\bibfnamefont{P.}~\bibnamefont{Bonderson}} \bibnamefont{and}
  \bibinfo{author}{\bibfnamefont{J.~K.} \bibnamefont{Slingerland}},
  \bibinfo{journal}{Phys. Rev. B} \textbf{\bibinfo{volume}{78}},
  \bibinfo{pages}{067836} (\bibinfo{year}{2008}), \eprint{arXiv:0711.3204}.

\bibitem[{\citenamefont{Bonderson et~al.}(2012)\citenamefont{Bonderson,
  Feiguin, M\"{o}ller, and Slingerland}}]{Bonderson12a}
\bibinfo{author}{\bibfnamefont{P.}~\bibnamefont{Bonderson}},
  \bibinfo{author}{\bibfnamefont{A.~E.} \bibnamefont{Feiguin}},
  \bibinfo{author}{\bibfnamefont{G.}~\bibnamefont{M\"{o}ller}},
  \bibnamefont{and} \bibinfo{author}{\bibfnamefont{J.~K.}
  \bibnamefont{Slingerland}}, \bibinfo{journal}{Phys. Rev. Lett.}
  \textbf{\bibinfo{volume}{108}}, \bibinfo{pages}{036806}
  (\bibinfo{year}{2012}), \eprint{arXiv:0901.4965}.

\bibitem[{\citenamefont{Read and Rezayi}(1999)}]{Read99}
\bibinfo{author}{\bibfnamefont{N.}~\bibnamefont{Read}} \bibnamefont{and}
  \bibinfo{author}{\bibfnamefont{E.}~\bibnamefont{Rezayi}},
  \bibinfo{journal}{Phys. Rev. B} \textbf{\bibinfo{volume}{59}},
  \bibinfo{pages}{8084} (\bibinfo{year}{1999}), \eprint{cond-mat/9809384}.

\bibitem[{\citenamefont{Sondhi et~al.}(1993)\citenamefont{Sondhi, Karlhede,
  Kivelson, and Rezayi}}]{Sondhi93}
\bibinfo{author}{\bibfnamefont{S.~L.} \bibnamefont{Sondhi}},
  \bibinfo{author}{\bibfnamefont{A.}~\bibnamefont{Karlhede}},
  \bibinfo{author}{\bibfnamefont{S.~A.} \bibnamefont{Kivelson}},
  \bibnamefont{and} \bibinfo{author}{\bibfnamefont{E.~H.}
  \bibnamefont{Rezayi}}, \bibinfo{journal}{Phys. Rev. B}
  \textbf{\bibinfo{volume}{47}}, \bibinfo{pages}{16419} (\bibinfo{year}{1993}).

\bibitem[{\citenamefont{Sondhi and Kivelson}(1992)}]{Sondhi92}
\bibinfo{author}{\bibfnamefont{S.~L.} \bibnamefont{Sondhi}} \bibnamefont{and}
  \bibinfo{author}{\bibfnamefont{S.~A.} \bibnamefont{Kivelson}},
  \bibinfo{journal}{Phys. Rev. B} \textbf{\bibinfo{volume}{46}},
  \bibinfo{pages}{13319} (\bibinfo{year}{1992}).

\bibitem[{\citenamefont{Simon}(2008)}]{Simon08}
\bibinfo{author}{\bibfnamefont{S.~H.} \bibnamefont{Simon}},
  \bibinfo{journal}{Phys. Rev. Lett.} \textbf{\bibinfo{volume}{100}},
  \bibinfo{pages}{116803} (\bibinfo{year}{2008}), \eprint{arXiv:0710.1088}.

\bibitem[{\citenamefont{Aharonov and Bohm}(1959)}]{Aharonov59}
\bibinfo{author}{\bibfnamefont{Y.}~\bibnamefont{Aharonov}} \bibnamefont{and}
  \bibinfo{author}{\bibfnamefont{D.}~\bibnamefont{Bohm}},
  \bibinfo{journal}{Phys. Rev.} \textbf{\bibinfo{volume}{115}},
  \bibinfo{pages}{485} (\bibinfo{year}{1959}).

\bibitem[{\citenamefont{Arovas et~al.}(1984)\citenamefont{Arovas, Schrieffer,
  and Wilczek}}]{Arovas84}
\bibinfo{author}{\bibfnamefont{D.}~\bibnamefont{Arovas}},
  \bibinfo{author}{\bibfnamefont{J.~R.} \bibnamefont{Schrieffer}},
  \bibnamefont{and} \bibinfo{author}{\bibfnamefont{F.}~\bibnamefont{Wilczek}},
  \bibinfo{journal}{Phys. Rev. Lett.} \textbf{\bibinfo{volume}{53}},
  \bibinfo{pages}{722} (\bibinfo{year}{1984}).

\bibitem[{\citenamefont{Bonderson et~al.}(2011)\citenamefont{Bonderson,
  Gurarie, and Nayak}}]{Bonderson11b}
\bibinfo{author}{\bibfnamefont{P.}~\bibnamefont{Bonderson}},
  \bibinfo{author}{\bibfnamefont{V.}~\bibnamefont{Gurarie}}, \bibnamefont{and}
  \bibinfo{author}{\bibfnamefont{C.}~\bibnamefont{Nayak}},
  \bibinfo{journal}{Phys. Rev. B} \textbf{\bibinfo{volume}{83}},
  \bibinfo{pages}{075303} (\bibinfo{year}{2011}), \eprint{arXiv:1008.5194}.

\bibitem[{\citenamefont{Bishara and Nayak}(2009)}]{Bishara09b}
\bibinfo{author}{\bibfnamefont{W.}~\bibnamefont{Bishara}} \bibnamefont{and}
  \bibinfo{author}{\bibfnamefont{C.}~\bibnamefont{Nayak}},
  \bibinfo{journal}{Phys. Rev. B} \textbf{\bibinfo{volume}{80}},
  \bibinfo{pages}{155304} (\bibinfo{year}{2009}), \eprint{arXiv:0906.0327}.

\bibitem[{\citenamefont{Rosenow et~al.}(2009)\citenamefont{Rosenow, Halperin,
  Simon, and Stern}}]{Rosenow09}
\bibinfo{author}{\bibfnamefont{B.}~\bibnamefont{Rosenow}},
  \bibinfo{author}{\bibfnamefont{B.~I.} \bibnamefont{Halperin}},
  \bibinfo{author}{\bibfnamefont{S.~H.} \bibnamefont{Simon}}, \bibnamefont{and}
  \bibinfo{author}{\bibfnamefont{A.}~\bibnamefont{Stern}},
  \bibinfo{journal}{Phys. Rev. B} \textbf{\bibinfo{volume}{80}},
  \bibinfo{pages}{155305} (\bibinfo{year}{2009}), \eprint{arXiv:0906.0310}.

\bibitem[{\citenamefont{{Clarke} and {Shtengel}}(2011)}]{Clarke11}
\bibinfo{author}{\bibfnamefont{D.~J.} \bibnamefont{{Clarke}}} \bibnamefont{and}
  \bibinfo{author}{\bibfnamefont{K.}~\bibnamefont{{Shtengel}}},
  \bibinfo{journal}{New Journal of Physics} \textbf{\bibinfo{volume}{13}},
  \bibinfo{pages}{055005} (\bibinfo{year}{2011}), \eprint{arXiv:1102.2016}.

\bibitem[{\citenamefont{Bonderson}(2007)}]{Bonderson07b}
\bibinfo{author}{\bibfnamefont{P.~H.} \bibnamefont{Bonderson}}, Ph.D. thesis,
  \bibinfo{school}{California Institute of Technology} (\bibinfo{year}{2007}).

\bibitem[{\citenamefont{{Volovik}}(1999)}]{Volovik99}
\bibinfo{author}{\bibfnamefont{G.~E.} \bibnamefont{{Volovik}}},
  \bibinfo{journal}{Soviet Journal of Experimental and Theoretical Physics
  Letters} \textbf{\bibinfo{volume}{70}}, \bibinfo{pages}{792}
  (\bibinfo{year}{1999}), \eprint{cond-mat/9911374}.

\bibitem[{\citenamefont{Read and Green}(2000)}]{Read00}
\bibinfo{author}{\bibfnamefont{N.}~\bibnamefont{Read}} \bibnamefont{and}
  \bibinfo{author}{\bibfnamefont{D.}~\bibnamefont{Green}},
  \bibinfo{journal}{Phys. Rev. B} \textbf{\bibinfo{volume}{61}},
  \bibinfo{pages}{10267} (\bibinfo{year}{2000}), \eprint{cond-mat/9906453}.

\bibitem[{\citenamefont{Ivanov}(2001)}]{Ivanov01}
\bibinfo{author}{\bibfnamefont{D.~A.} \bibnamefont{Ivanov}},
  \bibinfo{journal}{Phys. Rev. Lett.} \textbf{\bibinfo{volume}{86}},
  \bibinfo{pages}{268} (\bibinfo{year}{2001}), \eprint{cond-mat/0005069}.

\bibitem[{\citenamefont{{Hansson} et~al.}(2004)\citenamefont{{Hansson},
  {Oganesyan}, and {Sondhi}}}]{Hansson04}
\bibinfo{author}{\bibfnamefont{T.~H.} \bibnamefont{{Hansson}}},
  \bibinfo{author}{\bibfnamefont{V.}~\bibnamefont{{Oganesyan}}},
  \bibnamefont{and} \bibinfo{author}{\bibfnamefont{S.~L.}
  \bibnamefont{{Sondhi}}}, \bibinfo{journal}{Annals of Physics}
  \textbf{\bibinfo{volume}{313}}, \bibinfo{pages}{497} (\bibinfo{year}{2004}),
  \eprint{cond-mat/0404327}.

\bibitem[{\citenamefont{{Chen} et~al.}(2011)\citenamefont{{Chen}, {Liu}, and
  {Wen}}}]{Chen11a}
\bibinfo{author}{\bibfnamefont{X.}~\bibnamefont{{Chen}}},
  \bibinfo{author}{\bibfnamefont{Z.-X.} \bibnamefont{{Liu}}}, \bibnamefont{and}
  \bibinfo{author}{\bibfnamefont{X.-G.} \bibnamefont{{Wen}}},
  \bibinfo{journal}{\prb} \textbf{\bibinfo{volume}{84}},
  \bibinfo{pages}{235141} (\bibinfo{year}{2011}), \eprint{arXiv:1106.4752}.

\bibitem[{\citenamefont{{Chen} et~al.}()\citenamefont{{Chen}, {Gu}, {Liu}, and
  {Wen}}}]{Chen11b}
\bibinfo{author}{\bibfnamefont{X.}~\bibnamefont{{Chen}}},
  \bibinfo{author}{\bibfnamefont{Z.-C.} \bibnamefont{{Gu}}},
  \bibinfo{author}{\bibfnamefont{Z.-X.} \bibnamefont{{Liu}}}, \bibnamefont{and}
  \bibinfo{author}{\bibfnamefont{X.-G.} \bibnamefont{{Wen}}},
  \bibinfo{note}{arXiv:1106.4772}.

\bibitem[{\citenamefont{Kitaev}()}]{Kitaev-unpub}
\bibinfo{author}{\bibfnamefont{A.~Y.} \bibnamefont{Kitaev}},
  \bibinfo{note}{unpublished}.

\bibitem[{\citenamefont{{Lu} and {Vishwanath}}(2012)}]{Lu12}
\bibinfo{author}{\bibfnamefont{Y.-M.} \bibnamefont{{Lu}}} \bibnamefont{and}
  \bibinfo{author}{\bibfnamefont{A.}~\bibnamefont{{Vishwanath}}},
  \bibinfo{journal}{\prb} \textbf{\bibinfo{volume}{86}},
  \bibinfo{pages}{125119} (\bibinfo{year}{2012}), \eprint{arXiv:1205.3156}.

\bibitem[{\citenamefont{Das~Sarma et~al.}(2006)\citenamefont{Das~Sarma, Nayak,
  and Tewari}}]{DasSarma06a}
\bibinfo{author}{\bibfnamefont{S.}~\bibnamefont{Das~Sarma}},
  \bibinfo{author}{\bibfnamefont{C.}~\bibnamefont{Nayak}}, \bibnamefont{and}
  \bibinfo{author}{\bibfnamefont{S.}~\bibnamefont{Tewari}},
  \bibinfo{journal}{Phys. Rev. B} \textbf{\bibinfo{volume}{73}},
  \bibinfo{eid}{220502} (\bibinfo{year}{2006}), \eprint{cond-mat/0510553}.

\bibitem[{\citenamefont{Fu and Kane}(2008)}]{Fu08}
\bibinfo{author}{\bibfnamefont{L.}~\bibnamefont{Fu}} \bibnamefont{and}
  \bibinfo{author}{\bibfnamefont{C.~L.} \bibnamefont{Kane}},
  \bibinfo{journal}{Phys. Rev. Lett.} \textbf{\bibinfo{volume}{100}},
  \bibinfo{pages}{096407} (\bibinfo{year}{2008}), \eprint{arXiv:0707.1692}.

\bibitem[{\citenamefont{{Sau} et~al.}(2010)\citenamefont{{Sau}, {Lutchyn},
  {Tewari}, and {Das Sarma}}}]{Sau10}
\bibinfo{author}{\bibfnamefont{J.~D.} \bibnamefont{{Sau}}},
  \bibinfo{author}{\bibfnamefont{R.~M.} \bibnamefont{{Lutchyn}}},
  \bibinfo{author}{\bibfnamefont{S.}~\bibnamefont{{Tewari}}}, \bibnamefont{and}
  \bibinfo{author}{\bibfnamefont{S.}~\bibnamefont{{Das Sarma}}},
  \bibinfo{journal}{Phys. Rev. Lett.} \textbf{\bibinfo{volume}{104}},
  \bibinfo{pages}{040502} (\bibinfo{year}{2010}), \eprint{arXiv:0907.2239}.

\bibitem[{\citenamefont{{Kitaev}}(2001)}]{Kitaev01}
\bibinfo{author}{\bibfnamefont{A.~Y.} \bibnamefont{{Kitaev}}},
  \bibinfo{journal}{Physics Uspekhi} \textbf{\bibinfo{volume}{44}},
  \bibinfo{pages}{131} (\bibinfo{year}{2001}), \eprint{arXiv:cond-mat/0010440}.

\bibitem[{\citenamefont{{Lutchyn} et~al.}(2010)\citenamefont{{Lutchyn}, {Sau},
  and {Das Sarma}}}]{Lutchyn10}
\bibinfo{author}{\bibfnamefont{R.~M.} \bibnamefont{{Lutchyn}}},
  \bibinfo{author}{\bibfnamefont{J.~D.} \bibnamefont{{Sau}}}, \bibnamefont{and}
  \bibinfo{author}{\bibfnamefont{S.}~\bibnamefont{{Das Sarma}}},
  \bibinfo{journal}{Phys. Rev. Lett.} \textbf{\bibinfo{volume}{105}},
  \bibinfo{pages}{077001} (\bibinfo{year}{2010}), \eprint{arXiv:1002.4033}.

\bibitem[{\citenamefont{{Oreg} et~al.}(2010)\citenamefont{{Oreg}, {Refael}, and
  {von Oppen}}}]{Oreg10}
\bibinfo{author}{\bibfnamefont{Y.}~\bibnamefont{{Oreg}}},
  \bibinfo{author}{\bibfnamefont{G.}~\bibnamefont{{Refael}}}, \bibnamefont{and}
  \bibinfo{author}{\bibfnamefont{F.}~\bibnamefont{{von Oppen}}},
  \bibinfo{journal}{Phys. Rev. Lett.} \textbf{\bibinfo{volume}{105}},
  \bibinfo{pages}{177002} (\bibinfo{year}{2010}), \eprint{arXiv:1003.1145}.

\bibitem[{\citenamefont{{Alicea} et~al.}(2011)\citenamefont{{Alicea}, {Oreg},
  {Refael}, {von Oppen}, and {Fisher}}}]{Alicea11}
\bibinfo{author}{\bibfnamefont{J.}~\bibnamefont{{Alicea}}},
  \bibinfo{author}{\bibfnamefont{Y.}~\bibnamefont{{Oreg}}},
  \bibinfo{author}{\bibfnamefont{G.}~\bibnamefont{{Refael}}},
  \bibinfo{author}{\bibfnamefont{F.}~\bibnamefont{{von Oppen}}},
  \bibnamefont{and} \bibinfo{author}{\bibfnamefont{M.~P.~A.}
  \bibnamefont{{Fisher}}}, \bibinfo{journal}{Nature Physics}
  \textbf{\bibinfo{volume}{7}}, \bibinfo{pages}{412} (\bibinfo{year}{2011}),
  \eprint{arXiv:1006.4395}.

\bibitem[{\citenamefont{Cirac and Zoller}(1995)}]{Cirac95}
\bibinfo{author}{\bibfnamefont{J.~I.} \bibnamefont{Cirac}} \bibnamefont{and}
  \bibinfo{author}{\bibfnamefont{P.}~\bibnamefont{Zoller}},
  \bibinfo{journal}{Phys. Rev. Lett.} \textbf{\bibinfo{volume}{74}},
  \bibinfo{pages}{4091} (\bibinfo{year}{1995}).

\bibitem[{\citenamefont{Nielsen and Chuang}(2000)}]{Nielsen00}
\bibinfo{author}{\bibfnamefont{M.~A.} \bibnamefont{Nielsen}} \bibnamefont{and}
  \bibinfo{author}{\bibfnamefont{I.~L.} \bibnamefont{Chuang}},
  \emph{\bibinfo{title}{Quantum Computation and Quantum Information}}
  (\bibinfo{publisher}{Cambridge University Press},
  \bibinfo{address}{Cambridge}, \bibinfo{year}{2000}).

\bibitem[{\citenamefont{Bouchiat et~al.}(1998)\citenamefont{Bouchiat, Vion,
  Joyez, Esteve, and Devoret}}]{Bouchiat98}
\bibinfo{author}{\bibfnamefont{V.}~\bibnamefont{Bouchiat}},
  \bibinfo{author}{\bibfnamefont{D.}~\bibnamefont{Vion}},
  \bibinfo{author}{\bibfnamefont{P.}~\bibnamefont{Joyez}},
  \bibinfo{author}{\bibfnamefont{D.}~\bibnamefont{Esteve}}, \bibnamefont{and}
  \bibinfo{author}{\bibfnamefont{M.~H.} \bibnamefont{Devoret}},
  \bibinfo{journal}{Phys. Scr.} \textbf{\bibinfo{volume}{T76}},
  \bibinfo{pages}{165} (\bibinfo{year}{1998}).

\bibitem[{\citenamefont{Koch et~al.}(2007)\citenamefont{Koch, Yu, Gambetta,
  Houck, Schuster, Majer, Blais, Devoret, Girvin, and Schoelkopf}}]{Koch07}
\bibinfo{author}{\bibfnamefont{J.}~\bibnamefont{Koch}},
  \bibinfo{author}{\bibfnamefont{T.~M.} \bibnamefont{Yu}},
  \bibinfo{author}{\bibfnamefont{J.}~\bibnamefont{Gambetta}},
  \bibinfo{author}{\bibfnamefont{A.~A.} \bibnamefont{Houck}},
  \bibinfo{author}{\bibfnamefont{D.~I.} \bibnamefont{Schuster}},
  \bibinfo{author}{\bibfnamefont{J.}~\bibnamefont{Majer}},
  \bibinfo{author}{\bibfnamefont{A.}~\bibnamefont{Blais}},
  \bibinfo{author}{\bibfnamefont{M.~H.} \bibnamefont{Devoret}},
  \bibinfo{author}{\bibfnamefont{S.~M.} \bibnamefont{Girvin}},
  \bibnamefont{and} \bibinfo{author}{\bibfnamefont{R.~J.}
  \bibnamefont{Schoelkopf}}, \bibinfo{journal}{Phys. Rev. A}
  \textbf{\bibinfo{volume}{76}}, \bibinfo{pages}{042319}
  (\bibinfo{year}{2007}), \eprint{cond-mat/0703002}.

\bibitem[{\citenamefont{Bonderson
  et~al.}(2008{\natexlab{a}})\citenamefont{Bonderson, Shtengel, and
  Slingerland}}]{Bonderson07c}
\bibinfo{author}{\bibfnamefont{P.}~\bibnamefont{Bonderson}},
  \bibinfo{author}{\bibfnamefont{K.}~\bibnamefont{Shtengel}}, \bibnamefont{and}
  \bibinfo{author}{\bibfnamefont{J.~K.} \bibnamefont{Slingerland}},
  \bibinfo{journal}{Annals of Physics} \textbf{\bibinfo{volume}{323}},
  \bibinfo{pages}{2709} (\bibinfo{year}{2008}{\natexlab{a}}),
  \eprint{arXiv:0707.4206}.

\bibitem[{\citenamefont{Bonderson
  et~al.}(2008{\natexlab{b}})\citenamefont{Bonderson, Freedman, and
  Nayak}}]{Bonderson08a}
\bibinfo{author}{\bibfnamefont{P.}~\bibnamefont{Bonderson}},
  \bibinfo{author}{\bibfnamefont{M.}~\bibnamefont{Freedman}}, \bibnamefont{and}
  \bibinfo{author}{\bibfnamefont{C.}~\bibnamefont{Nayak}},
  \bibinfo{journal}{Phys. Rev. Lett.} \textbf{\bibinfo{volume}{101}},
  \bibinfo{pages}{010501} (\bibinfo{year}{2008}{\natexlab{b}}),
  \eprint{arXiv:0802.0279}.

\bibitem[{\citenamefont{Bonderson et~al.}(2009)\citenamefont{Bonderson,
  Freedman, and Nayak}}]{Bonderson08b}
\bibinfo{author}{\bibfnamefont{P.}~\bibnamefont{Bonderson}},
  \bibinfo{author}{\bibfnamefont{M.}~\bibnamefont{Freedman}}, \bibnamefont{and}
  \bibinfo{author}{\bibfnamefont{C.}~\bibnamefont{Nayak}},
  \bibinfo{journal}{Annals Phys.} \textbf{\bibinfo{volume}{324}},
  \bibinfo{pages}{787} (\bibinfo{year}{2009}), \eprint{arXiv:0808.1933}.

\bibitem[{\citenamefont{Bravyi and Kitaev}()}]{Bravyi00}
\bibinfo{author}{\bibfnamefont{S.~B.} \bibnamefont{Bravyi}} \bibnamefont{and}
  \bibinfo{author}{\bibfnamefont{A.~Y.} \bibnamefont{Kitaev}},
  \bibinfo{note}{unpublished}.

\bibitem[{\citenamefont{Freedman et~al.}(2006)\citenamefont{Freedman, Nayak,
  and Walker}}]{Freedman06}
\bibinfo{author}{\bibfnamefont{M.}~\bibnamefont{Freedman}},
  \bibinfo{author}{\bibfnamefont{C.}~\bibnamefont{Nayak}}, \bibnamefont{and}
  \bibinfo{author}{\bibfnamefont{K.}~\bibnamefont{Walker}},
  \bibinfo{journal}{Phys. Rev. B} \textbf{\bibinfo{volume}{73}},
  \bibinfo{pages}{245307} (\bibinfo{year}{2006}), \eprint{cond-mat/0512066}.

\bibitem[{\citenamefont{Bonderson et~al.}()\citenamefont{Bonderson, Das~Sarma,
  Freedman, and Nayak}}]{Bonderson10}
\bibinfo{author}{\bibfnamefont{P.}~\bibnamefont{Bonderson}},
  \bibinfo{author}{\bibfnamefont{S.}~\bibnamefont{Das~Sarma}},
  \bibinfo{author}{\bibfnamefont{M.}~\bibnamefont{Freedman}}, \bibnamefont{and}
  \bibinfo{author}{\bibfnamefont{C.}~\bibnamefont{Nayak}},
  \bibinfo{note}{arXiv:1003.2856}.

\bibitem[{\citenamefont{Freedman et~al.}(2002)\citenamefont{Freedman, Larsen,
  and Wang}}]{Freedman02b}
\bibinfo{author}{\bibfnamefont{M.~H.} \bibnamefont{Freedman}},
  \bibinfo{author}{\bibfnamefont{M.~J.} \bibnamefont{Larsen}},
  \bibnamefont{and} \bibinfo{author}{\bibfnamefont{Z.}~\bibnamefont{Wang}},
  \bibinfo{journal}{Commun. Math. Phys.} \textbf{\bibinfo{volume}{228}},
  \bibinfo{pages}{177} (\bibinfo{year}{2002}), \eprint{math/0103200}.

\bibitem[{\citenamefont{Bravyi and Kitaev}(2002)}]{Bravyi00b}
\bibinfo{author}{\bibfnamefont{S.}~\bibnamefont{Bravyi}} \bibnamefont{and}
  \bibinfo{author}{\bibfnamefont{A.}~\bibnamefont{Kitaev}},
  \bibinfo{journal}{Annals of Physics} \textbf{\bibinfo{volume}{298}},
  \bibinfo{pages}{210} (\bibinfo{year}{2002}), \eprint{quant-ph/0003137}.

\bibitem[{\citenamefont{Bonderson et~al.}(2010)\citenamefont{Bonderson, Clarke,
  Nayak, and Shtengel}}]{Bonderson10a}
\bibinfo{author}{\bibfnamefont{P.}~\bibnamefont{Bonderson}},
  \bibinfo{author}{\bibfnamefont{D.~J.} \bibnamefont{Clarke}},
  \bibinfo{author}{\bibfnamefont{C.}~\bibnamefont{Nayak}}, \bibnamefont{and}
  \bibinfo{author}{\bibfnamefont{K.}~\bibnamefont{Shtengel}},
  \bibinfo{journal}{Phys. Rev. Lett.} \textbf{\bibinfo{volume}{104}},
  \bibinfo{pages}{180505} (\bibinfo{year}{2010}), \eprint{arXiv:0911.2691}.

\bibitem[{\citenamefont{{Jiang} et~al.}(2011)\citenamefont{{Jiang}, {Kane}, and
  {Preskill}}}]{Jiang11}
\bibinfo{author}{\bibfnamefont{L.}~\bibnamefont{{Jiang}}},
  \bibinfo{author}{\bibfnamefont{C.~L.} \bibnamefont{{Kane}}},
  \bibnamefont{and}
  \bibinfo{author}{\bibfnamefont{J.}~\bibnamefont{{Preskill}}},
  \bibinfo{journal}{Physical Review Letters} \textbf{\bibinfo{volume}{106}},
  \bibinfo{eid}{130504} (\bibinfo{year}{2011}), \eprint{arXiv:1010.5862}.

\bibitem[{\citenamefont{{Bonderson} and {Lutchyn}}(2011)}]{Bonderson11}
\bibinfo{author}{\bibfnamefont{P.}~\bibnamefont{{Bonderson}}} \bibnamefont{and}
  \bibinfo{author}{\bibfnamefont{R.~M.} \bibnamefont{{Lutchyn}}},
  \bibinfo{journal}{Physical Review Letters} \textbf{\bibinfo{volume}{106}},
  \bibinfo{eid}{130505} (\bibinfo{year}{2011}), \eprint{arXiv:1011.1784}.

\bibitem[{\citenamefont{Bravyi and Kitaev}(2005)}]{Bravyi05}
\bibinfo{author}{\bibfnamefont{S.}~\bibnamefont{Bravyi}} \bibnamefont{and}
  \bibinfo{author}{\bibfnamefont{A.}~\bibnamefont{Kitaev}},
  \bibinfo{journal}{Phys. Rev. A} \textbf{\bibinfo{volume}{71}},
  \bibinfo{pages}{022316} (\bibinfo{year}{2005}), \eprint{quant-ph/0403025}.

\bibitem[{\citenamefont{Bishara et~al.}(2008)\citenamefont{Bishara, Fiete, and
  Nayak}}]{Bishara08c}
\bibinfo{author}{\bibfnamefont{W.}~\bibnamefont{Bishara}},
  \bibinfo{author}{\bibfnamefont{G.~A.} \bibnamefont{Fiete}}, \bibnamefont{and}
  \bibinfo{author}{\bibfnamefont{C.}~\bibnamefont{Nayak}},
  \bibinfo{journal}{Phys. Rev. B} \textbf{\bibinfo{volume}{77}},
  \bibinfo{pages}{241306} (\bibinfo{year}{2008}), \eprint{arXiv:0804.1960}.

\bibitem[{\citenamefont{{Fidkowski} et~al.}(2011)\citenamefont{{Fidkowski},
  {Lutchyn}, {Nayak}, and {Fisher}}}]{Fidkowski11}
\bibinfo{author}{\bibfnamefont{L.}~\bibnamefont{{Fidkowski}}},
  \bibinfo{author}{\bibfnamefont{R.~M.} \bibnamefont{{Lutchyn}}},
  \bibinfo{author}{\bibfnamefont{C.}~\bibnamefont{{Nayak}}}, \bibnamefont{and}
  \bibinfo{author}{\bibfnamefont{M.~P.~A.} \bibnamefont{{Fisher}}},
  \bibinfo{journal}{\prb} \textbf{\bibinfo{volume}{84}}, \bibinfo{eid}{195436}
  (\bibinfo{year}{2011}), \eprint{arXiv:1106.2598}.

\bibitem[{\citenamefont{{Hassler} et~al.}(2011)\citenamefont{{Hassler},
  {Akhmerov}, and {Beenakker}}}]{Hassler11}
\bibinfo{author}{\bibfnamefont{F.}~\bibnamefont{{Hassler}}},
  \bibinfo{author}{\bibfnamefont{A.~R.} \bibnamefont{{Akhmerov}}},
  \bibnamefont{and} \bibinfo{author}{\bibfnamefont{C.~W.~J.}
  \bibnamefont{{Beenakker}}}, \bibinfo{journal}{New Journal of Physics}
  \textbf{\bibinfo{volume}{13}}, \bibinfo{pages}{095004}
  (\bibinfo{year}{2011}), \eprint{arXiv:1105.0315}.

\bibitem[{\citenamefont{Mulligan et~al.}(2010)\citenamefont{Mulligan, Nayak,
  and Kachru}}]{Mulligan10}
\bibinfo{author}{\bibfnamefont{M.}~\bibnamefont{Mulligan}},
  \bibinfo{author}{\bibfnamefont{C.}~\bibnamefont{Nayak}}, \bibnamefont{and}
  \bibinfo{author}{\bibfnamefont{S.}~\bibnamefont{Kachru}},
  \bibinfo{journal}{Phys. Rev. B} \textbf{\bibinfo{volume}{82}},
  \bibinfo{pages}{085102} (\bibinfo{year}{2010}), \eprint{arXiv:1004.3570}.

\bibitem[{\citenamefont{Bakalov and Kirillov}(2001)}]{Bakalov01}
\bibinfo{author}{\bibfnamefont{B.}~\bibnamefont{Bakalov}} \bibnamefont{and}
  \bibinfo{author}{\bibfnamefont{A.}~\bibnamefont{Kirillov}},
  \emph{\bibinfo{title}{Lectures on Tensor Categories and Modular Functors}},
  vol.~\bibinfo{volume}{21} of \emph{\bibinfo{series}{University Lecture
  Series}} (\bibinfo{publisher}{American Mathematical Society},
  \bibinfo{year}{2001}).

\bibitem[{\citenamefont{Turaev}(1994)}]{Turaev94}
\bibinfo{author}{\bibfnamefont{V.~G.} \bibnamefont{Turaev}},
  \emph{\bibinfo{title}{Quantum Invariants of Knots and 3-Manifolds}}
  (\bibinfo{publisher}{Walter de Gruyter}, \bibinfo{address}{Berlin, New York},
  \bibinfo{year}{1994}).

\bibitem[{\citenamefont{MacLane}(1963)}]{MacLane63}
\bibinfo{author}{\bibfnamefont{S.}~\bibnamefont{MacLane}},
  \bibinfo{journal}{Rice Univ. Studies} \textbf{\bibinfo{volume}{49}},
  \bibinfo{pages}{28} (\bibinfo{year}{1963}).

\bibitem[{\citenamefont{Teo and Kane}(2010)}]{Teo10}
\bibinfo{author}{\bibfnamefont{J.~C.~Y.} \bibnamefont{Teo}} \bibnamefont{and}
  \bibinfo{author}{\bibfnamefont{C.~L.} \bibnamefont{Kane}},
  \bibinfo{journal}{Phys. Rev. Lett.} \textbf{\bibinfo{volume}{104}},
  \bibinfo{pages}{046401} (\bibinfo{year}{2010}), \eprint{arXiv:0909.4741}.

\bibitem[{\citenamefont{Freedman
  et~al.}(2011{\natexlab{a}})\citenamefont{Freedman, Hastings, Nayak, Qi,
  Walker, and Wang}}]{Freedman11a}
\bibinfo{author}{\bibfnamefont{M.}~\bibnamefont{Freedman}},
  \bibinfo{author}{\bibfnamefont{M.~B.} \bibnamefont{Hastings}},
  \bibinfo{author}{\bibfnamefont{C.}~\bibnamefont{Nayak}},
  \bibinfo{author}{\bibfnamefont{X.-L.} \bibnamefont{Qi}},
  \bibinfo{author}{\bibfnamefont{K.}~\bibnamefont{Walker}}, \bibnamefont{and}
  \bibinfo{author}{\bibfnamefont{Z.}~\bibnamefont{Wang}},
  \bibinfo{journal}{Phys. Rev. B} \textbf{\bibinfo{volume}{83}},
  \bibinfo{pages}{115132} (\bibinfo{year}{2011}{\natexlab{a}}),
  \eprint{arXiv:1005.0583}.

\bibitem[{\citenamefont{Freedman
  et~al.}(2011{\natexlab{b}})\citenamefont{Freedman, Hastings, Nayak, and
  Qi}}]{Freedman11b}
\bibinfo{author}{\bibfnamefont{M.}~\bibnamefont{Freedman}},
  \bibinfo{author}{\bibfnamefont{M.~B.} \bibnamefont{Hastings}},
  \bibinfo{author}{\bibfnamefont{C.}~\bibnamefont{Nayak}}, \bibnamefont{and}
  \bibinfo{author}{\bibfnamefont{X.-L.} \bibnamefont{Qi}},
  \bibinfo{journal}{Phys. Rev. B} \textbf{\bibinfo{volume}{84}},
  \bibinfo{pages}{245119} (\bibinfo{year}{2011}{\natexlab{b}}),
  \eprint{arXiv:1107.2731}.

\bibitem[{\citenamefont{{Halperin} et~al.}(2012)\citenamefont{{Halperin},
  {Oreg}, {Stern}, {Refael}, {Alicea}, and {von Oppen}}}]{Halperin12}
\bibinfo{author}{\bibfnamefont{B.~I.} \bibnamefont{{Halperin}}},
  \bibinfo{author}{\bibfnamefont{Y.}~\bibnamefont{{Oreg}}},
  \bibinfo{author}{\bibfnamefont{A.}~\bibnamefont{{Stern}}},
  \bibinfo{author}{\bibfnamefont{G.}~\bibnamefont{{Refael}}},
  \bibinfo{author}{\bibfnamefont{J.}~\bibnamefont{{Alicea}}}, \bibnamefont{and}
  \bibinfo{author}{\bibfnamefont{F.}~\bibnamefont{{von Oppen}}},
  \bibinfo{journal}{\prb} \textbf{\bibinfo{volume}{85}},
  \bibinfo{pages}{144501} (\bibinfo{year}{2012}), \eprint{arXiv:1112.5333}.

\end{thebibliography}

\end{document}